\documentclass[preprint2]{pp7}

\usepackage{float}
\usepackage{graphicx}
\usepackage{amstext}
\usepackage[fleqn]{amsmath}
\usepackage{amssymb}	
\usepackage{url}
\usepackage{tabularx}
\usepackage{mathtools}
\usepackage{mathrsfs}
\usepackage{color}
\usepackage{graphics}
\usepackage{multirow}
\usepackage{soul}
\usepackage{booktabs} 
\usepackage{colortbl} 
\usepackage{xfrac}
\usepackage{xspace}
\usepackage{verbatim}
\usepackage{longtable}
\usepackage[outercaption]{sidecap}    
\usepackage{rotating}
\usepackage{lastpage} 
\usepackage[english]{babel} 
\usepackage{microtype} 
\usepackage[svgnames]{xcolor}
\usepackage{enumitem} 
\setlist{noitemsep} 
\usepackage{sectsty} 
\usepackage{multicol} 
\usepackage{natbib}
\usepackage{hyperref}
\hypersetup{colorlinks=true,linkcolor=Blue,citecolor=Blue,filecolor=Blue,urlcolor=Blue}

\usepackage[a4paper, total={172mm,252mm}]{geometry} 

\usepackage[switch]{lineno} 

\mathchardef\mhyphen="2D

\newcommand{\lsun}{\ensuremath{\mathrm{L}_\odot}\xspace}
\newcommand{\msun}{\mbox{M$_\odot$}\xspace}
\newcommand{\msunyr}{\mbox{M$_\odot$yr$^{-1}$}}
\newcommand{\msunpc}{\mbox{M$_\odot$pc$^{-2}$}}

\newcommand{\msunyrkpc}{\mbox{M$_\odot$yr$^{-1}$kpc$^{-2}$}}

\newcommand{\yr}{\mbox{${\rm yr}$}}

\newcommand{\ergs}{\mbox{${\rm erg}~{\rm s}^{-1}$}}
\newcommand{\myr}{\mbox{${\rm Myr}$}}
\newcommand{\gyr}{\mbox{${\rm Gyr}$}}
\newcommand{\pc}{\mbox{${\rm pc}$}}

\newcommand{\kpc}{\mbox{${\rm kpc}$}}
\newcommand{\kms}{\mbox{${\rm km}~{\rm s}^{-1}$}\xspace}
\newcommand{\cmc}{\mbox{${\rm cm}^{-3}$}}

\newcommand{\hii}{H{\sc ii}\xspace}
\newcommand{\uchii}{UCH{\sc ii}\xspace}

\newcommand{\be}{\begin{equation}}
\newcommand{\ee}{\end{equation}}

\newcommand{\degree}{\ensuremath{^{\circ}}\xspace}
\renewcommand{\deg}{\degree}
\newcommand{\percc}{\ensuremath{\textrm{cm}^{-3}}\xspace}
\newcommand{\persc}{\ensuremath{\textrm{cm}^{-2}}\xspace}
\newcommand{\peryr}{\mbox{${\rm yr^{-1}}$}}
\newcommand{\um}{\,\textmu m\xspace}

\usepackage{times}
\usepackage{amsmath}

\begin{document}    

\title{\textbf{\LARGE Star formation in the Central Molecular Zone of the Milky Way }}

\author {\textbf{\large Jonathan~D.~Henshaw}}
\affil{\small\em Max Planck Institut f\"{u}r Astronomie, K\"{o}nigstuhl 17, D-69117 Heidelberg, DE}
\affil{\small\em Astrophysics Research Institute, Liverpool John Moores University, 146 Brownlow Hill, Liverpool L3 5RF, UK}
\author {\textbf{\large Ashley~T.~Barnes}}
\affil{\small\em Argelander-Institut f\"{u}r Astronomie, Universit\"{a}t Bonn, Auf dem H\"{u}gel 71, 53121 Bonn, DE}
\author {\textbf{\large Cara Battersby}}
\affil{\small\em Department of Physics, 196 Auditorium Road, Unit 3046, University of Connecticut, Storrs, CT 06269-3046, USA}
\author {\textbf{\large Adam Ginsburg}}
\affil{\small\em Department of Astronomy, University of Florida, Bryant Space Science Center, Gainesville FL 32611, USA}
\author {\textbf{\large Mattia C. Sormani}}
\affil{\small\em Institut f{\"u}r Theoretische Astrophysik, Zentrum f{\"u}r Astronomie, Universit{\"a}t Heidelberg, Albert-Ueberle-Str. 2, 69120 Heidelberg, DE}
\author {\textbf{\large Daniel L. Walker}}
\affil{\small\em Department of Physics, 196 Auditorium Road, Unit 3046, University of Connecticut, Storrs, CT 06269-3046, USA \vspace{1cm}}


\begin{abstract}
\baselineskip = 11pt
\leftskip = 0.65in 
\rightskip = 0.65in
\parindent=1pc
 {\small The Central Molecular Zone (CMZ) is a ring-like accumulation of molecular gas in the innermost few hundred parsecs of the Milky Way, generated by the inward transport of matter driven by the Galactic bar. The CMZ is the most extreme star-forming environment in the Galaxy. The unique combination of large-scale dynamics and extreme interstellar medium conditions, characterised by high densities, temperatures, pressures, turbulent motions, and strong magnetic fields, make the CMZ an ideal region for testing current star and planet formation theories. 
We review the recent observational and theoretical advances in the field, and combine these to draw a comprehensive, multi-scale and multi-physics picture of the cycle of matter and energy in the context of star formation in the closest galactic nucleus.
 \\~\\~\\~}
\end{abstract}  

\section{Introduction}

\begin{figure*}[ht]
    \centering
	\includegraphics[trim=0 0cm 0 0cm, clip, width=0.9\textwidth]{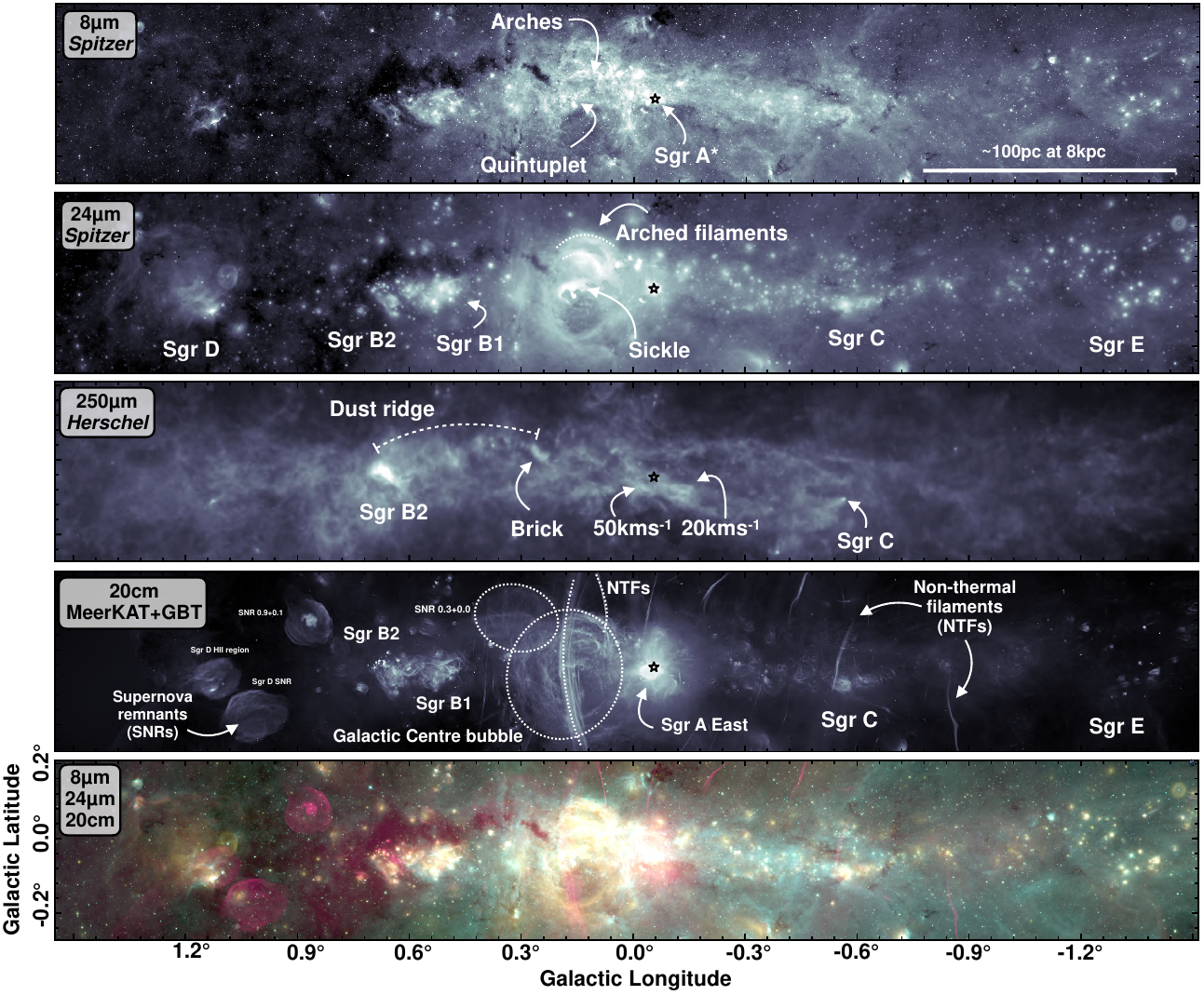}
    \caption{A multi-wavelength view of the CMZ. From top to bottom: 8\micron\ emission from the {\em Spitzer} GLIMPSE survey \citep{Churchwell2009}, 24\micron\ emission from the {\em Spitzer} MIPSGAL survey \citep{Carey2009}, 250\micron\ emission from the {\em Herschel} Hi-GAL survey \citep{Molinari2010}, and 20\,cm emission observed by MeerKAT \citep{Heywood2019, Heywood2022} and the Green Bank Telescope (GBT; \citealp{Law2008}). The bottom panel is a three-colour composite of the 8\micron\ (green), 24\micron\ (yellow) and 20\,cm (red) emission. Overlaid are labels highlighting several features of interest. 
    }
    \label{fig:rgb_main}
\end{figure*}

Understanding the formation process of stars and planets is one of the most prominent unsolved problems in contemporary astrophysics.
In particular, establishing what role, if any, environment plays in controlling important quantities, such as the rate and efficiency of star formation, as well as the properties of formed stars, such as their mass distribution at birth and how they cluster together in space and time, has profound implications for our understanding of how galaxies evolve across cosmic time. 

Much of our detailed knowledge of star and planet formation comes from molecular clouds located within $\sim500$\,pc of the Sun. Because these clouds are nearby, their embedded star formation can be studied in exquisite detail, from the cloud-scale down to the scales of individual protoplanetary discs. As a result, molecular clouds in the Solar neighbourhood have been instrumental in shaping our understanding of the structure of the cold interstellar medium (ISM) and of the process of (low-mass) star formation \citep{Ward-Thompson2007,Andre2014, Padoan2014}. As already documented throughout the Protostars \& Planets series, under the typical conditions found in the Solar neighbourhood and, more generally, the disc of the Milky Way, stars typically form in filaments \citep[][see also Pineda et al. and Hacar et al. this volume]{Andre2014}, and the mass distribution of their embedded core populations (i.e. the core mass function, CMF) and the emergent stellar initial mass function \citep[IMF;][]{Offner2014} show remarkably little variation. However, the narrow range of ISM conditions found in nearby clouds is a limiting factor in the development of a general theory for star formation. 

In the context of Galactic star formation, the Central Molecular Zone (i.e. the region within a Galactocentric radius of $R\simeq300$\,pc, hereafter, the CMZ), is an environment like no other. The CMZ hosts the nearest supermassive black hole, some of the closest and most massive young star clusters, the largest number of supernovae per unit volume, complex dynamics driven by the non-axisymmetric gravitational field of the Galactic bar, and it is the largest concentration of dense molecular gas in the Galaxy (see Fig.~\ref{fig:rgb_main}).
The ISM conditions at the centre of the Galaxy are extreme compared to the Solar neighbourhood. Turbulent motions, magnetic field strengths, gas densities, pressures, and temperatures are orders of magnitude greater than those measured locally. 
Many of these properties are more similar to those found at earlier epochs in the Universe \citep{Kruijssen2013}. 
Beyond the Milky Way, the centres of galaxies play an important role in galaxy evolution. They can contribute anywhere from $\sim$a few \% to $>80$\% of the overall star formation of their host galaxy \citep{Kormendy2004}, and the large-scale outflows launched by either central starbursts or active galactic nuclei (AGN) can drive the evolution of their host via the inside-out cessation (or ``quenching'') of star formation \citep{Veilleux2020}. 

The centre of the Milky Way is currently the only galactic nucleus in which it is possible to resolve the multi-scale physics of star formation down to the scales of protoplanetary discs. 
In a Galactic context, the CMZ is a unique region where current theories of star and planet formation can be bench-marked and tested.
In an extragalactic context, the CMZ is a \emph{rosetta stone}: a template to understand extragalactic nuclei and a potential window into earlier epochs in the Universe. Our review focuses on the recent advances in the field of star formation in the CMZ. It is organised as follows. In \S\ref{sec:IG} we describe the landscape of the inner Galaxy and the dynamical origin of the CMZ. In \S\ref{sec:global}, we review the global picture and evolution of CMZ star formation. \S\ref{sec:macroevolution} covers the macro-evolution of the CMZ, and the balance between inflow, outflow, and the central gas reservoir. In \S\ref{sec:cloudtodisc} and \S\ref{sec:environmentofstarformation}, we zoom in, and review the recent body of work dedicated to studying the population of extreme molecular clouds, and the process of star formation embedded within them. Finally, in \S\ref{sec:summary}, we discuss critical open questions and highlight important avenues for future studies.


\section{The inner Galactic landscape} \label{sec:IG}

The CMZ is not an isolated system. It is in a constant state of flux due to the inflow of matter from the Galactic disc. Understanding how stars and planets form in this environment requires an understanding of the dynamic landscape within which the CMZ is set. We therefore begin our review with a taxonomy of the inner Galaxy, its key components, and the dynamical features that provide essential context for our understanding of star formation in this environment. 

\subsection{Mapping the landscape: dynamical features observed in the cold ISM} \label{sec:cartography}

The centre of the Milky Way has been extensively mapped across the electromagnetic spectrum for over half a century. Observations of H\,{\sc i} in the 1950s and 60s unveiled its complex dynamics, shaping our understanding of the large-scale structure of the Milky Way \citep[e.g.][]{Oort1958,Rougoor1960,deVaucouleurs1964}. Later, observations of CO in the 1970s and 80s, free from the confusion in H{\sc i} due to the near-ubiquity of atomic gas along the line-of-sight, delineated the structure of the molecular gas associated with the CMZ \citep{Bania1977, Liszt1978, Bally1987}. For extensive discussion on the history of these observations we recommend the reviews by \citet{Oort1977}, \cite{Combes1991}, and \citet{Morris1996}, and for a more recent update, \citet{Bryant2021}. In the following sections, we describe several of the key features of the inner Galaxy that are foundational to our understanding of the origin of the CMZ and the star formation occurring within it. We highlight these features in Fig.~\ref{fig:lbv_main}, which depicts the distribution of molecular gas within  $|l|<10^{\circ}$ as it appears on the plane of the sky in longitude-latitude, $\{l,b\}$ space as well as in longitude-velocity, $\{l,v\}$ space.

\subsubsection{The dust lanes}\label{sec:dustlanes}

A key component of the star formation process in the Galactic Centre is the inflow from kiloparsec scales. We will discuss the theoretical understanding of the inflow process in \S\ref{sec:generaldynamics} and quantify the inflow rate towards the centre of the Milky Way in \S\ref{sec:barinflow}. Here, we deal with the observational manifestation of the inflow process in the Milky Way: the inner Galactic dust lanes. 

The dust lanes associated with galactic bars can be clearly seen in images of face-on spiral galaxies \citep{Sandage1961,Knapen2002, Comeron2009, Lee2022}. However, our embedded view through the Galactic plane makes identifying the Milky Way's dust lanes more challenging. The identification of the features believed to correspond to the Milky Way's dust lanes has therefore mostly relied on the emission from spectroscopic data (often H\,{\sc i} and CO observations) viewed in $\{l,v\}$-space, such as the diagrams shown in the lower panels Fig.~\ref{fig:lbv_main} \citep[e.g.][]{Fux1999}. The name ``dust lanes'' has been retained for historical reasons, though they have also been subsequently identified in three-dimensional dust extinction maps \citep{Marshall2008}. The two ``primary'' dust lanes are marked in Fig.~\ref{fig:lbv_main} as the near and far side lanes, respectively, and are further highlighted in the face-on schematic presented in Fig.~\ref{fig:sketch}. The near side dust lane is readily identified in H{\sc i} emission, and was originally dubbed the ``connecting arm'' \citep[][]{Cohen1976, McClure-Griffiths2012}. The vast majority of the emission associated with the near (far) side dust lane is located at positive (negative) longitudes and velocity, and negative (positive) latitudes, respectively, although note that the far side dust lane does extend to positive velocity \citep[][]{Fux1999, Liszt2006, Liszt2008, Rodriguez-Fernandez2006, Sormani2019b}. The detection of the dust lanes in opposite quadrants in $\{l,b\}$ suggests that they lie on a common plane that is tilted with respect to the Galactic plane by $\sim2\mhyphen3^{\circ}$ \citep{Sormani2019b,Tress2020}. A tilt of the inner Galactic gas layer is also observed in the ionised and neutral gas \citep[e.g.][]{Krishnarao2020b}. A secondary family of less prominent and less extended dust lanes that are roughly parallel to the primary dust lanes in the $\{l,v\}$ plane, was identified by \citet{Liszt2008}. Two of these secondary dust lanes are highlighted in Fig.~\ref{fig:lbv_main}. These secondary dust lanes have been interpreted as shorter lane segments parallel to the main dust lanes in real space \citep[Fig.~\ref{fig:sketch};][]{Rodriguez-Fernandez2006,Sormani2019a}. 

\subsubsection{Extended velocity features}\label{sec:EVFs}

Another phenomenon that may be related to the inflow process from kiloparsec scales is a class of object that we will refer to as Extended Velocity Features (EVFs; \citealp{Sormani2019a}). EVFs are immediately identifiable in the $\{l,v\}$ diagram as near-vertical features (Fig.~\ref{fig:lbv_main}). We distinguish EVFs from another class of features known as High Velocity Compact Clouds (HVCCs, not to be confused with HVC or High Velocity Clouds found in the halo of the Milky Way, e.g.~\citealt{Wakker1997}), despite some overlap in the original classification \citep[HVCCs are similar in that many appear as near-vertical emission features in $\{l,v\}$ diagrams;][]{Oka1998b}. HVCCs have been speculated to be associated with phenomena such as supernova explosions \citep{Oka1999} and even with the presence of intermediate-mass black holes \citep[IMBHs;][]{Oka2016, Oka2017, Takekawa2017b, Tokuyama2019}. In contrast, the connection (in $\{l,b,v\}$-space) between EVFs and the dust lane features described above implies a physical connection between the two \citep{Liszt2008}, indicating that EVFs more likely result from larger-scale dynamical processes. The most prominent of the EVFs are highlighted in Fig.~\ref{fig:lbv_main} at $l\approx5.4^{\circ}$ and $l\approx3.2^{\circ}$. This latter feature is also known as Bania's Clump 2 \citep{Bania1977}. It has a total velocity extent of the order $\sim150$\,\kms \ \citep{Stark1986, Longmore2017}. Similar features appear at $l\approx1.3^{\circ}$, $-0.5^{\circ}$, and $-1.0^{\circ}$ \citep{Liszt2006}. Generally, the EVFs are compact in longitude, but can be extended in latitude \citep[in some cases by over 100\,pc;][]{Liszt2006}. Their velocity extents can be as large as 200\,\kms. 

Following their discovery, EVFs were interpreted as streams of clouds falling into the CMZ \citep{Stark1986, Boyce1989, Baba2010}. However, \citet{Stark1986} acknowledged that this explanation relies on the near-perfect line-of-sight alignment of the clouds (the so-called ``fingers of God'' effect). A more recent interpretation, which has emerged thanks to the advancement of numerical simulations that describe the flow of gas in a barred potential (\S\ref{sec:generaldynamics}), is that the EVFs represent the rapid deceleration of gas resulting from collisions between gas travelling along the dust lanes towards the CMZ and either: 1) gas that was formerly travelling along the opposite dust lane but has `overshot' the CMZ; 2) gas that belongs to the CMZ itself \citep{Fux1999, Liszt2006, Liszt2008, Rodriguez-Fernandez2008, Bally2010, Sormani2019a, Akhter2021}. This interpretation naturally explains the apparent association between the EVFs and the inner Galactic dust lanes (\S\ref{sec:dustlanes}), and further explains why no known EVF extends beyond the terminal velocity curve at its observed Galactic longitude \citep{Sormani2019a}. That said, the EVFs have been broadly interpreted.  An alternative view is that the EVFs represent the footprints of ``giant molecular loops'' generated by the Parker instability \citep{Parker1966}. These footprints are speculated to be characterised by large velocity gradients and strong shocks caused by gas that is accelerated along magnetic field lines due to gravity, colliding with the Galactic plane \citep{Fukui2006, Fujishita2009, Machida2009, Torii2010, Riquelme2018, Enokiya2021}. They have also been suggested to arise from collisions between the gas in the CMZ and an expanding ring of molecular gas \citep{Uchida1994,Oka2020}, driven by an explosive event near the centre of the Galaxy (see \S\ref{sec:cmz}). 

\subsubsection{The Central Molecular Zone}\label{sec:cmz}

\begin{figure*}
    \centering
	\includegraphics[trim=0 0cm 0 0cm, clip, width=0.9\textwidth]{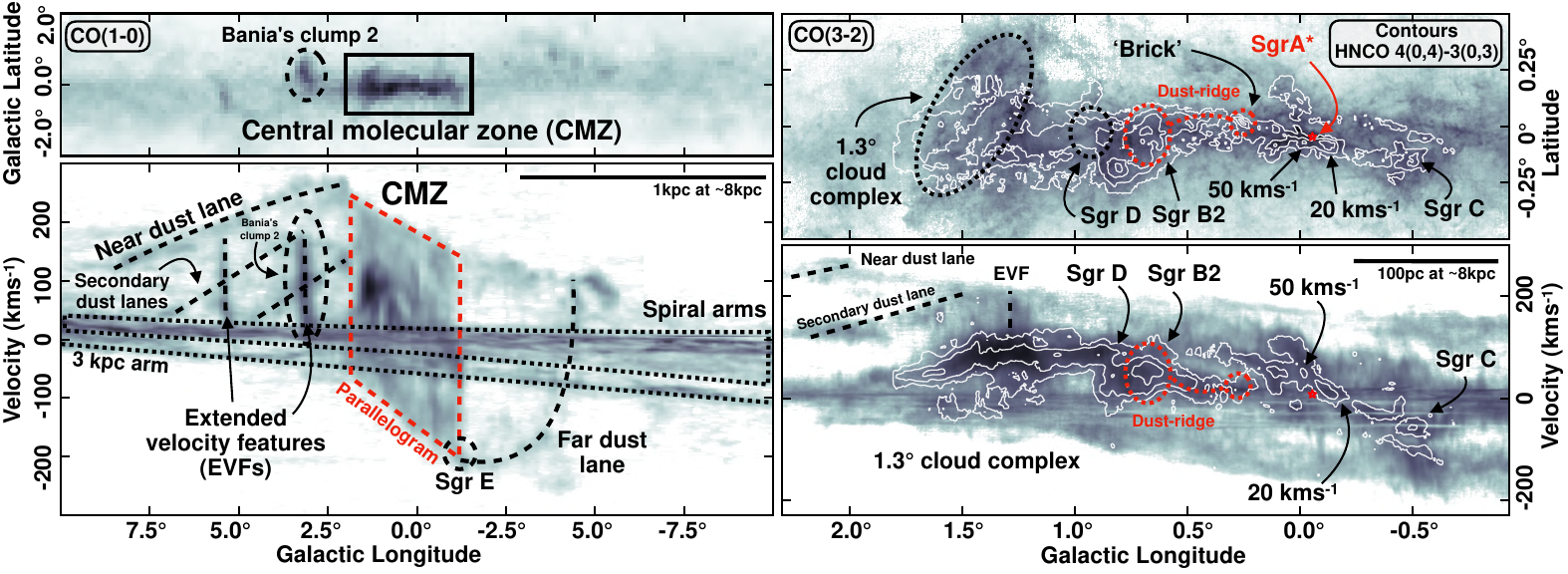}
    \caption{Atlas of the position-position-velocity ($\{l,b,v\}$ space) structure within the centre of the Milky Way. 
    The left panels show the integrated intensity of the $^{12}$CO $J=1-0$ line across the central few degrees ($\sim\,$ few kpc) of the Galaxy \citep{Dame2001}.
    The right panels show the integrated intensity of the $^{12}$CO $J=3-2$ line across the inner degree ($\sim\,$~300\,pc) taken as part of the CHIMPS2 survey \citep{Eden2020}.
    Overlaid as white contours is the HNCO\,$4(0,4)\mhyphen3(0,3)$ emission, a tracer of denser gas, taken as part of the Mopra CMZ survey \citep{Jones2012}.
    Labelled on both sets of panels are the prominent features discussed throughout this review (see in particular \S\ref{sec:IG} for the taxonomy of the inner Galaxy, and \S\ref{sec:massreservoir} for a detailed discussion of structures in the CMZ). The horizontal stripes in the CO emission shown in bottom-right panel are absorption from foreground spiral arms.}
    \label{fig:lbv_main}
\end{figure*}

``The CMZ'' is the concentration of molecular material observed within Galactocentric radius $R\simeq300$\,pc \citep[][]{Morris1996}. In CO emission, the CMZ is brightest over the inner $\sim4^{\circ}$ \citep{Bally1988, Oka1998b, Eden2020}. However, the gas distribution is asymmetric about the Galactic centre, with roughly three-quarters of the emission located at positive Galactic longitudes and velocities (see the right panels of Fig.~\ref{fig:lbv_main} and \S\ref{sec:gasstardistrib}). The total mass of molecular gas within $R\simeq300$\,pc is estimated to be $2\mhyphen6\times10^{7}$\,\msun \citep{Dahmen1998, Ferriere2007}. This corresponds to $3\mhyphen10\%$ of all the molecular gas in the Galaxy \citep[$6.5\times10^{8}$\,\msun; ][]{Roman-Duval2016} despite only accounting for $0.1\%$ of the surface area. 

Historically, the CMZ has been divided into two main components \citep{Morris1996,Fux1999}. Each component is distinguishable by its mass and kinematics, and we describe each of them below. 

\paragraph{The $\{l,v\}$ parallelogram vs. the ``expanding molecular ring''.}\label{subsubsec:180pcring} The first of the two components associated with the CMZ has been broadly interpreted. It is recognisable in $\{l,v\}$-space as a parallelogram-shaped feature (highlighted in Fig.~\ref{fig:lbv_main}). It traces highly non-circular motions between $-1.3^{\circ}\lesssim l \lesssim 2.0^{\circ}$, $b=0.0^{\circ}$ and $|v|\lesssim250$\,\kms. 

The original interpretation of this feature was that it represents a radially expanding ring of molecular gas \citep[sometimes referred to as the expanding molecular ring, EMR, or the ``$180$\,pc expanding ring'';][]{Scoville1972, Kaifu1972, sofue1995b, Oka2020}, although the geometry has evolved over time \citep{Sofue2017a}. It has been speculated that the EMR is driven by some highly energetic and expansive event (or a combination of events) occurring near the centre, namely extensive central star formation activity, many supernovae, or past activity from the central supermassive black hole, Sgr\,A* \citep{Sofue2017a}. 

There are two main persistent criticisms of this interpretation. First, the required energy input is very high. The molecular gas mass of the EMR is $\sim0.5\mhyphen1.0\times10^{7}$\,\msun, which, when combined with the estimated expansion velocity of $160$\,\kms, corresponds to a kinetic energy of $10^{54}\mhyphen10^{55}$\,erg \citep[roughly the equivalent of $10^{3}\mhyphen10^{4}$ supernovae][]{Scoville1972, sofue1995b, Sofue2017a, Oka2019}. Second, while there is evidence for outflows being driven vertically out of the Galactic plane from the central region \citep[][see \S\ref{sec:feedback}]{Bland-Hawthorn2003, Su2010, Heywood2019, Heywood2022}, there is little direct evidence for any in-plane impact of such an event \citep[][]{Combes1991}.

The more widely accepted interpretation of the parallelogram is that it arises in the context of the non-circular motions driven by the Galactic bar (see \S\ref{sec:generaldynamics}). This interpretation is attractive in that it does not need to invoke a large-scale expansive event, nor does it require ad-hoc assumptions since the presence of the Galactic bar is well established. However, the details have significantly evolved over time. \citet{Binney1991} initially interpreted the parallelogram as the result of gas following the self-intersecting ``cusped'' $x_1$ orbit (a type of elongated orbit that exists in bar potentials; \S\ref{sec:generaldynamics}). However, we now know that the Galactic bar is much larger than hypothesised by \cite{Binney1991}, and, as a consequence, the cusped orbit is located at a much larger Galactocentric radius than in their model (\S\ref{sec:potential}). Recent numerical simulations have helped to refine this scenario. The key modification is that the gas associated with the parallelogram does not trace the cusped orbit \citep[][]{Fux1999, Sormani2018b}. Instead, the parallelogram can be understood in the context of the dust lanes described above (\S\ref{sec:dustlanes}). As the dust lanes deliver gas to the central region, not all of this gas will merge directly with the CMZ. Recent simulations have demonstrated that as much as 50-70\% of the infalling gas can overshoot the CMZ \citep{Hatchfield2021}. When observed in $\{l,v\}$-space, the gas overshooting the CMZ populates the top and bottom of the parallelogram (and some of the emission connected to it). 
The lateral sides of the parallelogram represent EVF-like features caused by the rapid deceleration of gas following the dust lane that is colliding with either gas in the CMZ itself or with gas located on the opposite dust lane.

\begin{figure}[!t]
    \centering
 	\includegraphics[trim=2.5cm 2.cm 5cm 2.cm, clip, width=1.0\columnwidth]{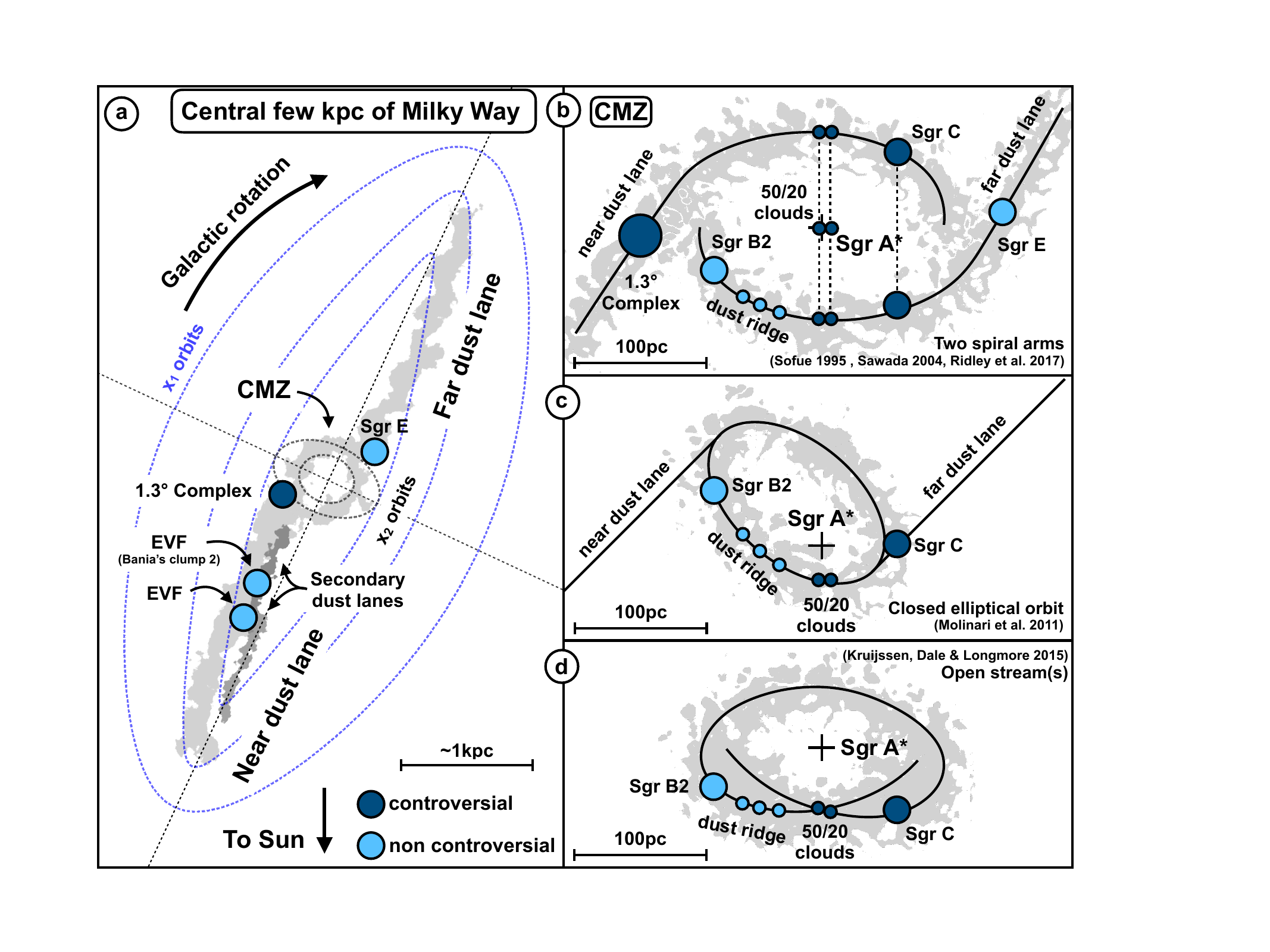}
    \caption{Schematic face-on view of the Galactic centre. On the left is a schematic of the inner few kiloparsecs of the Galaxy. It shows the dust lanes, EVFs, and CMZ discussed in \S\ref{sec:cartography}. Also included are the two main families of orbits ($x_1$, $x_2$) in a barred potential (\S\ref{sec:generaldynamics}). On the right are three different interpretations for the geometry of the CMZ. These are discussed in detail in \S\ref{sec:3d}. The features shown here are also labelled in Fig.~\ref{fig:lbv_main}. }
    \label{fig:sketch}
\end{figure}

\paragraph{``The CMZ'':}\label{subsubsec:cmz} The second component is mass-dominant, and is typically what we think of when we refer to ``the CMZ''. It is enclosed within the parallelogram described above. A map of this gas is shown in the right panels of Fig.~\ref{fig:lbv_main}, along with the corresponding $\{l,v\}$ diagram. For the remainder of this review, we make a distinction between this mass-dominant component, which we will henceforth refer to as ``the CMZ'', and the dust lanes described above and in \S\ref{sec:dustlanes}. The mass-dominant component extends over a Galactic longitude range of $-1.0^{\circ}\lesssim l \lesssim 1.7^{\circ}$, $|b|\lesssim0.5$, and $|v|\lesssim 150$\,\kms, and includes all of the major Galactic centre cloud complexes (Fig.~\ref{fig:lbv_main}): the 1.3$^{\circ}$ cloud complex, the so-called ``dust ridge'' clouds (including Sgr\,B2, G0.253+0.016 or ``the Brick'', and the clouds in between), the Sgr\,A clouds (consisting of the 20 and 50\,\kms \ clouds), and Sgr\,C. 
We further introduce here what we will refer to throughout the text as the ``100\,pc stream'' \citep{Kruijssen2015}. The 100\,pc stream contains all of the aforementioned clouds except for the $1.3^{\circ}$ complex. The 100\,pc stream is of particular interest for this review since it happens to be where the vast majority of present-day star formation is occurring (\S\ref{subsec:SFR:current}). The distinction between ``the CMZ'' and the ``100\,pc stream'' has been made in several works seeking to describe the 3D geometry of gas in the Galactic Centre \citep[\S\ref{sec:3d};][]{Rodriguez-Fernandez2006, Bally2010, Molinari2011, Kruijssen2015, Henshaw2016a}. However, the 3-D geometry of the CMZ remains unclear, and the $1.3^{\circ}$ cloud complex is a particular point of contention. This uncertainty is highlighted in Fig.~\ref{fig:sketch}, where we show schematic representations of the top-down view of the CMZ. For the purposes of this discussion, we shall state simply that the general consensus is that the gas in the CMZ is organised into a ring-like structure, the precise details of which are uncertain. We will revisit this topic, and the models presented in Fig.~\ref{fig:sketch}, in \S\ref{sec:macroevolution}.

\subsubsection{The circumnuclear disc}\label{sec:cnd} 

The final feature that we wish to discuss here is the circumnuclear disc (hereafter, CND). It remains uncertain how the CND relates to the larger scale features described above, but it is important in that it is the closest large reservoir of molecular gas to the central supermassive black hole Sgr\,A* \citep{Genzel1985,Guesten1987, Jackson1993}.  
It has an inner radius of $\sim1\mhyphen1.5$\,pc and an outer radius of $3\mhyphen7$\,pc, filling a ring-like structure with total mass $\sim3\times10^4$ \msun and densities $n\sim10^5\mhyphen10^7$\,cm$^{-3}$ \citep{Etxaluze2011,Oka2011,Mills2013,Tsuboi2018}. 
The exact mechanism by which gas is transported from the CMZ to the inner $10\mhyphen20$ parsecs is an open question (\S\ref{sec:inwardmassflow}). 
However, there is some evidence that the CND is built up from the tidal disruption of molecular clouds located within the inner $\sim10\mhyphen20$\,pc \citep{Ho1991,Coil2000, McGary2001, Martin2012,Mapelli2016,Hsieh2017,Hsieh2019, Tsuboi2018, Ballone2019}. 
Several filamentary molecular gas structures, or ``streamers'', have been detected surrounding the CND. These streamers have spatial extents $\sim$2\,pc and are speculated to act as channels that deliver gas to the CND \citep{Montero-Castano2009, Hsieh2017, Takekawa2017b}, possibly connected to ambient gas clouds located at a Galactocentric radius of 20\,pc (namely the Sgr\,A clouds; see \S\ref{sec:macroevolution}). 

The conditions found in the CND are extreme, even by CMZ standards.
The gas is highly excited, with most gas exceeding $T>500$ K and dust temperatures $\gtrsim 100$ K \citep{Bradford2005,Lau2013,Mills2013,Mills2017a,James2021}. The gas also exhibits extremely broad line-widths of $10\mhyphen40$\,\kms, reflecting a combination of high temperatures, elevated levels of turbulence \citep{Goicoechea2018, Tsuboi2018, Hsieh2021}, and considerable rotational velocity ($\sim100$\,\kms). 
Contrasting measurements of the gas density have led to conflicting views on whether the gas is gravitationally bound or not \citep{Christopher2005, Montero-Castano2009, Requena-Torres2012}. 
It has been suggested that the CND is forming stars based on the detection of water and shock-excited methanol masers, candidate outflows traced by SiO (5-4), and compact, highly-excited and broad linewidth SiO emission interior to the CND \citep{Yusef-Zadeh2013a,Yusef-Zadeh2015b}. 
However, other explanations for these signatures exist and conclusive evidence of ongoing star formation in the CND is still lacking \citep{Mills2017a}.

\subsection{The origin of the CMZ} \label{sec:generaldynamics}

The very existence of the CMZ is a direct consequence of the inflow driven by the Galactic bar. The gravitational potential of the bar, and of the other components in the Galactic Centre, controls the size of the CMZ, influences the distribution, structure, and evolution of molecular clouds, and may help to create preferred locations for star formation. 
This section reviews our current understanding of the origin of the CMZ, and provides theoretical context to interpret the key observational features described in \S\ref{sec:cartography}. 

\subsubsection{The gravitational field in the inner Galaxy} \label{sec:potential}

The potential in the inner Galaxy is dominated by the following components in order of increasing Galactocentric radius of influence: (i) The central black hole Sgr A*. This generates a Keplerian potential with a mass of $M=4.15 \times 10^6\,\msun$ \citep{Ghez2008, Gillessen2009, GravityCollaboration2019} that dominates gravity in the central parsec ($R\lesssim 1\pc$); (ii) The nuclear stellar cluster (NSC). The NSC is a dense, massive ($M \simeq 2.5 \times 10^7\, \msun$) and slightly flattened assembly of stars centred on Sgr\,A$^{*}$ \citep{Genzel2010,Schodel2014b,Feldmeier-Krause2017}. It dominates the potential in the range $1<R\lesssim30\,\pc$; (iii) The nuclear stellar disc (NSD). This is a flattened stellar structure with a mass of $M\simeq 1.05 \times 10^9\, \msun$ \citep{Launhardt2002,Sormani2020a, Sormani2021} that dominates the potential at Galactocentric radii $30 \lesssim R \lesssim 300\pc$. The NSD is co-spatial with the CMZ and generates most of the background gravitational field in which the gas in the CMZ flows. This co-spatiality is probably not coincidental. Gas in the CMZ and stars in the NSD rotate with similar velocities \citep{Schonrich2015,Schultheis2021}, and star formation in the CMZ contributes to the build-up of the NSD over secular timescales \citep{Baba2020}. Current observational constraints are consistent with the NSD being axisymmetric \citep{Gerhard2012,Valenti2016,Sormani2021}, although it cannot be ruled out that it is a nuclear bar \citep{Alard2001,Rodriguez-Fernandez2008}. Collectively, Sgr A*, the NSC and the NSD are often referred to as the nuclear bulge \citep{Launhardt2002}.

Finally, as with roughly 2/3 of all spiral galaxies in the local universe, the Milky Way hosts a stellar bar \citep{Blitz1991,Wegg2013}. The Galactic bar is a strongly non-axisymmetric structure whose major axis lies in the Galactic plane, with its nearer end at positive longitudes (for a review see \citealt{Bland-Hawthorn2016}). It has a mass of $M\simeq 1.9 \times 10^{10} \, \msun$ and provides the main contribution to the gravitational field in the range $0.3\lesssim R \lesssim4\,\kpc$ \citep{Portail2017}. The bar transports matter and energy from the Galactic disc to the CMZ and therefore plays a key role its evolution (\S\ref{sec:barinflow}). The contribution of dark matter to the potential in the inner Galaxy is probably small \citep{Portail2017,Li2020b}.

\subsubsection{Gas flow in a barred potential and the formation of nuclear rings} \label{sec:gasdynamics}

The first step to understanding the flow of gas in a barred galaxy like the Milky Way is to examine the orbital structure of the underlying gravitational potential \citep{Prendergast1983,Sellwood1993}. 
In the limit in which pressure forces (intended here broadly to include all non-gravitational forces) are completely negligible, the gas streamlines must coincide exactly with the orbits of non-interacting particles (also called ``ballistic'' or ``stellar'' orbits). Since the sound speed of the gas (both thermal and turbulent) is usually small compared to the orbital speed, pressure forces are often negligible, and in many cases the gas will follow closely ballistic orbits. 
However, unlike stellar orbits, gas streamlines cannot cross; the gas must have a unique stream velocity at each point in the flow. 
As a consequence, gas can follow ballistic orbits only when a sequence of closed orbits can be nested within one another without intersection. 
In certain situations (see below) crossings are inevitable, and pressure forces become important irrespective of the sound speed.

The two most important families of stable closed orbits in a barred potential are $x_1$ and $x_2$ orbits \citep[see Fig.~\ref{fig:sketch};][]{Contopoulos1989,Athanassoula1992a}. $x_1$ orbits exist inside the bar corotation radius and are highly elongated in the direction of the bar major axis. $x_2$ orbits are present only if the potential possesses an inner Lindblad resonance (ILR).
They are located inside the ILR (for a single ILR) or in between the ILRs (if there are two of them), and are mildly elongated in the direction perpendicular to the bar major axis. In fact, the extent of the $x_2$ orbits can be used to generalise the \emph{definition} of Lindblad resonance to strongly barred potentials \citep{VanAlbada1982}. The Milky Way bar has corotation at $R_{\rm CR}\simeq6\,\kpc$ and it is believed to have a single ILR located at $R_{\rm ILR}\simeq1\,\kpc$ in the epicyclic approximation \citep{Sormani2015a,Portail2017}.

Hydrodynamical simulations of gas flow in barred potentials show that the gas streamlines tend to follow closely $x_1$ and $x_2$ orbits when possible \citep{Athanassoula1992b,Sormani2015c}. However, when both families are present there are regions where orbit crossings are unavoidable, and the gas streamlines must shift from one family of closed orbit to another. In a potential with a single ILR (as in the Milky Way), the gas located towards the outer parts of the bar roughly follows $x_1$ orbits, while dissipation processes cause it to slowly drift inwards along a sequence of such orbits. As the centre is approached, $x_1$ orbits become more and more elongated until they become self-intersecting. Where these self-intersections occur, the gas transitions onto the $x_2$ orbits lying deeper within the potential.
The transition happens through large-scale shocks, which correspond to the ``dust lanes'' often observed in external barred galaxies \citep{Athanassoula1992b} and to the analogous features seen in the Milky Way (\S\ref{sec:dustlanes}). The shocked gas then plunges from $x_1$ orbits towards the centre in a dynamical time, where it piles up and organises into a mildly elliptical ring or disc-like structure where the gas follows $x_2$ orbits (Fig.~\ref{fig:sketch}). 

In reality, gas streamlines will not coincide precisely with $x_1$ and $x_2$ orbits. This is because physical agents such as pressure, turbulence, stellar feedback (particularly supernovae), cloud collisions, or external perturbations will produce deviations from periodic orbits. These deviations can be random and transient, causing, for example, viscous-driven inward drifting of gas, or organised in such a way to collectively generate regular spiral patterns \citep{Sormani2015b}. Indeed, the gas in the CMZ is unlikely to follow exactly closed $x_2$ orbits \citep{Kruijssen2015,Tress2020}. Because most non-closed orbits can be understood as librations around an underlying stable closed orbit \citep{Binney2008}, it is often useful to decompose the gas motion into two components: the motion of a guiding centre, which follows a closed $x_1$ or $x_2$ orbit, and excursions with respect to the guiding centre.

\subsubsection{What controls the size of the CMZ?} \label{sec:insights}

The framework presented in \S\ref{sec:gasdynamics} tells us that, in the presence of a barred potential, a nuclear ring-like structure forms in the region where $x_2$ orbits exist. 
However, there is no theoretical consensus on what determines the exact radius of the ring, despite several theories being proposed. 
It therefore remains an open question as to why the Milky Way's CMZ has a radius of $R\simeq100$-$200\pc$ rather than, for example, twice or half this value.

It is clear that to support a ring-like structure at all, the gravitational potential must possess $x_2$ orbits, and therefore that the ring must lie within the radial range where such orbits exist \citep{Athanassoula1992b,Regan2003,Kim2012b}.  The $x_2$ family is smaller (and can even disappear completely) for stronger bars, larger bar pattern speeds, larger bar axial ratios and less centrally concentrated mass distributions, i.e.\ rotation curves that rise more gently in the centre \citep{Athanassoula1992a}. The radius of the ring is, therefore, expected to correlate with these quantities. 
The Milky Way potential is currently constrained well enough that the existence of the $x_2$ family appears established. However, there are large uncertainties on the extent of the $x_2$ family. The radius of the \emph{largest} $x_2$ orbit could be anything between $R=200~\pc$ up to almost $R=1~\kpc$, although the real value is probably around halfway between these two (see Section \ref{sec:gasdynamics} and references therein).

However, even with a perfect knowledge of the gravitational potential, it is not clear which $x_2$ orbits will be actually populated by gas and which ones will be empty. Hydrodynamical simulations show that, for a fixed gravitational potential, the radius of the ring decreases if the sound speed (and so, the pressure) of the gas is increased \citep{Englmaier1997,Patsis2000,Sormani2015c,Li2015}. Increasing magnetic pressure can also decrease the radius of the ring \citep{Kim2012c}. Turbulence on the other hand increases the width of the ring, but does not seem to significantly affect its radius \citep{Salas2020}, suggesting a difference in the effects of ``real'' vs. turbulent pressure, although more studies are needed in this direction.

\cite{Lesch1990} and more recently \cite{Krumholz2015} and \cite{Krumholz2017} proposed that the ring forms where the shear calculated from the rotation curve reaches its minimum. The idea is that the $x_2$ disc behaves similarly to an axisymmetric viscous accretion disc \citep{Lynden-Bell1974}, so that gas piles up and forms a ring where the viscous transport becomes less efficient, i.e.\ where shear is lowest. In these models, the radius of the nuclear ring depends only on the shape of the rotation curve. However, this is in tension with results of simulations that show that the radius of nuclear rings varies with the bar pattern speed, the quadrupole of the potential, and the sound speed of the gas, \emph{even with a fixed rotation curve} \citep{Patsis2000,Kim2012a,Sormani2015a}. \citet{Armillotta2019} propose that this discrepancy might be resolved if the ring initially builds up at the shear minimum, and it later moves inward due to turbulence-driven angular momentum transport. However, the simulations of \cite{Sormani2020c} show that nuclear rings form even when there is no shear minimum. The presence of a shear minimum is therefore not a necessary condition for the formation of a nuclear ring. On the observational side, uncertainties on the Milky Way rotation curve are currently too large to determine whether the radius of the nuclear ring in the Milky Way coincides with the shear minimum \citep{Sormani2020a}.

\cite{Combes1988} and \cite{Buta1996} argued that nuclear rings form at Lindblad resonances. This theory is equivalent to the statement that the radius of the ring coincides with that of the largest $x_2$ orbit, because the ILR in both the weak and strong bar limits can be defined by the radius of the largest $x_2$ orbit \citep{VanAlbada1982}. While this conclusion may be correct in the limit of vanishing sound speed, it will tend to overestimate the radius of the ring in general \citep{Sormani2020c}. Simulations show that when the sound speed and pressure forces are larger, the radius of the ring decreases, while the underlying orbital structure remains the same \citep[e.g.][]{Englmaier1997,Patsis2000,Sormani2015c}. In the Milky Way, the ILR \emph{in the epicyclic approximation} is at $R_{\rm ILR}\simeq 1\kpc$, the extent of the largest $x_2$ orbit (which generalises the notion of ILR for strong bars) is uncertain but is in the range $R=200\pc\mhyphen1\kpc$, while the radius of the CMZ ring-like structure is $R\simeq100$-$200\pc$.

\cite{Sormani2018a} proposed a mechanism for the confinement of the ring (i.e.\ how it can survive without spreading even in the presence of viscosity) and argued that its radius coincides with the smallest radius, $R_{\rm c}$, at which the bar potential can excite density waves in the gas that are strong enough to turn into spiral shocks. This radius depends on the sound speed of the gas (and, more generally, on pressure and magnetic forces) in a non-trivial way. In the limit of vanishing sound speed, $R_{\rm c}$ coincides with the largest $x_2$ orbit that does not cross adjacent orbits. When the sound speed is larger, $R_{\rm c}$ decreases, as does the radius of the ring. At the moment, there is no simple analytic way of calculating $R_{\rm c}$ other than running a full hydrodynamical simulation.

In summary, we can predict that the radius of CMZ must be somewhere in the radial range where $x_2$ orbits exist (see the beginning of this section), but we currently do not have a simple way of predicting its exact location within this range. The latter remains an interesting theoretical challenge in astrophysical fluid dynamics.


\section{The global view of star formation}
\label{sec:global}

In the following section, we summarise the global view of star formation in the CMZ. 
Due to the highly embedded nature of forming stars, determining star formation rates (SFR) relies on the emission produced either directly from, or indirectly by, star formation over the lifetime of the emission mechanism.
In \S \ref{sec:SFR:current} we review the recent body of literature dedicated to measuring the SFR in the CMZ. As we will see, the SFRs independently derived from various methods, including source counting and integrated light measurements, point to a SFR that has been more or less constant at a value of $\simeq0.07$\,\msunyr \ for at least the last 5\,Myr (Table\,\ref{tab:SFR}).
In \S\ref{subsec:SFR:starformationhistory} and \S\ref{subsec:SFR:timeevolution} we consider the star formation history of the CMZ and theories seeking to explain the time evolution of star formation in galactic nuclei, respectively. 

\subsection{Current star formation} \label{sec:SFR:current}

\subsubsection{Star formation rate from source-counting}
\label{subsec:SFR:current}

The best estimates of the ``present-day'' SFR are obtained by counting sources. This involves identifying individual young stellar objects (YSOs), \hii\ regions, or supernovae, and then, by assigning some representative mass and age to each source, determining the SFR. The mean value of the SFR derived from various source counting methods across the CMZ is $0.07^{+0.08}_{-0.02}$\,\msunyr.
A significant contribution to the scatter of these results is the different definitions of the CMZ area over which the measurements are made.
This average and scatter does not account for the uncertainties in the individual measurements such as the assumed timescale of the tracers and the assumed IMF, both of which have a generally unknown degree of uncertainty. We summarise the details of each method below and in Table\,\ref{tab:SFR}.

\citet{Yusef-Zadeh2009} photometrically identified a sample of 559 potential YSO sources within the central $|l|<1.3$\degree\ and $|b|<0.17$\degree\ by using the 24\,\micron\ colour excess relative to 8\,\micron, which is thought to reasonably trace YSOs (under a number of key assumptions). They classified the YSOs according to three evolutionary stages, namely Stage I, II, and III \citep{Robitaille2006}, by fitting the 1.24\,\micron\ to 24\,\micron\ spectral energy distribution (SED) for 360 of these sources \citep{Robitaille2006,Robitaille2007}. 
They identified a population of 213 Stage I YSOs within the CMZ, which have a total stellar mass (assuming a Kroupa IMF) of $\sim1.4\times10^{4}$\msun.
\citet{Yusef-Zadeh2009} also identified a population of 30 very young Stage I YSOs that showed the 4.5\,\micron\ excess (referred to also as ``green fuzzies'' or ``Extended Green Objects''; e.g. \citealp{Cyganowski2008,Chambers2009}), which has a total stellar mass of $\sim$\,500\msun.
The canonical age for a Stage I phase for low-mass stars is thought to be $\sim$\,0.1\,Myr, and the 4.5\,\micron\ excess phase is 0.05 to 0.1\,Myr \citep{Evans2009}. 
Assuming these representative ages, \citet{Yusef-Zadeh2009} estimated SFRs from the Stage I and pre-Stage I sources of $\sim$\,0.14\,\msunyr and $\sim0.01$\,\msunyr, respectively. 

\begin{table}[!ht]
    \caption{Summary of star formation rate (SFR) measurements in the literature (see \S\,\ref{subsec:SFR:current} and \ref{subsec:SFR:past}). Columns show the timescale ($t_\mathrm{sf}$) when the stars relating to the measured SFR formed, the area (galactic longitude and latitude) over which the measurement is made, and the measured SFR. Adapted from the literature summaries provided in \citet{Barnes2017} and \citet{Nandakumar2018}}\vspace{1mm}
    
    \centering
    \label{tab:SFR}
  
    \begin{tabular}{lcc} 
    \hline \hline \noalign{\vspace{0.5mm}}
     Timescales, $t_\mathrm{sf}$ [Myr] & ($|l|,|b|$) [$^{\circ}$]  & SFR [M$_\odot$yr$^{-1}$] \\
    \hline
    \multicolumn{3}{l}{\small YSO counting} \\ 
    \cline{0-0}
    \noalign{\smallskip} 
$\sim$\,0.01$^{(a)}$       & (1.3, 0.17)  & 0.01 \\
$\sim$\,0.1$^{(b)}$        & (1.3, 0.17)  & 0.06 \\
$\sim$\,0.1$^{(c)}$        & (1.3, 0.17)  & 0.07 \\
$\sim$\,1$^{(d)}$          & (1.5, 0.5)   & 0.08 \\
$\sim$\,0.75$^{(e)}$       & (1.5, 0.5)   & 0.05 \\
$\sim$\,0.3$^{(f)}$        & (1, 0.5)     & $>$0.025 \\

    \multicolumn{3}{l}{\small \hii\ region counting} \\ 
    \cline{0-0}
    \noalign{\smallskip} 
$\sim$\,4$^{(g)}$          & (1, 0.5)     & $>$0.012\\
$\sim$\,0.75$^{(h)}$       & (1.5, 0.5)   & 0.02 - 0.07 \\

    \multicolumn{3}{l}{\small SNR counting} \\ 
    \cline{0-0}
    \noalign{\smallskip} 
$\sim$\,0.01$\mhyphen$0.04 ($\sim$\,3)$^{(i)}$ & (1.5, 0.5)   & 0.035 - 0.15  \\

    \multicolumn{3}{l}{\small Integrated light} \\
    \cline{0-0}
    \noalign{\smallskip} 
$\sim$5$\mhyphen$100$^{(j)}$   & (1.3, 0.17)   & 0.07 \\
$\sim$5$\mhyphen$100$^{(k)}$   & (0.8, 0.3)    & 0.08 \\
$\sim$5$\mhyphen$100$^{(l)}$   & (1, 0.5)      & 0.09 \\ 

    \multicolumn{3}{l}{\small Averages} \\ 
    \cline{0-0}
    \noalign{\smallskip} 
$\sim$\,1$\mhyphen$5$^{(m)}$            & (1, 0.5)  & $\sim$\,0.07 \\
$\sim$\,5$\mhyphen$100$^{(n)}$          & (1, 0.5)  & $\sim$\,0.09 \\

    \hline
    \end{tabular}

    \begin{minipage}{\columnwidth}\small
    \vspace{1mm} {\bf Notes:} 
    $(a)$ Pre-Stage I (4.5\micron\ excess) YSOs from \citet{Yusef-Zadeh2009}.
    $(b)$ Stage I (24\micron\ excess) YSOs from \citet{Yusef-Zadeh2009}, corrected by \citet{Koepferl2015}. 
    $(c)$ Stage I YSOs from \citep{Yusef-Zadeh2009}, corrected by \citet{An2011}.
    $(d)$ \citet{Immer2012b}.
    $(e)$ \citet{Nandakumar2018}.
    $(f)$ \citet{Lu2019a,Lu2019b}. 
    $(g)$ Estimate from \citet{Longmore2013b} represents a lower limit due to the source identification routine and flux estimation method from \citet{Lee2012}.
    $(h)$ \citet{Nguyen2021}. 
    $(i)$ Estimate from \citet{Ponti2015}. The $t_\mathrm{sf}$ is the supernova ages used to calculate to SFR, yet this range is representative of the SFR when these stars formed $\sim$\,3\,Myr ago \citep{Leitherer2014}. 
    $(j)$ \citet{Yusef-Zadeh2009}. 
    $(k)$ \citet{Crocker2011a}.   
    $(l)$ \citet{Barnes2017}.   
    $(m)$ Star formation in the last $\sim$\,1$\mhyphen$5\,Myr (\S\,\ref{subsec:SFR:current}).
    $(n)$ Star formation averaged over 5$\mhyphen$100\,Myr (\S\,\ref{subsec:SFR:past}).
    \end{minipage}
    \vspace{-2mm}
\end{table}

Several works have investigated potential contaminants to the \citet{Yusef-Zadeh2009} Stage I YSO sample, which may cause an overestimation of the SFR. \citet{Koepferl2015} demonstrated that main-sequence objects can mimic YSOs' 24\,\micron\ emission, suggesting that the fraction of misclassified YSOs is at least 60\,\%, and that the SFR estimated by \citet{Yusef-Zadeh2009} is likely to be at least a factor of three too high; the corrected value is therefore around $\sim$0.05\,\msunyr. Similarly, \citet{An2011} aimed at spectroscopically confirming these YSO candidates. These authors identified 35 YSOs from an initial sample of 107 that showed a 15.4\,micron\ shoulder feature in the spectra. The presence of a 15.4\,micron\ shoulder on the absorption profile of CO$_2$ ice is suggestive of a mixture of CO$_2$ ice and CH$_3$OH ice on grains, which is observed in both high- and low-mass YSOs and allows for differentiation from contaminants (e.g. AGB stars; \citealp{An2009}). Comparing to the sample of \citet{Yusef-Zadeh2009}, they showed that 50\,\% of their sources can be spectroscopically confirmed as YSOs, suggesting that the \citet{Yusef-Zadeh2009} estimate is around a factor of two too high (corrected value around $\sim$0.07\,\msunyr).

\citet{Immer2012b} analysed the mid-IR (5 to 38\,\micron) spectra of bright IR sources to define selection criteria for YSOs across the central $|l|<1.5$\degree\ and $|b|<0.5$\degree\ region. 
They used the spectroscopic classification of 68 sources to determine a [7\,\micron]-[15\,\micron] colour excess and define a spatial extent parameter, which together reliably singles out young objects. 
They then apply these parameters to point-source catalogues from MSX and ISOGAL. 
They estimated that the 1141 young object candidates have a total (IMF corrected) mass of 77,000\,\msun, which assuming an average lifetime of $\sim$1\,Myr equates to a SFR of $\sim$\,0.08\,\msunyr. 

\citet{Nandakumar2018} obtained KMOS spectra (between 2 to 2.5\micron) towards a sample of 91 photometrically selected YSO candidates across the central $|l|<1.5$\degree\ and $|b|<0.5$\degree\ region (\citealp{Nishiyama2006, Ramirez2008}). 
These authors separated out 23 YSOs that did not show a CO absorption feature at 2.3\micron, yet had Br$\gamma$ emission at 2.17\micron. 
This sample was used to define a photometric selection criterion in the H, K$_\mathrm{S}$ and 8\micron\ bands, which were then used to identify a larger sample of 334 YSO candidates from the SIRUS survey \citep{Nishiyama2006}. 
They estimated these young object candidates have a total (IMF corrected) mass of 35,000\,\msun, which assuming an average lifetime of $\sim$0.75\,Myr equates to a SFR of $\sim$\,0.05\,\msunyr. 

As part of the GLOSTAR Galactic plane survey, \citet{Nguyen2021} studied the properties of radio counterparts (4 to 8\,GHz) to the sample of 334 YSO candidates identified by \citet{Nandakumar2018}. 
A sub-sample of 26 sources displayed spectral indices consistent with thermal free-free emission from \hii\ regions (see \S\,\ref{sec:hiiregions}). 
The zero-age main sequence masses of the stars generating these \hii\ regions are estimated to be in the range of $10 \mhyphen 40\,\msun$, which after correcting for an IMF, gives a total mass of 30,000\,\msun. 
Assuming an average lifetime of $\sim$0.75\,Myr (from \citealp{Nandakumar2018}) these authors estimate an SFR of 0.04$\pm$0.02\msunyr. 
Of all possible YSOs, these authors only selected those that have associated \hii\ regions and, therefore, only the stars that are radio bright. 
To correct for this bias, they related the SFR to the number of \hii\ regions \citep[following][]{Kauffmann2017b}. 
They determined the maximum stellar mass within each cluster, extrapolated the number of cluster members assuming an IMF, and estimated the total mass of stars associated with each \hii\ region assuming a mean stellar mass. 
Assuming a timescale of 1.1\,Myr, they determined a total SFR of 0.068\msunyr \ (including the SFR estimates from \citealp{Kauffmann2017b}).

\citet{Longmore2013b} used integrated ionizing radiation flux measurements made at cm-wavelengths from the Wilkinson Microwave Anisotropy Probe observations ({\em WMAP}; e.g. \citealp{Lee2012}), to estimate the total mass of embedded (high-mass) stars (e.g. \citealp{Murray2010}).
They then estimated a SFR of 0.012 to 0.018\,\msunyr \ within the central $|l|<1$\degree\ and $|b|<0.5$\degree\ region (or $\sim$\,0.06\,\msunyr \ including $|b|<1$\degree), assuming an ionization-weighted stellar lifetime of $\sim$\,4\,Myr (e.g. \citealp{Murray2010}). 
This SFR estimate sits factors of several below the aforementioned YSO counting SFR measurements, possibly resulting from the \citealp{Lee2012} source extraction routine. 
As noted by \citealp{Lee2012}, in this catalogue the ionising luminosity of the Arches cluster could be underestimated by a factor of four, and the most actively star-forming regions within the CMZ, Sgr B2, is entirely missed.
The SFR from \citet{Longmore2013b} should, thus, be taken as a lower limit.

\citet{Lu2019b} estimated the SFR using H$_2$O maser detections across the CMZ (e.g. \citealp{Walsh2011}). 
Masers are associated with the early stages of star formation, when YSOs are still heavily embedded within their host environment (see \S\,\ref{sec:masers}). Using a sample of 112 masers across the central $|l|<1$\degree\ and $|b|<0.5$\degree, and assuming that a single high-mass star is responsible for each, they estimate a total stellar mass of $\sim$\,11,000\,\msun. 
Assuming a maser lifetime of $\sim$\,0.3\,Myr, these authors then estimate a SFR of $\sim$\,0.04\,\msunyr. 
\citet{Lu2019a} estimated the SFR from a sample of 20 class II CH$_3$OH masers and 47 ultra-compact HII regions. They estimate a total stellar mass of $\sim$\,7,500\,\msun, again assuming a maser lifetime of $\sim$\,0.3\,Myr, then estimate a SFR of $\sim$\,0.025\,\msunyr. These authors caution that this estimate is a lower limit, however, since not all young stars produce masers.

Lastly, an estimate of the SFR can be inferred from counting supernova remnants (also see \S\,\ref{sec:feedback:outflowmechanisms}). 
\cite{Ponti2015} analysed deep {\it XMM–Newton} observations across $\sim$\,1\degree of the CMZ, and, in combination with 90-cm radio data \citep{LaRosa2000}, estimated a supernova rate of $\sim3.5\mhyphen15\times$10$^{-4}$\,yr$^{-1}$ (also see \S\,\ref{sec:feedback:outflowmechanisms} for additional estimates).
Assuming each supernova originates from a high-mass star, and accounting for an IMF, \cite{Ponti2015} estimated a SFR of 0.035 to 0.15\msunyr.

\subsubsection{Star formation rate from integrated light measurements}
\label{subsec:SFR:past}

In contrast to the source counting methods described in \S\ref{subsec:SFR:current}, integrated light measurements use the total observed luminosity within a given wavelength range to infer the total SFR across a region (see e.g. \citealp{Kennicutt2012}, and references therein).
This method assumes continuous star formation over the past 100\,Myr and that the mean age of the stellar population contributing to the measured luminosity is $\sim$\,5\,Myr (see e.g. \citealp{Kennicutt1998}). 
However, there can be a substantial contribution from stars that have been forming over a much longer timescale, e.g., $\sim$\,10\% of the emission could come from stars with ages $>$\,100\,Myr \citep{Kennicutt2012}.
Therefore, the methods discussed below probe star formation averaged over 5\,$\mhyphen$\,100\,Myr. 

\citet{Yusef-Zadeh2009} measured a total extinction-corrected 24\,\micron\ luminosity of $\sim$\,9\,$\times$\,10$^{7}$\,\lsun\ within $|l|<1.3$\degree\ and $|b|<10$\arcmin. Assuming the conversion factor from \citet{Rieke2009}, this luminosity implies a total SFR of $\sim$\,0.07\,\msunyr. \citet{Crocker2011a} measured the total infrared luminosity using {\em IRAS} observations across the central $|l|<0.8$\degree\ and $|b|<0.3$\degree\ region. Assuming a conversion factor from \citet{Kennicutt1998}, these authors estimated a total SFR of $\sim$\,0.08\,\msunyr. 

\citet{Barnes2017} measured the luminosity of 24\,\micron\ from {\it Spitzer} observations, along with the 70\,\micron\ luminosity from {\it Herschel} and total infrared luminosity determined from the combined {\it Spitzer} and {\it Herschel} observations (5 to 500\,\micron), across the central $|l|<1$\degree\ and $|b|<0.5$\degree\ region. 
Using a large sample of SFR prescriptions, which are commonly adopted within nearby to high-$z$ extragalactic systems, these authors determined an average SFR across all diagnostics of $\sim$\,0.09\,\msunyr \ (Table\,\ref{tab:SFR}). 
This value from \citet{Barnes2017} is taken as a representative value for SFR within the CMZ over the last $\lesssim$\,100\,Myr. 

As with the present-day SFRs, there is some scatter between the longer timescale averaged SFR discussed above due to different definitions of the $\{l,b\}$ area corresponding to the CMZ. Moreover, the integrated light measurements are subject to several systematic uncertainties. A discussion of these uncertainties is beyond the scope of this review (see \citealp{Kennicutt2012}), but the main sources include: 1) the star-formation history in the last $>$100\,Myr; 2) the level of dust attenuation as a function of stellar age (the models assume complete dust attenuation, such that all emission contributes to dust heating); 3) metallicity; 4) contamination from older stellar populations; 5) contamination from other sources not associated with star formation, such as Sgr A$^*$. 

\subsubsection{Is the CMZ currently under-producing stars?} 
\label{subsec:SFR:context}

\begin{figure*}[ht]
    \centering
 	\includegraphics[trim=0 0cm 0 0cm, clip, width=0.85\textwidth]{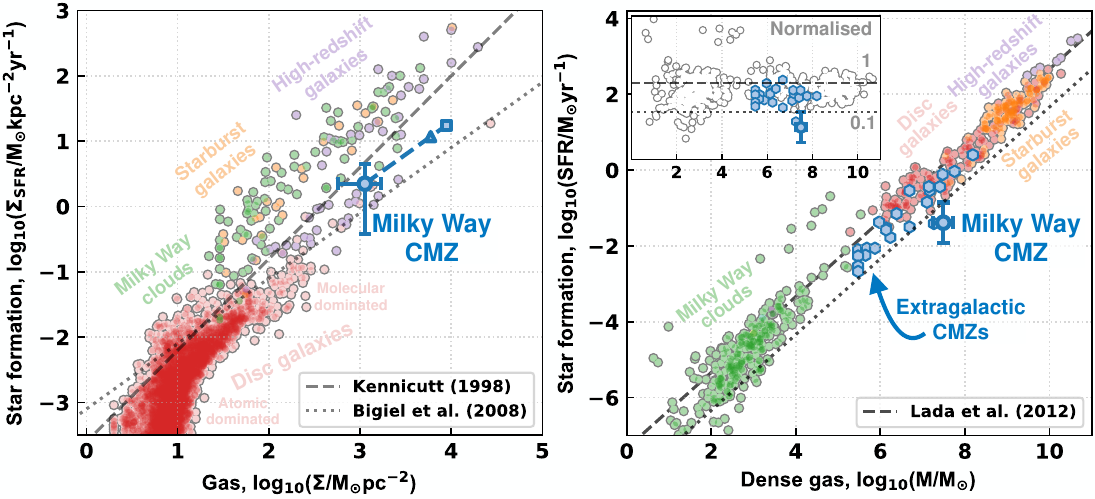}
    \caption{The CMZ star-forming properties relative to several commonly used scaling relations. {\em Left:} The SFR surface density ($\Sigma_\mathrm{SFR}$) as a function of the gas surface density ($\Sigma_{\rm gas}$). The CMZ is shown by the large blue circle with error bars, which spans the range of $\Sigma_\mathrm{SFR}$ and $\Sigma_{\rm gas}$ determined in the literature (see Table\,\ref{tab:properties_overview}). This point is determined assuming a face-on disc geometry, whereas the connected points show the results of assuming either face-on ring (triangle) and edge-on geometry (square; see \S\,\ref{subsec:SFR:context}). Shown as grey-outlined coloured points are measurements of systems from the literature (see \citealp{Krumholz2014a} and references therein). Overplotted are the scaling relations from \citet{Kennicutt1998} and \citet{Bigiel2008}. {\em Right:} The relation between the dense molecular gas mass ($M_\mathrm{dense}$) and SFR. The CMZ is shown by the large blue circle with error bars, which span the range of $M_\mathrm{dense}$ and SFR determined in the literature (see Table\,\ref{tab:properties_overview}). Shown as grey-outlined coloured points are the values for systems taken from the literature (see \citealp{Jimenez-Donaire2019} for references). Shown as blue hexagons are extragalactic CMZs (\citealp{Querejeta2019,Jimenez-Donaire2019,Jiang2020,Beslic2021}). The dashed horizontal line shows the scaling relation of \citet{Lada2012}, and the dotted line shows a factor of ten below this relation. {\em Right (inset panel):} The $M_\mathrm{dense}$ and SFR/$M_\mathrm{dense}$, normalised to the \citet{Lada2012} relation.}
    \label{fig:sfr_main}
\end{figure*}

\begin{table*}[ht]
    \caption{Overview of bulk properties of the CMZ compared to the (order of magnitude) properties determined for the solar neighborhood, nearby extragalactic CMZs, and high-{\em z} Milky Way-like environments. 
    \label{tab:properties_overview}
    } \vspace{2mm}
    
    \centering
    \begin{tabular}{lcccc}
    \hline\hline
Physical Quantity                                  & CMZ                            & Solar Neighbourhood & Extragalactic CMZs & $z\sim2$  \\
\hline
Distance [kpc]$^{(a)}$                             & 8.2                            & 0.1 - 0.5           & 3500 - 20000       & $\sim$\,10$^6$ ({\em z}\,$\sim$\,2) \\
SFR [\msunyr]$^{(b)}$                              & 0.07 (0.012$\mhyphen$0.14)     & 0.002               & $0.001\mhyphen0.08$& 1-100\\
$\Sigma_{\rm gas}$ [log$_{10}$(\msunpc)]$^{(c)}$   & 3.1 (2.8$\mhyphen$3.2)         & 1.5                 & 0.6$\mhyphen$3     & 1.5$\mhyphen$3.5\\
$\Sigma_{\rm SFR}$ [log$_{10}$(\msunyrkpc)]$^{(d)}$& 0.3 ($-$0.4$\mhyphen$0.6)      & -2.5                & -3$\mhyphen$0      & -1.5$\mhyphen$1.5\\
$\Sigma_{*}$ [log$_{10}$(\msunpc)]$^{(e)}$         & 3.9                            & 1.5                 & 3.4$\mhyphen$3.9   & 1$\mhyphen$4 \\
$t_{\rm dep}$ [Gyr]$^{(f)}$             & 0.5 (0.4$\mhyphen$1.5)         & 1                   & $0.3\mhyphen2.6$          & 0.2$\mhyphen$1 \\
$t_{\rm dyn}$ [Myr]$^{(g)}$                        & 5                              & 220                 & 4-40               & ? \\
$B [\mu \mathrm{G}]$$^{(h)}$                       & 10$\mhyphen$1000               & 1$\mhyphen$100      & ?                  & ? \\
Metallicity, $Z$$^{(i)}$                           & 2                              & 1                   & $\sim$2            & 0.2$\mhyphen$0.6\\
CRIR [log$_{10}$(s$^{-1}$)]$^{(j)}$                & $-$15 to $-$13                 & $-$17 to $-$15      &  ?                 & ? \\
Linewidth, $\sigma(10 \mathrm{pc})$ [\kms]$^{(l)}$  & 12                            & 3                   & 10                 &  20$\mhyphen$70 \\
Linewidth scaling, $b$$^{(m)}$                      & 0.7                           & 0.5                 & ?                  &  ? \\
IMF slope, $\alpha$$^{(n)}$                         & $\leq$2.35                    & 2.35                & ?                  &  ? \\
DGMF, $f(n>10^4)$$^{(o)}$                           & 0.95                          & 0.03                & ?                  &  ? \\
$T_\mathrm{gas}$ [K]$^{(p)}$                        & 50$\mhyphen$100               & 10$\mhyphen$30      & 50$\mhyphen$250    & ? \\
$T_\mathrm{dust}$ [K]$^{(q)}$                       & 20$\mhyphen$50                & 10$\mhyphen$30      & 30$\mhyphen$45     & ? \\
$P_\mathrm{ext}/k_\mathrm{B}$ [K cm$^{-3}$]$^{(r)}$ & $\gtrsim10^7$                 & $\gtrsim10^5$       & $10^6\mhyphen10^8$  & ? \\ \hline
    \end{tabular}
    
    \begin{minipage}{\textwidth}\small
    \vspace{1mm} {\bf Notes:} References: $\{\mathrm{CMZ,\,Solar\ Neighbourhood,\,Extragalactic\ CMZ,\,}z\sim2\}$. The $\dagger$ symbol indicates values that are inferred from other properties in the table.
    $(a)$ Distance: $\{$\citet{GravityCollaboration2019}, --, --, -- $\}$.
    $(b)$ Star formation rate: $\{$\S~\ref{subsec:SFR:current}, \citet{Spilker2021}, \citet{Pessa2021}, \citet{Leslie2020}$\}$.
    $(c)$ Gas mass surface density: $\{$\S\ref{subsec:SFR:context}, \citet{Spilker2021}, Sun et al. (in prep.), \citet{Bolatto2015} $\}$.
    $(d)$ Star formation rate surface density: $\{$\S\ref{subsec:SFR:context}, \citet{Spilker2021}, \citet{Pessa2021}, \citet{Freundlich2013} $\}$.
    $(e)$ Stellar surface density: $\{$\citet{Sormani2021}, \citet{McKee2015}, \citet{Querejeta2015}, \citet{vanDokkum2010}$\}$. For the CMZ, we have considered the average value in a 100 pc radius circle.
    $(f)$ Depletion time, $\Sigma_{\rm gas}/\Sigma_{\rm SFR}$: $\{$$\dagger$, $\dagger$, Sun et al. (in prep.), \citet{Tacconi2020}$\}$.
    $(g)$ Dynamical or orbital time: $\{$\citet{Kruijssen2015}, \cite{Bland-Hawthorn2016}, \cite{Comeron2010}, -- $\}$.
    $(h)$ Magnetic field strength: $\{$\S~\ref{sec:magneticfields}, \citet{Chapman2011}, --, -- $\}$.
    $(i)$ Metallicity: $\{$\citet{Balser2011}, --, --, \citet{Maiolino2008} $\}$.
    $(j)$ Cosmic ray ionization rate: $\{$\S~\ref{sec:cosmicrays}, \citet{Neufeld2017}, --, -- $\}$.
    $(l)$ Linewidth: $\{$\citet{Shetty2012}, \citet{Heyer2015}, --, \citet{Swinbank2015} $\}$.
    $(m)$ Size-Linewidth relationship scaling: $\{$\S~\ref{sec:velocitystructure}, \citet{Heyer2015}, --, -- $\}$.
    $(n)$ Stellar initial mass function slope: $\{$\S~\ref{sec:starclusters}, \citet{Offner2014}, --, -- $\}$.
    $(o)$ Dense gas fraction, or fraction of mass above $n>10^{3\mhyphen4}$\,\cmc: $\{$\citet{Longmore2013b}, \citet{Spilker2021}, --, -- $\}$.
    $(p)$ Molecular gas temperature: $\{$\citet{Krieger2017}, \citet{Friesen2017}, \citet{Mangum2013}, -- $\}$. 
    $(q)$ Dust temperature: $\{$\citet{Tang2021a}, \citet{Chen2016}, \citet{Mangum2013}, -- $\}$.
    $(r)$ External pressure: $\{$\citet{Myers2022}, \citet{Field2011}, \citet{Sun2020b}, -- $\}$
    \end{minipage}
    
\end{table*}

A key development that emerged around the time of Protostars \& Planets VI was that the current SFR of \linebreak $\simeq0.07$\,\msunyr \ (\S\ref{subsec:SFR:current} and \S\ref{subsec:SFR:past}; Table\,\ref{tab:SFR}) is an order of magnitude below that expected for the available reservoir of dense molecular gas in the CMZ \citep{Longmore2013b}.
This expectation comes from so-called star formation scaling relations, which are empirical correlations describing the relationship between the rate at which stars form and the properties of the ISM out of which they are born.

Perhaps the most familiar of these scaling relations is the ``Schmidt–Kennicutt'' relation \citep[][hereafter SK relation]{Schmidt1959,Kennicutt1998}, which describes the correlation between the SFR surface density, $\Sigma_\mathrm{SFR}$, and the total gas surface density, $\Sigma_{\rm gas}$ ($\Sigma_\mathrm{SFR}\propto\Sigma_{\rm gas}^{n}$, where $n=1.4$; see the dashed line in the left panel of Fig.~\ref{fig:sfr_main}). Because of the non-linearity of the SK relation, the degree to which the CMZ agrees (or disagrees) with it depends on geometrical assumptions. \citet{Yusef-Zadeh2009} argued that the central 400\,pc of the Galaxy appears to be forming stars in accordance with the SK relation if the gas surface density is calculated using the area projected on the sky and using their measured value of the SFR (which was probably overestimated, see \S \ref{subsec:SFR:current}). \citet{Kruijssen2014a} instead calculated the surface densities assuming that the gas and star formation in the CMZ are distributed throughout a 100-pc ring/annulus with a width of 10\,pc, i.e. $\Sigma_{\rm gas}=M_\mathrm{gas}\pi^{-1}(R_1^2-R_2^2)^{-1}$, where $R_1-R_2$ is the width of the annulus \citep[see also][]{Longmore2013b}. Under this assumption, they found that the 100-pc ring is forming stars at a rate that is $\sim$ an order of magnitude below the SK relation. We illustrate the effect of different geometrical assumptions for the CMZ (a 100\,pc ring/annulus, a 100\,pc disc, and the projected area on the sky) in the left panel of Fig.~\ref{fig:sfr_main}. 
In conclusion, depending on the assumed geometry, the CMZ appears to be roughly consistent, or at most marginally inconsistent, with the SK relation, taking into account the considerable typical scatter around this relation.

\citet{Kruijssen2014a} showed that the CMZ is consistent with the \citet{Bigiel2008} relation, which describes the linear relationship between the molecular (as opposed to total) gas surface density and the SFR ($\Sigma_\mathrm{SFR}\propto\Sigma_\mathrm{mol}$; see the dotted line in the left panel of Fig.~\ref{fig:sfr_main}).
They commented that this agreement is surprising, since dynamical evolution should proceed faster under the influence of self-gravity when the surface density is higher (i.e. $n>1$). The long depletion time implied by the \cite{Bigiel2008} relation ($t_\mathrm{dep}\sim1\mhyphen2$\,Gyr) perhaps indicates that some mechanism is inhibiting star formation in the CMZ (also see e.g. \citealp{Sofue2017b} and \citealp{Sofue2017c} for varying $\Sigma_\mathrm{SFR}\mhyphen\Sigma$ relations within the Galaxy). 

A star formation relation that the CMZ is clearly inconsistent with is the so-called dense gas star formation relation (e.g. see \citealp{Longmore2013b}), which is a linear relation between the total SFR and the mass of dense molecular gas (i.e.\ gas denser than that traced by CO), $M_\mathrm{dg}$. This relation was first proposed by \citet{Gao2004}, using global measurements of external galaxies in which the amount of dense gas was inferred from the luminosity of HCN. \citet{Lada2010,Lada2012} extended this relationship to Milky Way clouds, and found that SFR is directly proportional to the mass of gas above a constant surface density threshold of $\Sigma_\mathrm{gas} \approx 116$\,\msun~pc$^{-2}$. This tight correlation appears to hold well both for nearby molecular clouds and for external galaxies (see the dashed line in the right panel of Fig.~\ref{fig:sfr_main}). 

The vast majority of the gas in the CMZ is above the density threshold proposed by 
\citet{Lada2010,Lada2012}. \citet{Longmore2013b} showed that the CMZ is presently under-producing stars by an order of magnitude relative to the SFR that would be predicted by the dense gas scaling relation (see inset panel in Fig.~\ref{fig:sfr_main}).
This is a strong counterargument against the notion of a constant surface density threshold for star formation. 
Clearly, star formation in the CMZ does not only depend on the amount of available dense gas, and other environmental factors are probably important \citep[][we will return to this topic in the context of individual clouds in \S\ref{sec:sfthreshold}]{Longmore2013b, Kruijssen2014a, Rathborne2014a, Ginsburg2018a}.

Observations of extragalactic nuclei may help to provide some further insight. 
Studies of the central $\sim$\,1\,kpc regions of external galaxies have found higher dense gas fractions and lower dense gas star formation efficiencies ($\mathrm{SFR}/M_\mathrm{dense}$), similar to the CMZ (see e.g.\ \citealp{Usero2015,Bigiel2016,Jimenez-Donaire2019,Querejeta2019,Beslic2021,Eibensteiner2022}). 
However, the current sample of resolved observations of extragalactic CMZs is very small. Further resolved observations of nearby galactic nuclei are needed to better understand whether the CMZ represents an outlier, or if other CMZs are similarly inconsistent with the dense gas star formation relation.

\subsection{Star formation history}
\label{subsec:SFR:starformationhistory}

The star formation history (SFH) of the CMZ is encoded within the stellar populations of the Nuclear Bulge (NB, $R<300\pc$) and in particular of the Nuclear Stellar Disc (NSD; \citealt{Launhardt2002}). 
For a long time it was believed that the NSD has been built up by quasi-continuous star formation occurring over the last $\sim 10\, \gyr$ \citep{Figer2004}. 
A quasi-continuous SFR may seem natural. 
After all, the different methods outlined in \S\ref{sec:SFR:current} suggest that the ``present-day'' SFR is comparable to that averaged over the past $5\mhyphen100$\,Myr within a factor of two \citep{Barnes2017}. 
And if we assume that the SFR has remained roughly constant at the present value of $0.07$ \msunyr \ over this $\sim 10\, \gyr$ time frame, we obtain a total mass of $7 \times 10^8$ \msun, which is strikingly similar to the current mass of the NSD \citep{Sormani2021}.
Furthermore, \citet{Matsunaga2011}, based on the presence of classical cepheids with ages well-determined from their pulsation periods, found that the SFR in the CMZ between 20-30 \myr \ ago was $0.075^{+0.15}_{-0.05}$\,\msunyr \ , very close to the present value.

However, \cite{Nogueras-Lara2020b} recently challenged the view of a quasi-continuous SFR using data from the GALACTICNUCLEUS survey. 
They modelled the extinction-corrected K-band colour-magnitude diagram over extended regions of the NSD as a superposition of star formation events occurring at different times. 
They concluded that $\sim 80 \%$ of the stars in the NSD formed more than $8\, \gyr$ ago, followed by a drop in star formation activity between $8\,\gyr$ and $1\, \gyr$ ago. 
They also found that the SFR has been variable during the past $1\, \gyr$, with periods of more intense activity (SFR$\sim 0.5\, \msun \peryr$). They found a SFR of $\sim0.1$ \msunyr \ averaged over the past 100 \myr, in accordance with the estimates in \S\ref{subsec:SFR:past}, and of $0.2\mhyphen0.8\, \msun \peryr$ in the past 30 \myr, i.e.\ a factor of a few higher than the present-day value (\S\ref{subsec:SFR:current}).

The CMZ also contains two young massive clusters known as the Arches and Quintuplet, which both formed $\lesssim5$\,Myr ago (\S\ref{sec:starclusters}), and the central parsec contains of the order 200 young ($\sim3\mhyphen6$\,Myr) high-mass stars \citep{Genzel2010, Lu2013}. In the context of present-day star formation, some work \citep[e.g.][]{Lu2019b} suggests the possibility that the SFR is currently increasing (\S\ref{sec:sfinaction}), though this is not strongly evidenced in full CMZ studies \citep{Hatchfield2020, Battersby2020}. Taken together, this evidence could suggest that the CMZ has experienced short bursts of star formation activity which are averaged out in integrated light measurements (\S\ref{subsec:SFR:past}).

Though the sample size is small, there is some evidence suggesting that star formation in extragalactic nuclei may also proceed episodically in discrete bursts. \cite{Allard2006} analysed the SFH in the nuclear ring of the barred galaxy NGC 4321 and found that star formation in the last $500\, \myr$ proceeded in a succession of bursts separated by roughly $\sim 100\, \myr$. \cite{Sarzi2007} extended this analysis to a sample of 8 nearby galaxies and found similar results, concluding that star formation is more likely to proceed in episodic bursts rather than continuously. \cite{Prieto2019} derived ages for 171 star clusters in the nuclear ring of NGC 1097 
by fitting SEDs in the UV and IR range, and concluded that the ring has been subject to intermittent bursts of star formation spread over the last few $100\, \myr$ with a time separation of about $20$-$30\,\myr$ (finer time separations are not resolved). \cite{Callanan2021} combined ALMA observations with ages of massive clusters derived from \citet{Harris2001} to argue that the nuclear ring in M83 experienced a starburst $5\mhyphen7\, \myr$ ago which was highly localised in time. On longer timescales, \cite{Gadotti2019} derived the SFHs in the nuclear regions of two selected galaxies in the sample of the TIMER survey, NGC 1097 and NGC 4643, with a temporal resolution of $\sim 700 \, \myr$, and found that NGC 1097 exhibits two starbursts episodes $\sim 0.5$\,Gyr and $\sim 2.5$\,Gyr ago respectively (in addition to the one that is ongoing now), while NGC 4643 does not show signs of burst activity at the temporal resolution allowed by the data. 

In conclusion, there is now clear evidence that the SFR in the CMZ has varied significantly over the past few tens to hundreds of Myrs, and that the SFRs of galactic nuclei, in general, vary considerably as a function of time.

\subsection{What drives the time evolution of star formation in galactic nuclei?}
\label{subsec:SFR:timeevolution}

Although there is evidence to suggest that star formation in the CMZ has varied significantly in the past, and that bursts of star formation activity may occur in galactic nuclei more generally, it is currently debated what may drive such variations in the SFR. 
If star formation in the CMZ is episodic, this could explain the present discrepancy between the CMZ and some of the star formation relations discussed in \S\ref{subsec:SFR:context}. 
In this context, the \emph{time-averaged} SFR of the CMZ may be consistent with star formation scaling relations, even if the CMZ is currently at a low point in its star formation history \citep{Kruijssen2014a, Krumholz2015}. 

One way to explain the variable SFR, based on semi-analytical or simple 1D hydrodynamical models, is that stellar feedback drives a recurrent cycle of star formation \citep{Loose1982,Krugel1993,Elmegreen1994,Morris1996,Stark2004,Kruijssen2014a,Krumholz2015,Krumholz2017,Torrey2017}. The common idea of all these models is that gas accumulates in the central region until a starburst occurs. This starburst releases a great amount of energy that temporarily halts star formation, and once that turbulent energy dissipates, the cycle repeats. The predicted time intervals between the bursts range between $\sim 20$\,Myr \citep{Krumholz2017} to $\sim 100$\,Myr \citep{Loose1982}. 

Various authors have recently tested this picture using full 3D simulations, finding contrasting results. \cite{Torrey2017} ran numerical simulations of isolated galaxies with the FIRE feedback model and found that the SFR within the central $\sim100$\,pc goes through dramatic, oscillatory cycles. \cite{Armillotta2019} performed simulations of the Milky Way in the rigidly-rotating barred gravitational potential of \cite{Ridley2017}. They found that the CMZ SFR varies by $\approx1.5$ dex over a timespan of $\sim 500\,\myr$, even though the gas mass in the CMZ stays relatively constant (implying large variations in the depletion time). This variability occurs in cycles of $50\,\myr$ and is driven by stellar feedback, in accordance with the expectations of the 1D models. 

However, \cite{Sormani2020b}, using a similar setup to that of \citet{Armillotta2019}, found that stellar feedback does not drive cycles of star formation in their simulations. Rather, they found that the SFR is quite steady and directly proportional to the time-varying CMZ mass (implying a roughly constant depletion time). \cite{Moon2021} performed semi-global hydrodynamic simulations of nuclear rings subject to constant mass inflow rates (to avoid the time-variability that naturally arises in fully global simulations) and found that the SFR exhibits only modest (within a factor of $\sim$2) temporal fluctuations, similar to \cite{Sormani2020b}. 
Finally, \citet{Orr2021} perform cosmological zoom simulations of seven Milky Way mass galaxies from FIRE-2 and find a mix of the two pictures described above. They identify two classes of galaxy centres: one with highly asymmetric gas and SFR distributions, which undergo rapid SFR variations on 10 Myr timescales, and another with smooth gas and SFR distributions. 

The origin of the diverse results obtained from the simulations mentioned above is not clear. The absence of the boom/bust behaviour in the simulations of \cite{Sormani2020b} and \cite{Moon2021} may be because they only include supernova feedback but lack early feedback such as stellar winds, photoionization, and radiation pressure from young stars \citep[which can disperse gas to parsec scales surrounding sites of star formation before the onset of the first supernova;][]{Barnes2020b}. Alternatively, the simulations of \cite{Armillotta2019} may not be sufficiently resolved and thus produce numerical artefacts \citep{Sormani2020b,Moon2021}. Addressing the origin of the contrasting simulation results described above will require careful comparison of the different ways of implementing star formation and feedback in the simulations.

Alternatively (or in addition) to stellar feedback cycles, variations in the SFR may also reflect time variability in the mass inflow rate \citep[][]{Seo2019,Sormani2020b, Moon2022}. Observations of the Milky Way \citep{Sormani2019b} and external galaxies \citep{Leroy2021, Beslic2021} often show that the dust lanes, which transport the gas to the centre, contain dense clumps that may produce discrete large accretion events. 
\citet{Moon2022} used simulations to study SFR variability resulting from a controlled oscillating inflow rate. 
They found that if the inflow oscillations have sufficient amplitude they can cause large (a factor of $>5$) fluctuations of the SFR over timescales of $\Delta t > 50$ Myr. Finally, starbursts may be also produced by perturbations induced by a live stellar potential \citep[][as opposed to the fixed potential in the above simulations]{Emsellem2015} or minor mergers \citep{Mihos1994}, while AGN activity might quench star formation by blowing the gas out \citep{Combes2017a}.

In conclusion, the main driver of SFR variations in the CMZ is currently unclear, though the main candidates are stellar-feedback-driven star formation cycles and a variable bar-driven inflow rate. There are good prospects that numerical simulations will soon be able to determine whether the former is a physically viable mechanism, while insight on the latter might come from studying nearby barred galaxies.


\section{Macro-evolution of the CMZ} 
\label{sec:macroevolution}
The gas inwardly migrating from large-scales towards the CMZ may meet a number of fates: it may contribute to the central mass reservoir, form stars, or may be expelled from the centre via an outflow \citep{Morris1996}.
It is useful to think about these fates in the context of mass conservation in a cylindrical volume containing the CMZ, $R<200\, \pc$ and $|z|<100\, \pc$ \citep[e.g.][]{Crocker2012}:
\begin{equation}
    \dot{M}_\mathrm{IN} = {\rm SFR} +  \dot{M}_{\rm OUT} + \dot{M}_{\rm CMZ},
\end{equation}
where $\dot{M}_\mathrm{IN}$ is the mass inflow to the CMZ, SFR is the star formation rate, $\dot{M}_{\rm OUT}$ is the mass lost from the CMZ due to outflowing gas, and $\dot{M}_{\rm CMZ}$ is the rate of change of the total gas mass in the CMZ. Note that the rate of change of the central black hole hole mass is negligible compared to the other contributions ($\dot{M}_{\rm SgrA*}<10^{-8}\,\msun\,\yr^{-1}$, e.g.\ \citealt{Genzel2010}). Current estimates (quantified throughout this section) suggest $\dot{M}_\mathrm{IN}\simeq 0.8\,\msun\,\yr^{-1}$ and $\dot{M}_{\rm OUT}\gtrsim0.6\,\msun\,\yr^{-1}$. When combined with the present day SFR of $\sim$\,0.07\,\msunyr\ (\S\ref{subsec:SFR:context}), these values could imply that the gas mass in the CMZ is either in a quasi-steady state or presently increasing \citep{Crocker2012, Krumholz2015,Sormani2019b}, with potentially important implications for the time-variability of star formation (\S\ref{subsec:SFR:timeevolution}). We therefore begin the second half of the review by describing in detail the macro-evolution of the CMZ. 

\subsection{Gas flow towards the nucleus}
\label{sec:inwardmassflow}

Interstellar matter is transported from the Galactic disc (Galactocentric radii $R>3~\kpc$) down to the very centre of the Milky Way in a sequence of steps. 

\subsubsection{From the Galactic disc to the CMZ} \label{sec:barinflow}

The Galactic bar transports gas from the Galactic disc at $R\gtrsim3~\kpc$ down to the outskirts of the CMZ. However, there are different theories as to what happens to the gas once it approaches this location. 

One interpretation is that the Galactic bar drives gas inwards from $R\gtrsim3\kpc$ down to $R\sim$ several hundred pc, at which point it settles into a disc \citep{Krumholz2015, Krumholz2017}. 
\citet[][]{Kruijssen2014a} argued that acoustic instabilities \citep{Montenegro1999} drive both turbulence and angular momentum transport in this disc, causing it to flow inwards and accumulate into a ring of material at $R\simeq 100\pc$. 
However, \citet{Sormani2020c} used hydrodynamical simulations to show that the modes that are predicted to be acoustically unstable by \citet{Montenegro1999} are actually stable, and therefore that the acoustic instability does not drive turbulence or angular momentum transport in the gas. 
The alternative view is that gas flow onto the CMZ happens directly through the bar dust lanes \citep[\S\ref{sec:generaldynamics} and Fig.~\ref{fig:sketch};][]{Fux1999,Liszt2006,Liszt2008,Rodriguez-Fernandez2008,Tress2020}. 
Our view through the Galactic plane complicates the interpretation (see \S\ref{sec:3d}), though a direct connection between dust lanes and nuclear rings is commonly seen in both observations of nearby galactic nuclei and numerical simulations (\S\ref{sec:IG}). 

\cite{Gerhard1992} estimated an inflow rate into the CMZ of $\sim0.1\,\msun\,\yr^{-1}$ by dividing the mass contained in the contour of the $\{l,v\}$ parallelogram by the dynamical time of the cusped orbit (\S\ref{sec:generaldynamics}). \cite{Figer2004} arrived at a higher value of $\sim0.4\,\msun\,\yr^{-1}$ using the same method but assuming a larger mass in the $\{l,v\}$ parallelogram. Note however, as discussed in \S\ref{sec:cmz}, that the interpretation of the parallelogram has changed since these works were published. \cite{Crocker2012} estimated $0.4<\dot{M}_\mathrm{IN}/\msun\,\yr^{-1}<1.8$ by assuming that the CMZ is in a quasi-steady state.

\cite{Sormani2019b} recently revisited this question, and quantified the mass inflow rate by combining CO observations of the dust lanes with a simple geometrical model informed by simulations of gas flow in barred potentials. They report an inflow rate of $\dot{M}_\mathrm{IN} = 2.7^{+1.5}_{-1.7} \msun \yr^{-1}$ averaged over a timescale of $\sim 15\myr$. 
They find evidence that the inflow is time-variable since the dust lanes contain dense clumps that, when accreted, will produce spikes in the inflow rate. Discrete accretion events might have helped to fuel starburst episodes in the past (see \S\ref{subsec:SFR:starformationhistory} and \S\ref{subsec:SFR:timeevolution}). 

However, \cite{Sormani2019b} overestimate the inflow rate because of the assumption that all the gas on the dust lane will immediately accrete onto the CMZ when it reaches it. It is likely that some of the gas will overshoot the CMZ and get accreted at a later stage. \cite{Hatchfield2021} quantified this effect using hydrodynamical simulations, and estimated the overall efficiency of the inflow via the dust lanes to be of the order $30 \pm 12\%$. They found a corrected, instantaneous inflow rate of $\dot{M}_\mathrm{IN} = 0.8 \pm 0.6\, \msunyr$. 

In summary, over the past few decades, the inflow rate along the bar has been reported to be in the range $0.4\mhyphen2.7\,\msunyr$; we adopt a value of $\dot{M}_\mathrm{IN} = 0.8 \pm 0.6\, \msunyr$ in this review. Accretion at this rate implies that the entire gas content of the CMZ is completely renewed on a timescale of $M_{\rm CMZ}/\dot{M}_\mathrm{IN} \simeq 50\, \myr$. 

\subsubsection{From the CMZ to the nucleus} \label{sec:cndinflow}

How gas migrates from the CMZ towards the nucleus is unclear. 
Although the Galactic bar is very efficient at transporting the gas from the disc to the CMZ (\S\ref{sec:barinflow}), it is ineffective at driving the gas further inwards (e.g.\ \citealt{Shlosman1990}; and next paragraph). Several mechanisms have been suggested, including: (i) transport driven by stellar feedback \citep{Davies2007}; (ii) viscous mass transport driven by magnetic field instabilities  \citep{Balbus1998}; (iii) the presence of a nuclear bar, which could repeat on a smaller scale the process that causes the larger bar-driven inflow (\citealt{Shlosman1989}; as discussed in \S\ref{sec:potential}, it cannot be ruled out that the NSD is actually a nuclear bar); (iv) weak $m=2,4,6,\dots$ or external (e.g.\ mergers) perturbations which may trigger episodic inflow \citep{Combes2001,Kim2017}. 

\cite{Tress2020} compared two simulations of gas flow in a Milky Way barred potential that are identical except that one has gas self-gravity and supernova feedback and the other does not. They found that while the bar-driven inflow rate to the CMZ is identical in the two simulations at a rate of $\sim 1\, \msun\yr^{-1}$ (consistent with the measured value, see \S\ref{sec:barinflow}), the inflow from the CMZ inwards is zero in the simulation without supernova feedback (the gas simply piles up in the CMZ ring-like structure), while it is significant in the simulation with supernova feedback (see also \citealt{Salas2020}). They concluded that supernova feedback associated with the intense star formation activity in the CMZ can generate an inflow rate of $\sim 0.03\,\msun\,\yr^{-1}$ ($<5\%$ of the inflow rate from large-scales; \S\ref{sec:barinflow}), by stochastically launching parcels of gas towards the centre and/or by randomly changing their angular momentum. At this rate the CND would only take $\sim 3~\myr$ to build up, consistent with the view inferred from observations that the CND is a transient structure on this timescale \citep[][\S\ref{sec:cnd}]{Requena-Torres2012,Mapelli2016,Ballone2019,Dinh2021}. 

The gas continues its journey from the CND towards the centre through the so-called Galactic centre minispiral (also known as the Sgr\,A West H{\sc ii} region), a system of orbiting filamentary streamers that are photoionised by the high-mass stars located within the central parsec \citep{Lo1983, Ekers1983, Nitschai2020a, Heywood2022}. The minispiral streamers have an associated mass inflow rate of $\sim10^{-3}$\,M$_{\odot}$\,yr$^{-1}$ (or $\sim0.1\%$ of the inflow rate from large-scales; \S\ref{sec:barinflow}) into the central few arcseconds \citep{Jackson1993, Genzel1994}. Eventually, only a small fraction of this gas will reach the central black hole (see \citealt{Genzel2010} for a review).

How the contribution of supernova feedback compares to that of the other mechanisms listed above in driving gas from the CMZ towards the centre is an open question \citep{Combes2017b}. Insights from extragalactic observations may help in addressing this problem \citep{Garcia-Burillo2005,Hunt2008}. In the future, it will be important to address the difficult task of quantifying this more precisely.

\subsection{Gas expulsion from the Galactic Centre} 
\label{sec:feedback}

Gas inflowing from large scales that does not contribute directly to the central mass reservoir or to star formation may be expelled from the Galactic Centre. Here we review the evidence for gas expulsion from the centre of the Galaxy and the possible driving mechanisms. 

\subsubsection{Evidence for outbursts from the Galactic Centre}
\label{sec:feedback:largescaleoutflow}

Although the CMZ and Sgr\,A$^{*}$ are relatively quiescent compared to nuclear starburst galaxies and active galactic nuclei, there is substantial evidence for historic bursts of energetic phenomena from the Galactic Centre. The largest of these are the so-called ``bipolar hyper shells", detected in both X-ray and radio emission, which extend approximately 14\,kpc above and below the Galactic Plane \citep{Sofue2000, Carretti2013, Sofue2016, Predehl2020}.
Additionally, there are the Fermi bubbles: a bipolar structure identified in $\gamma$-ray emission that extends $\sim$\,10\,kpc out of the Galactic plane (e.g.\ \citealp{Su2010, Ackermann2014}). 
The total energy content of the hypershells and Fermi bubbles is of the order $10^{55\mhyphen56}$\,erg \citep{Sofue1987, Bland-Hawthorn2003, Su2010}. 
The Fermi Bubbles and the X-ray emission detected with eROSITA show remarkably similar morphology, leading to the suggestion that the two are causally related \citep{Predehl2020}. 
Despite their very large vertical extent, the narrow 100-200\,pc diameter waist of the $\gamma$-ray and X-ray bubbles indicates that they are driven from close to the nucleus \citep{Carretti2013}. 

On smaller scales, there are the radio lobes and the X-ray ``chimneys", which extend to a height of $\sim$\,400\,pc (e.g. \citealp{Sofue1984, Law2009, Law2010, Ponti2019, Heywood2019}), originating from within $\sim50$\,pc of Sgr\,A$^{*}$. 
Their energy content is $\sim10^{52\mhyphen53}$\,erg, which led \citet{Heywood2019} to speculate that they may be a lower energy analogue to the Fermi bubbles, with \citet[][see also \citealp{Ponti2021}]{Ponti2019} further suggesting that the chimneys may act as a channel that transports energy to larger Galactic latitudes. 

Centred on the Galactic nucleus, extending over $\sim1.8^{\circ}\times0.5^{\circ}$, there is an extended warm ($kT\sim1$~keV; $T\sim10^7$~K) and hot ($kT\sim6.5$~keV; $T\sim10^{8}$~K) plasma, producing a high background of soft and hard X-ray radiation \citep[see e.g.][]{Ponti2015}. 
The majority of the soft X-ray emission is from diffuse thermal emission (\citealp{Ebisawa2001, Wang2002}), but the origin of the hot component is debated (e.g. \citealp{Crocker2012, Ponti2013}).
At $1.5^{\circ}$ from the Galactic Centre, $\sim80\%$ of the hot component is resolved into point sources \citep[e.g.\ accreting white dwarfs and coronally active stars;][]{Revnivtsev2009}.
Nevertheless, it is not excluded that a diffuse hot-plasma component is present in the Galactic Centre (e.g. \citealp{Koyama2009, Uchiyama2013}). 
At temperatures close to $10^{7\mhyphen8}$\,K the gas would be unbound to the Galaxy \citep[though magnetic fields may help to confine it;][]{Nishiyama2013}, so the energy required to maintain a diffuse hot-plasma component is substantial, $E\sim10^{52-56}$\,erg (e.g. \citealp{Morris1996, Crocker2012, Uchiyama2013}).

\subsubsection{Outflow rate and driving mechanisms}
\label{sec:feedback:outflowmechanisms}

The rate at which gas is expelled from the CMZ is uncertain. \citet{Bordoloi2017} estimated an outflow rate in warm ionised gas associated with the Fermi bubbles of $\dot{M}_{\rm OUT, warm} \sim 0.4$\,\msunyr. 
\citet{DiTeodoro2018} estimated an outflow rate in neutral gas of $\dot{M}_{\rm OUT, HI} \sim 0.1$\,\msunyr \ and
\citet{Diteodoro2020} inferred a comparable contribution from cold molecular gas of $\dot{M}_{\rm OUT, H2} \sim 0.1$\,\msunyr.
There is also a possible contribution from the hot gas component, though this is far more uncertain \citep[upper limit of $\dot{M}_\mathrm{hot} < 1$\,\msunyr, and possibly much lower;][]{Miller2016, Sormani2019b}. 
Combining these estimates, the total CMZ outflow rate associated with the Fermi bubbles is of the order $\dot{M}_{\rm OUT, Fermi} \gtrsim 0.6 $\,\msunyr. 
Corresponding estimates for the mass outflow associated with the X-ray chimneys are scarce. 
\citet{Crocker2012} estimated an outflow rate of $\dot{M}_{\rm OUT, warm} \sim 0.3$\,\msunyr \ assuming a mass of $\sim$\,2$\times10^{5}$\,\msun\ for the warm ionised gas (\citealp{Law2010}), a canonical outflow speed of 100\,\kms, and a physical extent of $140$\,pc. 
The molecular gas mass is similar \citep[$\sim$\,3$\times10^{5}$\,\msun;][]{Bland-Hawthorn2003}, and hence $\dot{M}_{\rm OUT, H2} \sim 0.45$\,\msunyr.
Combining the two gives a total outflow rate of $\dot{M}_{\rm OUT, chimneys} \sim 0.8 $\,\msunyr.  The above estimates for the outflow rates within the Fermi bubble and radio lobes are, therefore, broadly consistent. Assuming that the chimneys act as a channel connecting to the base of the Fermi bubbles \citep{Ponti2021}, this points to a \emph{time averaged} outflow rate of $\dot{M}_{\rm OUT} \gtrsim 0.6\mhyphen0.8 $\,\msunyr, corresponding to a mass loading factor of $\eta_\mathrm{ml}\equiv \dot{M}_{\rm OUT}/\dot{M}_{\rm SFR} \simeq10$.

The driving source of the outflow(s) is contested \citep{Su2010}. Theories include an explosive outburst from Sgr\,A$^{*}$ \citep{Zubovas2011, Guo2012,Yang2022} or star formation activity \citep[e.g.,][]{Strickland2000,Law2010,Crocker2012, Lacki2014,DiTeodoro2018}. Focusing on the latter, individual mini starburst events, such as those that led to the formation of the Galactic Centre's young clusters (e.g. the Arches, Quintuplet, and the young nuclear cluster; \S\ref{sec:starclusters}), may power the outflows. Indeed, \citet{Heywood2019} speculated that the radio lobes may be associated with the young central cluster, noting the similarity between the dynamical time of the bubbles ($\sim$7\,Myr) and the estimated age of the population of 200 or so young ($\sim3\mhyphen6$\,Myr) high-mass stars located in the central parsec \citep{Genzel2010}. 

\citet{Law2010} advocated for a more secular picture. They demonstrated that energy required to drive the Galactic Centre Lobe ($10^{52\mhyphen53}$\,erg; \S\ref{sec:feedback:largescaleoutflow}) is consistent with the energy input by supernovae and stellar winds in the Galactic Centre region, and therefore suggested that a ``starburst'' may not be necessary. There are 18 known supernova remnants within $\ell\pm2\deg$ of the centre \citep{Green2019,Dokara2021}, 10-12 of which are within the inner $|\ell|<1\deg$ \citep{Ponti2015}, though the true number may be higher \citep[e.g.][]{Oka2007, Tsujimoto2018}. The estimated supernova rate is $\sim2\mhyphen15\times10^{-4}$\,yr$^{-1}$ \citep{Crocker2011a, Ponti2015}. Assuming a mechanical energy release per supernova of $10^{51}$\,erg, the corresponding power delivered by supernova is $\sim6\mhyphen50\times10^{39}$\,erg\,s$^{-1}$. Other sources of mechanical energy, e.g., stellar winds, could provide an additional contribution to the total power input. \cite{Crocker2015} argued that weak but sustained star formation (and the resulting supernovae), over the past $\sim$ few $\times10^{8}$\,yr may be driving the Fermi bubbles.

In summary, the gas outflowing from close to the centre plays a significant role in regulating the macro-evolution of the CMZ.
The primary driver of the outflow remains unclear; it may be driven either by star formation or by past activity from the now-dormant central supermassive black hole.

\subsection{The CMZ gas reservoir}\label{sec:massreservoir}

The gas that is not actively forming stars, and which is not ejected from the Galactic Centre (\S\ref{sec:feedback}), contributes to the central gas reservoir. Here we review our current understanding of the structure of the CMZ gas reservoir, as well as the impact that the physical processes discussed in \S\ref{sec:inwardmassflow} and \ref{sec:feedback} have on the ISM conditions.

\subsubsection{The projected distribution of gas and stars}
\label{sec:gasstardistrib}

For several decades it has been noted that the gas in the CMZ is asymmetric about the Galactic Centre. \citet{Bally1988} showed that roughly 3/4 of the emission from $^{13}$CO $(1-0)$ in the CMZ is found at $(l>0)$ \citep[see Fig.~\ref{fig:lbv_main} and][]{Dame2001, Lis2001, Bally2010, Molinari2011, Longmore2013a, Eden2020}.
The gas is also asymmetric in velocity, with most of the gas at positive longitudes also at positive velocity \citep{Bally1988, Henshaw2016b}. Roughly half of all of the negative velocity emission is located at positive longitudes, which is ``forbidden'' for gas on purely circular orbits (Fig.~\ref{fig:lbv_main}).

The gas asymmetry is reflected in the locations of prominent molecular clouds, which are also preferentially located at positive longitudes \citep{Bally2010,Longmore2013a}. 
This is evident in Figure~\ref{fig:rgb_main}, where the clouds stand out as dark features at 8~$\mu m$ and 24~$\mu m$ \citep{Churchwell2009, Carey2009}, and emit strongly at $\lambda\geq250$~$\mu m$ \citep{Molinari2010, Molinari2011}. 
From $-1^{\circ}\lesssim l\lesssim0^{\circ}$, the most prominent molecular clouds are Sgr\,C, and the 20\,\kms \ and 50\,\kms \ clouds (otherwise known as the Sgr\,A clouds).
At $l>0^{\circ}$ we find the highest column density clouds in the Galactic centre. G$0.253+0.016$, otherwise known as ``the Brick'' \citep{Lis1998}, represents the first in a sequence of molecular clouds known collectively as the ``dust ridge'', roughly spanning $0.2^{\circ} \mhyphen 0.8^{\circ}$ in longitude. 
The dust ridge is bookended by G$0.253+0.016$ (`cloud a') and Sgr\,B2, with a further five clouds, `clouds b -- f', in between (Fig.~\ref{fig:rgb_main}). Although the bulk of the high column density material (total mass of $\approx1.8\times10^{7}$\,\msun) is contained within the inner $|l| \leq 1.0$, $|b| \leq 0.5$, a comparable amount of mass is located at longitudes beyond Sgr\,B2, with large cloud complexes, including the $1.3^{\circ}$ cloud, extending from $l \sim$ 1.0$^{\circ}$ to 3.5$^{\circ}$ \citep{Molinari2011, Longmore2013b}.

Numerical simulations of Milky Way-like galaxies suggest that asymmetric gas distributions in nuclear rings are statistically likely. High-resolution 2D isothermal simulations without self-gravity show that the flow along dust lanes is subject to hydrodynamic instabilities, and is therefore unsteady \citep{Kim2012c,Sormani2015c}. \cite{Sormani2018b} used 3D hydrodynamical simulations to explicitly test the idea that this unsteady flow may explain the observed asymmetry in the CMZ even in the absence of self-gravity, stellar feedback, and magnetic fields. They demonstrated that an asymmetry develops spontaneously, even when starting with perfectly symmetric initial conditions. Comparing to the CMZ, they found that $>70\%$ of the CO lies at one side of the Galactic centre for $>30\%$ of the time in their simulation. 
The inclusion of additional physical mechanisms (e.g.\ stellar feedback, self-gravity, a live stellar potential, or pre-existing large-scale asymmetries) will create further asymmetries in the gas distribution \citep{Fux1999, Rodriguez-Fernandez2008, Emsellem2015, Torrey2017, Armillotta2020}.
Together, these simulations demonstrate that the asymmetry may be driven, at least partly, by physical processes acting on scales larger than the CMZ itself. It is also likely to be transient, such that in a dynamical timescale ($\sim5\mhyphen10$\,Myr) an asymmetry in the opposite sense is just as likely. 

It remains unclear whether there is any asymmetry among young stellar populations in the CMZ.
Most of the present-day embedded YSOs, as traced by dust cores, masers, and outflows (\S\ref{sec:sfinaction}), are located at positive longitudes, following the gas asymmetry \citep[\S\ref{sec:incipientsf}; ][]{Ginsburg2018b,Rickert2019,Hatchfield2020,Lu2019a,Lu2019b,Lu2020,Lu2021,Walker2021}. 
However, stars that formed $\gtrsim1$ Myr ago do not.
\citet{Yusef-Zadeh2009} suggested that there is a substantial excess of YSOs, including Class II (disc-only) objects that are no longer embedded, at negative longitudes.
However, at least some of these sources have been shown to be misclassified as YSOs, and are likely main sequence stars \citep[][see \S\ref{subsec:SFR:current}]{An2011, Koepferl2015}.
Among somewhat older stars detected in the near-infrared (1-2\um) window, which formed recently ($<10$ Myr), the spatial distribution appears relatively even with longitude \citep{Nandakumar2018,Clark2021}, though because of the challenges of classifying these sources, the samples remain incomplete. 
Adding to the confusion, there is a population of stars believed to be forming along the far side dust lane (see \S\ref{sec:dustlanes}), falling toward, but not necessarily into, the CMZ \citep{Anderson2020}.

In summary, there is clear evidence for an asymmetry in the gas distribution in the CMZ, and numerical simulations have demonstrated that such an asymmetry is statistically likely. However, while many models have been developed that can partly or completely account for the asymmetry, it remains extremely challenging to pinpoint which of the potentially important physical processes dominates in driving it. As for the young stellar populations in the CMZ, asymmetry has been proposed but remains inconclusive.

\subsubsection{The 3D geometry of the CMZ} 
\label{sec:3d}

Observations of extragalactic systems from the UV through to the (sub-)millimetre offer some qualitative insight into the 3D geometry of the CMZ. Extragalactic nuclei exhibit a range of structures, including rings and both tightly wound and chaotic nuclear spirals \citep{Peeples2006, Comeron2010, Pan2013, Viti2014, Audibert2019, Callanan2021, Lee2022}.
Broadly speaking, the overall consensus is that the gas in the CMZ is organised into an eccentric ring-like or toroidal structure (\S\ref{sec:cmz}). 
However, constraining the precise details of the geometry is challenging, necessarily multi-faceted, and the topic remains controversial.

\paragraph{Global models of the 3D geometry:}\label{sec:3dmodels} Attempts to determine the 3D geometry of the CMZ have typically involved translating the $\{l,b,v\}$ distribution of gas into a face-on morphological and kinematical description. 
Most of the molecular gas mass in the CMZ \citep[$\sim3\times10^{7}$\,\msun;][]{Sofue1995a} is distributed throughout two extended gas streams that are located between $-0.65^{\circ}<l<0.7$ and $-150\,{\mathrm \kms}<v<100\,{\mathrm \kms}$ \citep{Bally1987, Sofue1995a, Tsuboi1999, Henshaw2016b, Eden2020}.
Although the streams are difficult to distinguish on the plane of the sky due to confusion along the line-of-sight, the distinction becomes more clear in $\{l,v\}$-space (Fig.~\ref{fig:lbv_main}). 
They are separated in velocity by $\approx50$\,\kms \ and appear almost parallel in $\{l,v\}$, increasing in velocity in the direction of increasingly positive longitudes \citep{Henshaw2016b}. 
The gas associated with the most massive molecular clouds can be directly attributed to these contiguous streams. 
The modelling of the $\{l,b,v\}$ distribution of the molecular gas has led to three main interpretations for the true 3D distribution, which we illustrate in Fig.~\ref{fig:sketch}.

The first interpretation is that the gas in the CMZ is organised into a two-armed spiral (see panel B; Fig.~\ref{fig:sketch}), similar to the nuclear spirals that are commonly seen in the centres of barred spiral galaxies \citep{Schinnerer2002, Martini2003a, Martini2003b, vandeVen2010}. \citet{Sofue1995a} qualitatively interpreted the $\{l,b,v\}$-streams described in \S\ref{sec:3dobs} as two spiral arms centred on Sgr\,A$^{*}$ \citep[see also][]{Scoville1974, Johnston2014}. 
The precise details of this interpretation have evolved over time. 
\cite{Ridley2017}, for example, used simulations of gas flow in a Milky Way-like barred potential to develop this theory into a quantitative dynamical model. 
Within this interpretation, there are different views on where the individual clouds should be located (see \S~\ref{sec:3dobs}).

The second interpretation is that the gas is distributed throughout an approximately elliptical ring, with the streams described above representing the near- and far-side of this ellipse.
This interpretation is also inspired by observations of nearby barred galaxy centres, which often show a ring-like gaseous nuclear structure. 
\citet{Binney1991} initially interpreted the dense gas in the CMZ as following nearly-elliptical $x_2$ orbits (\S\ref{sec:gasdynamics}). 
The advent of \emph{Herschel} in the early 2010s provided a striking new view of the CMZ, inspiring a new geometrical model within this framework.
\citet{Molinari2011} noted that the dust continuum emission appears to form an $\infty$-shaped pattern on the plane of the sky (Fig.~\ref{fig:rgb_main}).
They modelled the $\{l,b,v\}$ streams as a single, vertically oscillating, elliptical orbit with a radius of $\approx100$\,pc (see panel C; Fig.~\ref{fig:sketch}). 
A notable difference between the \cite{Molinari2011} kinematic $x_2$-like orbits and the dynamical $x_2$ orbits of \cite{Binney1991} is that in the former the centre of the ring is displaced with respect to Sgr\,A$^{*}$, such that Sgr\,A$^{*}$ is closer to the front than the back of the ellipse. 
However, this displacement is problematic from a dynamical point of view since $x_2$ orbits are always centred on the bottom of the gravitational potential (\S\ref{sec:generaldynamics}).

The third interpretation is that the observed gas streams are different portions of a single ballistic open orbit (see panel D; Fig.~\ref{fig:sketch}). 
By using an order of magnitude more kinematic measurements than \cite{Molinari2011}, \citet[][]{Kruijssen2015} highlighted that the \citet{Molinari2011} model provides a poor fit to the $\{l,b,v\}$ distribution.
Building on some of the successes of the \citet{Molinari2011} model, namely that the orbit is eccentric and that it oscillates vertically, \citet{Kruijssen2015} integrated ballistic orbits in an axisymmetric gravitational potential and fitted them to the molecular gas distribution in $\{l,b,v\}$-space.
Their best-fitting model is an open, eccentric, \emph{pretzel}-shaped orbit.

\citet{Henshaw2016b} compared the \citet{Sofue1995a}, \citet{Molinari2011}, and \citet{Kruijssen2015} models. 
They concluded that the latter of these provides the best morphological match to the molecular gas distribution in $\{l,b,v\}$-space. 
Despite this, many questions remain. 
It is not clear, for example, how the \citet{Kruijssen2015} model relates to the larger-scale context discussed in \S\ref{sec:IG}, in particular to the dust lanes and the other features surrounding the CMZ. 
Furthermore, this geometry has yet to be replicated by large-scale numerical simulations in a bar potential \citep{Armillotta2020, Tress2020}. 

\paragraph{The line-of-sight location of clouds, points of contention, and open questions:}
\label{sec:3dobs} Several works have independently attempted to determine the line-of-sight locations of individual CMZ clouds, with mixed success. 
The different models described in \S\ref{sec:3dmodels} exhibit various levels of (dis)agreement both with these observations and with each other (Fig.~\ref{fig:sketch}). 
In what follows, we describe the effort to determine the line-of-sight location of individual clouds and the extent to which the models agree (or disagree) with each other. 
\medskip

\noindent\emph{Sgr\,B2 \& the dust ridge}: The cloud with the strongest constraints on its line-of-sight position so far is Sgr\,B2.
\citet{Reid2009} determined the distance to Sgr\,B2 using trigonometric parallax measurements, finding that the cloud resides in the foreground of Sgr\,A$^{*}$ at a distance of $R\simeq130\pm60$\,pc. 
\citet[][see also \citealp{Yan2017}]{Sawada2004} compared the relative strength of CO emission and OH absorption features, and found results that are consistent with Sgr\,B2 residing in the foreground, although the spatial resolution of the OH observations (12\,arcmin or $\approx30$\,pc) was insufficient to resolve the individual streams or clouds.
X-ray measurements that use the time delay of reflected X-ray radiation, due to flare-like events from Sgr\,A$^{*}$, to constrain cloud positions also support the view that Sgr\,B2 is in the foreground \citep{Ponti2010,Clavel2013,Walls2016, Churazov2017a, Churazov2017b, Chuard2018, Terrier2018}.
In particular, \citet{Chuard2018} find good agreement between the line-of-sight position of Sgr\,B2 determined using this method and that estimated by \citet{Reid2009}.
The Sgr B2 `cores' M, N, and S appearing as absorption features in [CII] maps further supports this orientation \citep{Harris2021}.
Finally, location of Sgr\,B2 in the foreground of Sgr\,A$^{*}$ is qualitatively consistent with the fact that the cloud appears in extinction in the mid-infrared (Fig.~\ref{fig:rgb_main}).
The remaining dust ridge clouds, belong to a contiguous stream in $\{l,b,v\}$-space that connects to Sgr\,B2, and, like Sgr B2, are observed in extinction, suggesting that they too are in the foreground of Sgr\,A$^{*}$. Although there has been some suggestion that the ``Brick'' (G0.253+0.016) is located outside of the Galactic Centre \citep{Zoccali2021}, this has been refuted \citep{Nogueras-Lara2021}.

As can be seen in Fig.~\ref{fig:sketch}, both the spiral arm (panel B) and the open stream (panel D) models agree that the dust ridge clouds are located in the foreground of Sgr\,A$^{*}$. 
Given the independent measurements described above, this conclusion seems reasonably uncontroversial. 
However, the elliptical orbit (panel C) places Sgr\,B2 behind Sgr\,A$^{*}$, in contrast to the parallax and X-ray measurements.
\medskip 

\noindent\emph{The Sgr\,A clouds}: The complexity imposed by projection effects in the Sgr\,A region means that the location of the 20 and 50\,\kms \ clouds are particularly controversial \citep[see ][for a comprehensive summary]{Ferriere2012}. 
There is circumstantial evidence to suggest that these clouds are interacting with the known supernova remnant Sgr\,A East, which is thought to envelop the CND (\S\ref{sec:cnd}) and the minispiral (\S\ref{sec:inwardmassflow}), located within the central 10\,pc \citep{Zylka1990,Ho1991,Serabyn1992,Coil1999, Coil2000, Yusef-Zadeh1999a, Sjouwerman2008, Hsieh2017, Hsieh2019, Tsuboi2018, Tanaka2021}. 
The general consensus among the studies seeking to describe the 3D structure of the gas in the immediate vicinity of Sgr\,A (i.e. not the CMZ as a whole) is that the 20\,\kms \ cloud lies in front of the Galactic centre, the CND, and Sgr\,A East, and that the majority of the gas associated with the 50\,\kms \ cloud lies adjacent to Sgr\,A East, although some note that part of the cloud may curve around into the foreground \citep{Herrnstein2005, Lee2008}.

The placement of the Sgr\,A clouds within 10-20\,pc of Sgr\,A$^{*}$ is somewhat problematic for each of the models described in \S\ref{sec:3dmodels}. 
While the \citet{Sofue1995a} model does not explicitly describe the location of the Sgr\,A clouds, this model is based on the interpretation that the two contiguous $\{l,b,v\}$-streams represent physically continuous spiral arms. 
Since the emission belonging to the 20 and 50\,\kms \ clouds is associated with the far-side arm in this model, it implicitly places them behind Sgr\,A$^{*}$, connecting to Sgr\,C.
\citet{Ridley2017} instead associated the 20 and 50\,\kms \ clouds with the near-side spiral arm in their dynamical model (though the location of the dust ridge clouds is modified as a result). 
The displacement of the elliptical ring relative to the location of Sgr\,A$^{*}$ in the \citet{Molinari2011} model means that the clouds are located close to the nucleus, while also remaining part of the contiguous $\{l,b,v\}$-stream. 
However, as described in \S\ref{sec:3dmodels}, this displacement comes at the expense of physical consistency with regard to the assumption that the gas is following $x_2$ orbits, and furthermore forces the model to place Sgr\,B2 behind Sgr\,A$^{*}$, in contention with several independent lines of evidence.
The \citet{Kruijssen2015} model places the 20\,\kms \ and 50\,\kms \ clouds $\gtrsim60$\,pc in the foreground of Sgr\,A$^{*}$.
These authors proposed that the only way for the clouds to be closer to Sgr\,A$^{*}$ is if they are physically unrelated to the $\{l,b,v\}$-stream with which the bulk of their molecular emission is associated.
They argued that this is unlikely, and instead speculated that the extension of the clouds along the line-of-sight may be a deciding factor in resolving this uncertainty. 
Finally, \citet{Tress2020} suggest that the conundrum could be resolved if the gas geometry is more complicated than assumed by each the models described in \S\ref{sec:3dmodels}.
They proposed that the $\{l,b,v\}$-stream containing the 20 and 50\,\kms \ clouds may bifurcate in physical space, with the Sgr\,A clouds following a path that takes them closer to Sgr\,A$^{*}$.

In conclusion, the line-of-sight location of the 20 and 50\,\kms \ clouds is controversial, and there is disagreement between models describing the CMZ as a whole and studies that have focused on the morphology and kinematics in the vicinity of Sgr\,A only.  
\medskip 

\noindent\emph{Sgr\,C}: The line-of-sight location of Sgr\,C is poorly determined. \citet{Chuard2018} constrained its position using X-ray echoes, but the method picks up multiple sources distributed over a line-of-sight distance of $\sim50$\,pc. 
Low-angular resolution absorption measurements consistently place Sgr\,C on the far-side of the Galactic Centre, behind Sgr\,A$^{*}$ \citep{Sawada2004, Yan2017}. 

Sgr\,C is part of the same contiguous $\{l,b,v\}$-stream that is associated with the 20 and 50\,\kms \ clouds.
Though all of the models described in \S\ref{sec:3dmodels} agree on this point, the precise location of Sgr\,C relative to Sgr\,A$^{*}$ is an open question. 
Some of the models place the cloud on the far-side of the Galactic Centre \citep{Sofue1995a, Molinari2011} and others on the near-side \citep{Kruijssen2015, Ridley2017}. 
In the absence of independent constraints, it is difficult to make a conclusive statement regarding the location of this cloud. 
\medskip

\noindent\emph{The 1.3$^{\circ}$ complex}: How the 1.3$^{\circ}$ cloud complex fits into the picture is highly debated, to the point that it is unclear whether it should even be considered as part of the main CMZ ring-like structure or not. 
A recurrent suggestion is that the $1.3^{\circ}$ complex is located at a contact point between the CMZ and the near-side dust lane \citep[][see also \S\ref{sec:dustlanes}]{Huettemeister1998,Fux1999,Rodriguez-Fernandez2006,Sormani2019a}. Theoretical models predict that the mass accretion occurring at the contact points should appear in observations as EVFs \citep[\S\ref{sec:EVFs};][]{Fux1999,Sormani2019a}. 
Consistent with this prediction, a clear example of an EVF is detected in $^{12}$CO data at the location of the $1.3^\circ$ complex between $v=100\mhyphen 200 \, \kms$ (\citealt{Liszt2006,Sormani2019a}, and Fig.~\ref{fig:lbv_main}). 
This led \citet{Tress2020} to argue that the $\{l,b,v\}$-contiguity between the $1.3^{\circ}$ cloud and the gas in the 100-pc stream implies a physical connection between the two. 
Therefore, in this interpretation the 1.3$^{\circ}$ complex is part of the CMZ ring-like structure, situated at its edge.

\cite{Krumholz2015} offer an alternative interpretation.
They suggest that the $1.3^{\circ}$ complex is part of a highly turbulent gas reservoir that will gradually lose angular momentum over the next $\sim5$\,Myr before entering the inner 100\,pc stream \citep[in contrast to the faster and more violent accretion events implied by the EVF interpretation;][]{Sormani2019a}. 
In this interpretation, the 1.3$^{\circ}$ complex is not part of the CMZ ring-like structure.

Projection effects play a considerable role in this debate. 
The scale height of the $1.3^{\circ}$ cloud complex is roughly $5$ times larger than that of the main $\{l,b,v\}$-streams \citep{Rodriguez-Fernandez2008, Henshaw2016b}. 
This can be interpreted as evidence both for and against the $1.3^{\circ}$ complex being part of the CMZ ring-like structure.
If the complex does reside at the contact point, collision-driven turbulence may explain its puffiness and extensive shocked gas emission \citep{Huettemeister1998, Rodriguez-Fernandez2006, Tress2020}. 
Alternatively, if the cloud is not physically connected to the main CMZ-ring like structure, a combination of elevated turbulence and projection may help to explain its extensive scale height relative to the $\{l,b,v\}$-streams \citep{Krumholz2017}.
\\

To summarise, we know that the CMZ is organised in a torus-like structure and there is some consensus on the position of a few molecular clouds, including the dust ridge molecular clouds. 
However, the position of several clouds as well as the detailed 3D geometry (nuclear spirals vs. ring vs. open stream) are open questions.
Contributing to this issue is the low-angular resolution of the data upon which all of the models described in \S\ref{sec:3dmodels} are based. 
This makes it difficult to distinguish between parts of the gas streams that lie close in $\{l,b,v\}$-space. 
Future higher-angular resolution observations of the CMZ (from e.g.\ ALMA; \S\ref{sec:summary}) will help disentangling such cases, and it may turn out that the distribution of gas around the Galactic centre is simply more complicated than that assumed in the models in \S\ref{sec:3dmodels}. 
A separate issue is that independent constraints on the location of individual clouds are in short supply. 
Additional constraints are likely to come from proper motion measurements of masers \citep[e.g.][]{Immer2020} and from X-ray emission \citep[e.g.][]{Chuard2018}. 
Combined, these new observations will help us to build a coherent picture of the 3D geometry of the CMZ in the future.

\subsubsection{Star formation ``hot spots''} 
\label{sec:sfhotspots}

One of the motivations for delineating the 3D structure of the CMZ is that morphology may be important in controlling where and when star formation occurs in the Galactic Centre (\S\ref{subsec:SFR:starformationhistory}). 
Evolutionary sequences in the ages of stars are sometimes observed in extragalactic nuclei, suggesting that star formation in nuclear rings may be triggered at preferred spatial locations \citep{Ryder2001, Allard2006, Mazzuca2008, Boker2008, Hennig2018, Callanan2021}. 

\citet{Longmore2013a} proposed that star formation in CMZ clouds may be triggered by tidal compression at pericentre in an eccentric orbit around the Galactic Centre (\S\ref{sec:3d}). 
This scenario was originally posed in relation to the dust ridge molecular clouds, where G$0.253+0.016$ shows very little evidence of widespread star formation and is interpreted as having passed through pericentre $\sim0.3$\,Myr ago, while Sgr\,B2, one of the most prodigiously star-forming regions in the Galaxy, resides $\sim0.75$\,Myr post pericentre \citep[Fig.~\ref{fig:sketch}][]{Kruijssen2015}. The observational evidence in support of this scenario includes increasing gas temperatures \citep{Ginsburg2016, Krieger2017}, possible star formation activity \citep{Immer2012a, Rathborne2014a, Ginsburg2018b, Walker2018, Barnes2019}, and the increasing evolutionary stage of H\,{\sc ii} regions \citep{Barnes2020b} downstream of the model-predicted location of pericentre \citep{Kruijssen2015}. If confirmed, such a sequence would help to place important constraints on the time evolution of star formation in the CMZ. 

It is clear however, that the star formation activity along the dust ridge is not strictly monotonic \citep{Walker2018}. Moreover, \citet{Kauffmann2017b} found no obvious trends in mass-size relation and SFRs of clouds as a function of orbital phase more generally throughout the CMZ. 
\citet[][see also \citealp{Kruijssen2017}]{Kruijssen2019} argued that the variation in the initial conditions of the clouds prior to the trigger event means that one would not necessarily expect strict downstream monotonicity.
Indeed, \citet{Henshaw2016a, Henshaw2020} identified a series of quasi-periodical cloudlets located $\sim0.3\mhyphen0.8$\,Myr upstream from the location of G$0.253+0.016$ in the \citet{Kruijssen2015} model.
Their $\{$masses, radii, volume densities, free-fall times$\}$ show variation of the order $\{$0.19, 0.09, 0.16, 0.08$\}$ dex \citep{Henshaw2017, Kruijssen2017}, providing a plausible explanation for the lack of monotonicity in the star formation sequence evident post pericentre in the dust ridge. 
However, \citet{Henshaw2020} argued that the clouds have formed via gravitational instabilities, suggesting that pericentre passage may not be an exclusive condition for collapse. 
This view is supported by \citet{Jeffreson2018a}, who found that cloud collapse due to tidal compression at pericentre occurs less often than that initiated by gravitational instabilities.

\citet{Hatchfield2021} used hydrodynamic simulations in a barred potential to investigate the interaction between gas infalling along the dust lanes and the CMZ, in the absence of self-gravity and star formation. 
They found sharp peaks in the cloud density that are strongly correlated with the location of orbital apocentre.
These density enhancements result from gas clouds slowing down at apocentre (which creates a ``traffic jam'' effect) and from collisions between the gas inflowing along the dust lanes and clouds in the nuclear ring. 
The inclusion of self-gravity and subgrid prescriptions for star formation and supernova feedback tells a similar, albeit more complicated story. 
\citet{Sormani2020b} found that the time-averaged surface density of very young ($t\leq0.25$\,Myr) stars increases just downstream of apocentre. 
However, they commented that the width of these surface-density enhancements are broad ($\leq$ half an orbit), and that they are not the only locations where star formation takes place in their simulation. 
\citet{Armillotta2020} similarly concluded that star formation sequences are more likely to occur downstream from the contact point with the dust lanes, rather than at pericentre. 

There is some evidence to suggest that orbital dynamics may influence star formation in Sgr\,B2.
In particular, the location of Sgr\,B2 close to apocentre in each of the models described in \S\ref{sec:3dmodels} suggests that the gas could pile up at this location \citep{Hatchfield2021}.
The molecular gas associated with Sgr\,B2 has a conical appearance in $\{l,b,v\}$-space. When integrated over discrete velocity intervals the cloud appears as a sequence of nested shells \citep{Henshaw2016b, Armijos-Abendano2020}. 
The largest of these shells (between $\sim20\mhyphen40$\,\kms) has a projected spatial extent of $\gtrsim30$\,pc \citep{Bally1988}. 
\citet[][see also \citealp{Sato2000}]{Hasegawa1994} cited the morphological similarity between a small hole evident near the apex of this cone (at $\sim40\mhyphen50$\,\kms) and a clump of molecular gas (at $\sim70\mhyphen80$\,\kms) as evidence that a small cloud has ``punched through'' the cloud. 
At the proposed collision site there is evidence for prodigious star formation activity and feedback \citep{Tsuboi2015b}, maser emission \citep{Sato2000}, chemical complexity \citep{Zeng2020, Colzi2022}, and an enhancement in shocked gas tracers \citep{Armijos-Abendano2020}.
However, the presence of these signatures may have multiple explanations \citep{Henshaw2016b, Kruijssen2019}, including star formation triggered at pericentre \citep{Longmore2013a}.
Therefore it remains debated whether the starburst in Sgr\,B2 has been triggered by a cloud-cloud collision.

Cloud-cloud collisions have been invoked to explain the physical, dynamic, and chemical properties of a number of Galactic Centre clouds in addition to Sgr\,B2. These include G$0.253+0.016$ \citep{Higuchi2014, Johnston2014}, the 50\,\kms \ cloud \citep{Tsuboi2015a}, M$0.014-0.054$ \citep{Tsuboi2021} (which is located close to the 50\,\kms \ cloud), and the EVFs discussed in \S\ref{sec:EVFs}. The evidence commonly cited in favour of collisions includes the identification of shells or cavities \citep{Hasegawa1994, Sato2000, Higuchi2014,Tsuboi2015a}, ``bridge features'' that connect multiple velocity components \citep{Johnston2014, Tsuboi2021}, and the prevalence of emission from shocked gas \citep{Zeng2020, Armijos-Abendano2020}, sometimes coupled to star formation events. Unambiguous signatures of cloud-cloud collisions are hard to come by \citep[though progress is being made;][]{Haworth2015,Fukui2021,Priestley2021}, particularly in an environment as dynamically complex as the CMZ. And the rate at which they occur in the CMZ may be low relative to other mechanisms dominating the cloud lifetime \citep[e.g. gravitational instability;][]{Jeffreson2018a}. An exception are the EVFs, which can be identified relatively clearly thanks to their extreme relative collision velocities (often $>100\kms$; \S\ref{sec:EVFs}).

\subsubsection{Turbulent Driving}
\label{sec:turbulentdriving}
The processes discussed throughout this section directly influence the physical state of the gas in the CMZ, contributing to the extreme ISM conditions.
The gas in the CMZ is characterised by velocity dispersions that are well above those measured in Galactic disc clouds, as has been noted since the earliest molecular observations of CMZ gas \citep{Bally1987}. This indicates an overall higher level of turbulence, and raises the question of what is driving it. Given the key role turbulence is expected to play in determining the star formation rate (\S\ref{sec:cloudtodisc} and \S\ref{sec:environmentofstarformation}), here we give a quantitative summary of CMZ turbulent driving.

To maintain the broad velocity dispersion, turbulence in the CMZ needs to be constantly driven.
The energy per unit time dissipated by the observed turbulent motions in the CMZ can be estimated as \citep[e.g.][]{MacLow2004} 

\setlength{\mathindent}{0pt}

\begin{footnotesize}

\begin{align} \label{eq:turb1}
    \dot{E} &= \frac{1}{2} \frac{M_\mathrm{CMZ} \sigma^{3}}{h} \nonumber \\
    & \simeq 2.8 \times 10^{39} \left(\frac{M_{\rm CMZ}}{5\times10^7\, \msun} \right) \left(\frac{\sigma}{12 \kms}\right)^3 \left(\frac{h}{20\pc}\right)^{-1}  \, \mathrm{erg\,s^{-1}},
\end{align}
\end{footnotesize}
where $M_{\rm CMZ}$ is the total gas mass of the CMZ, $\sigma$ is the typical velocity dispersion and $h$ is the CMZ scale-height.  

Several possible turbulent drivers have been evaluated, though no consensus has been reached on which, if any, dominates.
\cite{Kruijssen2014a} examined a range of possible turbulence driving mechanisms, including inflow along the bar, stellar feedback, gravitational instabilities, and acoustic instabilities. 
They found that bar inflow and acoustic instabilities could explain the observed turbulent energy, feedback makes a minor contribution, and gravitational instability, though perhaps important in the densest gas, is likely not important in the broader non-star-forming material.
However, \cite{Sormani2020c} demonstrated that the acoustic instability proposed by \citet{Montenegro1999} is a spurious result (\S\ref{sec:barinflow}), ruling out this latter mechanism as a source of turbulent driving.

Following the compilation of \citet[][and \citealp{MacLow2004}]{Kruijssen2014a}, we revisit the question of turbulence driving using the updated values presented in this review. 
Assuming that the kinetic energy is completely converted into turbulent motions, the energy injected by the bar inflow into the CMZ per unit time is

\setlength{\mathindent}{0pt}
\begin{footnotesize}
\begin{align}
    \dot{E}_\mathrm{IN} & =\frac{1}{2}\dot{M}_{\rm IN} v_{\rm IN}^2 \nonumber \\
    & \simeq 2.5\times 10^{39}\left(\frac{\dot{M}_{\rm IN}}{0.8\,\mathrm{M}_\odot\mathrm{yr}^{-1}}\right)\left(\frac{v_\mathrm{IN}}{100\,\mathrm{km\,s}^{-1}}\right)^{2}\, \mathrm{erg\,s^{-1}},
\end{align}

\end{footnotesize}
where $\dot{M}_{\rm IN}$ is the mass inflow rate and $v_\mathrm{IN}\simeq100$\,\kms \ is taken as a representative relative velocity between the inflowing gas and the gas already in the CMZ \citep[e.g.][]{Sormani2019b}. Although this simple estimate gives a value that is comparable to the dissipation rate (Eq.~\ref{eq:turb1}), it is unclear how much of the kinetic energy is converted into turbulent energy versus how much is lost to e.g. heat and radiated away \citep{Klessen2010}. Simulations that include only this source of turbulence \citep[e.g.][]{Sormani2019a} produce features in the $\{l,v\}$ plane that appear too narrow in velocity, inconsistent with the observations.
Thus, while turbulence driven by gas inflow may be important, it is probably not the only contributing factor.

The energy injected by supernovae can be estimated as

\setlength{\mathindent}{0pt}
\begin{footnotesize}

\begin{align}
    \dot{E}_{\rm SN} & =\sigma_\mathrm{SN}\eta_\mathrm{SN}E_\mathrm{SN} \nonumber \\
    & \simeq 5\times10^{39} \left( \frac{\sigma_\mathrm{SN}}{15\times10^{-4}\,\mathrm{yr}^{-1}} \right)\left( \frac{\eta_{\rm SN}}{0.1} \right)\left( \frac{E_\mathrm{SN}}{10^{51}\,\mathrm{erg}} \right)\ergs,
\end{align}

\end{footnotesize}
where $\sigma_\mathrm{SN}$ is the supernova rate (where $15\times10^{-4}\,\mathrm{yr}^{-1}$ is the upper limit; \S\ref{sec:feedback:outflowmechanisms}), $\eta_{\rm SN} \simeq 0.1$ is a (rather uncertain) factor that quantifies the efficiency of energy transfer from supernovae to the interstellar gas \citep{MacLow2004}, and $E_\mathrm{SN}$ is the energy delivered per supernova (assumed to be $10^{51}$\,erg). This estimate indicates that supernovae may make an important contribution to turbulence in the CMZ.

A further contribution to the turbulence may come from magnetorotational instabilities, which can extract rotational (orbital) energy and convert it into turbulent energy  at a rate given by \citep{Sellwood1999}:

\setlength{\mathindent}{0pt}

\begin{footnotesize}
\begin{align}
    & \dot{E}_{\rm MRI} = -T_{R\theta}\frac{d\Omega}{d\mathrm{ln}R}V \simeq T_{R\theta}\Omega V \simeq \frac{0.6}{8\pi}B^{2}\Omega V \nonumber \\
    & \simeq 2 \times 10^{38} \left( \frac{B}{100 \mu \mathrm{G}} \right)^2 \left(\frac{\Omega}{1.25 \myr^{-1}}\right) \left( \frac{V}{6 \times 10^5 \, \pc^3} \right) \ergs,
\end{align}

\end{footnotesize}
where $T_{R\theta}$ is the Maxwell stress tensor \citep[the approximation $T_{R\theta}\simeq \frac{0.6}{8\pi}B^{2}$ is determined from numerical models;][]{Hawley1995}, $B$ is the magnetic field strength, $\Omega$ is the orbital angular velocity, and $V=\pi R^{2}h$ is the volume of the CMZ, where we have taken $R=100$\,pc and $h=20$\,pc. 
This contribution is sensitive to the uncertain strengths of the magnetic fields. The B-field adopted in this estimate is in the middle of the measured range range (\S \ref{sec:magneticfields}).
While a stronger B-field would allow the MRI contribution to be dominant, such a B-field is disfavored by observations. Thus, while magnetorotational instabilities may provide a non-negligible contribution, they are unlikely to be dominant. 

Additional, smaller contributions may come from gravitational instabilities, protostellar outflows, and ionising radiation from massive stars \citep{MacLow2004,Kruijssen2014a}. However, these latter mechanisms have the problem that they cannot sustain turbulence on scales much larger than individual clouds or are limited to clouds that are already actively star-forming.

Overall, the elevated level of turbulent energy in the CMZ is likely dominated by the combined effects of supernova feedback, inflow along the bar, and possibly magnetorotational instabilities, though further work is needed to ascertain the relative contribution of each of these mechanisms. 

\section{Zooming in on star formation} 
\label{sec:cloudtodisc}

Having described the macro-evolution of the CMZ and the present-day distribution of gas and stars in \S\ref{sec:macroevolution}, we now  zoom in on individual regions of current and future star formation in the CMZ, from cloud ($\sim10$\,pc) to protostellar ($<1000$\,au) scales. In \S \ref{sec:extremeclouds} we discuss how global processes in the CMZ result in a population of molecular clouds with extreme properties, which provide the best Galactic analogues to high-redshift star-forming regions. In \S \ref{sec:sfinaction} we explore the growing body of work dedicated to identifying and characterising the incipient and current sites of star formation throughout the CMZ, and how the environmental conditions in the CMZ may inhibit the formation of compact substructures within these clouds.

\subsection{The extreme conditions in CMZ clouds}
\label{sec:extremeclouds}
The CMZ contains $\sim2\mhyphen6\times10^{7}$ \msun of gas at high density, yet the majority of this gas is not forming stars (\S\ref{sec:global}).  
As we expect to find star formation in the highest density gas, one of the key open questions about the CMZ is: Why is the dense gas forming stars inefficiently?

There are several ways in which molecular gas in the CMZ is qualitatively different from that in the rest of the Galaxy (Table \ref{tab:properties_overview}).
We summarise the significant effort that has been made to understand the global properties of the cloud population in the CMZ in this section.

\subsubsection{Observations at cloud scales}

Observations of star-forming gas are best done at far-infrared to millimetre wavelengths.
We begin by summarising these observations, and then describe some of their key results.
Many of the data sets mentioned in this section are publicly available and are indexed on the \href{https://github.com/CentralMolecularZone/DataSets}{CMZ Data Sets github repository}.

A number of continuum surveys targeting the CMZ have been conducted at far-IR to mm wavelengths, including the APEX Telescope Large Area Survey of the Galaxy \citep[ATLASGAL,][]{Schuller2009}, the Bolocam Galactic Plane Survey \citep[BGPS,][]{Bally2010, Rosolowsky2010, Aguirre2011, Ginsburg2013}, the Herschel Infrared Galactic Plane Survey \citep[Hi-GAL,][]{Molinari2010, Molinari2011}, the JCMT SCUBA-2 Galactic Centre Survey \citep{Parsons2018}, the IRAM-GISMO survey \citep{Arendt2019,Staguhn2019}, the MUSTANG Galactic Plane Survey \citep{Ginsburg2020}, the CARMA CMZ survey \citep{Pound2018}, the SMA CMZoom survey \citep{Battersby2020}, and the AzTEC survey of the CMZ \citep{Tang2021a, Tang2021b}.

These surveys achieve angular resolutions ranging from $3\mhyphen40^{\prime\prime}$, corresponding to physical scales $\sim0.1\mhyphen1.6$~pc. As they sample from the far-IR to mm wavelengths, the dust spectral energy distribution (SED) can be modelled to constrain the column density ($N_{\rm H_{2}}$), dust temperature ($T_d$), and opacity index ($\beta$) (see e.g. \citealt{Battersby2011, Molinari2011,Marsh2016,Tang2021a, Tang2021b} for details).  Typical column densities towards the CMZ clouds are $\sim$ 10$^{23}$~cm$^{-2}$, peaking at $>10^{24}$~cm$^{-2}$ towards Sgr B2. The cloud dust temperatures range from $\sim20\mhyphen25$~K, showing an inverse correlation with column density, with the lowest temperatures found towards the densest parts of the clouds. \citet{Tang2021a} report a dust opacity index $\beta$ in the range $1.8\mhyphen2.4$, which is steeper than seen in local clouds, even those contained within the field of view of the same CMZ data set.  The steeper index hints that dust properties in the CMZ may be fundamentally different than those in the rest of the Galaxy, which affects observational inferences at all wavelengths.

There have also been a number of molecular line surveys targeting the CMZ, including the H$_{2}$O southern Galactic Plane Survey \citep[HOPS,][]{Walsh2008, Walsh2011, Purcell2012, Longmore2017, Akhter2021}, which includes water maser and NH$_3$ observations at $\sim2.5\arcmin$ resolution, the ASTE [{C{\sc i}\xspace}] survey at 0.7mm and 34$\arcsec$ resolution \citep{Tanaka2011}, the MOPRA CMZ survey \citep{Jones2012, Jones2013} covering many lines in the 3mm band with $\sim$arcminute resolution, the APEX CMZ survey \citep{Ginsburg2016} covering H$_2$CO and $^{13}$CO in the 1.4mm band, the Survey of Water and Ammonia in the Galactic Centre \citep[SWAG,][]{Krieger2017} observing NH$_3$ and H$_2$O at $\sim30\arcsec$ resolution, and Nobeyama Radio Observatory surveys \citep{Tanaka2018,Tanaka2020} that cover many bands and transitions, particularly of HCN and HCO+.   Several surveys cover a range of CO transitions, including recent CO 3-2 observations with JCMT \citep{Parsons2018,Eden2020} at 15\arcsec resolution and many previous surveys at coarser resolution.

These extensive surveys, along with higher angular resolution follow-up observations, have led to the following important discoveries about CMZ gas density, velocity, temperature, and chemistry.

\paragraph{Molecular gas in the CMZ is denser on parsec scales than in the rest of the Galaxy:}\label{subsubsec:densitystructure}
Dust mass measurements yield average volume densities for all dust ridge clouds comparable to cluster-forming regions in the Galactic disc \citep{Walker2015,Walker2016}.
The masses of the molecular clouds in the CMZ range from $\sim$ 10$^{4\mhyphen5}$~M$_{\odot}$ in the dust ridge and the 20/50 km~s$^{-1}$ clouds, to \textgreater \ 10$^{6}$~M$_{\odot}$ in the most extreme case of Sgr B2 \citep[e.g.][]{Immer2012b,Kauffmann2017a}.
The clouds are very compact, with typical radii of a few parsecs.
Under a simplistic assumption of spherical symmetry and uniform distribution, this translates to volume densities $>10^{4}$ \percc for all of the prominent molecular clouds in the CMZ. 

Molecular line observations confirm these high inferred densities.
Studies using multiple transitions of molecular species that are not in local thermodynamic equilibrium (LTE) have been used to measure local volume densities from line ratios.
\citet{Mills2018c} analysed multi-transitional HC$_{3}$N data towards G0.253+0.016 (The Brick) and the 20/50~km~s$^{-1}$ clouds, and concluded that only $\sim$ 15\% of the gas is at $n>10^{4}$~cm$^{-3}$, with the majority residing at lower densities (10$^{3\mhyphen4}$~cm$^{-3}$). 
\citet{Tanaka2018} used a multi-species analysis with 20-30\arcsec\ ($\sim$\,1\,pc) resolution to find typical densities $n\sim10^{3.4}\mhyphen10^{4.5}$ \percc, somewhat higher than \citet{Mills2018c}.

More detailed modeling has only been performed for a few clouds, Sgr B2 and G0.253+0.016.
\citet{Schmiedeke2016} performed a full 3D modelling of the continuum emission in Sgr B2 to obtain constraints on the gas density. 
Their models include both stars and gas, and they find $\sim10^5$ \msun of gas at density $n>10^6$ \percc that is surrounded by $\sim10^7$ \msun at $n\gtrsim10^3$ \percc.
This high-density gas component, which is absent or a smaller mass fraction throughout most of the CMZ, helps to explain the high SFR of Sgr B2 \citep{Barnes2017,Ginsburg2018b} compared to other clouds.
The ambitious \citet{Schmiedeke2016} approach to modelling density structures, while successful, requires great effort and therefore has not been applied more generally.

Density substructure in G0.253+0.016 remains an open topic of study and takes on particular importance because of its low star formation rate (\S \ref{sec:sfinaction}).
\citet{Rathborne2014b} combined ALMA and JCMT-SCUBA data to measure the distribution of column density in G0.253+0.016 to determine that it is consistent with a turbulent-driven lognormal probability distribution with only a small gravitationally-dominated power-law tail. \citet{Johnston2014} reached the same conclusion for G0.253+0.016 using combined SMA and JCMT-SCUBA data, though they do not find a power-law tail, likely due to the lower resolution of the data.
However, \citet{Henshaw2019} showed, by decomposing the ALMA data from \citet{Rathborne2014b} into clusters of individual velocity components, that the cloud is not one coherent object, so the interpretation of the observed column density distribution is no longer straightforward.

Part of the problem driving this challenge is the shape of clouds in the CMZ.
Simulations by \citet{Dale2019} show that clouds orbiting the CMZ become flattened and elongated even if they start off as spherical  \citep[see also][]{Kruijssen2019,Petkova2021}.
\citet{Tress2020} note that clouds enter the CMZ already elongated and filamentary in their simulations.
G0.253+0.016 is the most prominently elongated CMZ cloud, with an aspect ratio $>2$ \citep{Rathborne2014a}, but other clouds may be significantly elongated along the line of sight.

The extent to which overdensity in Galactic centre clouds differs from that in Galactic disc clouds remains unclear, partly because of differences in analytical methods.
Systematic studies of clouds that are over-dense on parsec scales have been performed with the SMA to measure smaller-scale substructure \citep[\S\ref{sec:incipientsf},][]{Kauffmann2017b,Battersby2020,Hatchfield2020}. These studies find that most clouds have $\lesssim10\%$ of their gas in overdense substructures detectable with the SMA (generally with $n\sim10^4\mhyphen10^7$ \percc).
\citet{Lu2019b} used SMA data to measure the fraction of gas in bound objects, finding similar (9\%) overdensity in bound structures in the actively-star-forming Sgr C cloud, but much lower fraction  ($<1\%$) in non-star-forming clouds.
While \citet{Kauffmann2017b} argue that there is a difference in the mass at each density between CMZ and Galactic disc clouds, \citet{Parmentier2020} suggest that this distribution is similar.
The discrepancy in these results is in part because of the angular size sensitivity of the observations; in local clouds, full 3D modeling has enabled sensitive multi-scale characterization of gas density, while in the CMZ, we are presently limited to determining the amount of mass above a given column density.

While there are extensive observational constraints on gas density on parsec scales in the CMZ, the details of density structure on small (sub-pc) scales remains unclear and is an important topic for future study.

\paragraph{Dense molecular gas in the CMZ is warmer than in typical molecular clouds elsewhere in the Galaxy:}\label{subsubsec:tempstructure} 
The gas temperature has been measured using several independent molecular line tracers, such as NH$_{3}$ and H$_{2}$CO, with reported gas temperatures ranging from $T_\mathrm{gas}\,\sim$\,30 to \textgreater \ 100~K (for 218\,GHz formaldehyde, H$_{2}$CO, estimates see \citealp{Ao2013, Ginsburg2016, Immer2016}; for ammonia, NH$_{3}$, inversion transition estimates see \citealp{Krieger2017}).
All of these data sets had $\sim30\arcsec$ resolution, and measurements of gas temperature on smaller scales are limited.
There are hints from both single-dish observations of highly excited lines \citep{Mills2013} and interferometric observations \citep{Johnston2014} that higher-temperature molecular gas ($T_\mathrm{gas}\,>300$ K) is widespread throughout the CMZ, but these observations push the limits of the thermometers being used to infer these temperatures.  Temperature structure on $<30\arcsec$ scales has yet to be mapped in detail outside of selected ``hot cores'' \citep[e.g.][]{Sanchez-Monge2017,Bonfand2017, Walker2018, Walker2021}.

The measured gas temperatures are significantly higher than the dust temperatures of $T_\mathrm{dust}\,\sim$ 20~K towards the clouds \citep[e.g.][]{Marsh2016,Tang2021a}, suggesting that the gas and dust are not thermally coupled in CMZ clouds, despite their high densities of $>10^{3\mhyphen4}$~cm$^{-3}$.
\citet{Clark2013} investigated this using SPH simulations of a single cloud (G0.253+0.016), and found that under the gas conditions in the CMZ, the gas and dust remain thermally uncoupled below very high volume densities (\textgreater \ 10$^{7}$~cm$^{-3}$).
Similar gas-dust temperature differences are observed in other galaxy centres \citep[e.g., M83, NGC 253,][see Table \ref{tab:properties_overview}]{Mangum2013}, suggesting that this decoupling is a common feature of regions with a high concentration of star formation.

The processes responsible for these higher gas temperatures are debated, with arguments following the same lines as in external galaxies \citep[e.g.,][]{Meijerink2011}: is the gas mechanically heated or heated by cosmic rays?  
Direct heating of the gas via the elevated interstellar radiation field is inefficient, since the radiation couples poorly to the gas (\citealp{Clark2013, Ginsburg2016, Oka2019}).
In addition, the cooler dust temperature ensures that collisions between the gas and dust do not heat the gas, but rather contribute to the gas cooling.
At higher energies, X-ray radiation in the CMZ today lacks the energy to heat the gas to observed temperatures, even in the inner $\sim10$ pc \citep{Ao2013}.
The remaining candidates, mechanical heating (the transformation of kinetic energy into thermal energy via shocks as the end stage of the turbulent cascade) and cosmic ray heating are both plausible heat sources \citep{Immer2016,Ginsburg2016,Krieger2017}.
The ultimate source of that energy is, in both cases, uncertain; possible sources of mechanical heating are discussed below in \S\ref{sec:velocitystructure},
while cosmic rays are produced in supernovae, around the central black hole, and perhaps in other regions where the magnetic field is compressed (\S\ref{sec:cosmicrays}). 

While the dust temperatures are colder than gas temperatures, they are still substantially warmer than in local cold clouds. The lowest reported dust temperature in CMZ is $T_\mathrm{dust}\,\sim15$\,K, though most of the dust is at $T_\mathrm{dust}\,\sim20\mhyphen25$\,K \citep{Tang2021b,Marsh2016}.
This is a factor of $2\mhyphen3$ higher than in comparable-density cloud cores in the Galactic disc \citep[e.g.][]{Peretto2010}, suggesting that molecules are less often frozen onto grain surfaces, which likely has a significant effect on the gas chemistry (see \ref{sec:cloudchemistry}).

While several CMZ-spanning temperature maps have now been created with $\sim$pc resolution, the spatial dynamic range of temperature measurements is currently quite limited.
Future observations are needed to determine how substructured the temperature is, and on what scales the gas is heated.
Key open questions about the dense gas temperature remain:
Are there local, high-temperature shocks consistent with the mechanical driving model?
What is the cosmic ray ionisation rate (CRIR), and how uniform is it both in space or time?
Are CRs really heating the gas preferentially, and can the CRIR accommodate the range of observed temperatures?

\paragraph{Molecular gas appears to be more turbulent on parsec scales than in the rest of the Galaxy:}\label{sec:velocitystructure}
The elevated turbulent energy (\S \ref{sec:turbulentdriving}) in the CMZ is evident in the size-linewidth relation in the molecular gas, which is vertically offset from that seen in Galactic disc clouds \citep{Heyer2015}, indicating a greater dispersion on scales of $2\mhyphen20$\,pc \citep{Shetty2012}. 
The relation may converge to that of local gas at scales smaller than $\sim0.1\mhyphen1$ pc \citep{Kauffmann2017c}, though this remains debated.
A contributing factor to the controversy on small physical scales is that the intrinsically broad velocity coverage required to observe all CMZ gas at once often requires, because of technical limitations, that the gas be observed at low spectral resolution.
Future observations focused on the size-linewidth relation should use high (better than 0.1 \kms) spectral resolution to advance this debate.

The size-linewidth relation is the observational manifestation of the turbulent energy cascade, in which most energy is present on the largest scales.
The size-linewidth relation in the CMZ follows approximately  $\sigma(R) = \sigma_{0} R^{0.7}$ \citep{Shetty2012,Kauffmann2017c,Tanaka2020,Krieger2020}, where most authors agree on the slope of the relation, but disagree on the intercept ($\sigma_{0}$).
Most authors have derived a slope $b$ slightly steeper than that of the typical Galactic disc relation, $b=0.5$, while \citet{Henshaw2020} find a slope slightly shallower at $b\approx0.37$.  The difference may in part be attributed to the adopted technique, since, of these authors, only \citet{Henshaw2020} present the size-linewidth relationship within a single cloud (G0.253+0.016), whereas the others use some variant of a clump cataloguing algorithm to build an inter-cloud relationship.
Additionally, \citet{Henshaw2019} used a Gaussian decomposition of G0.253+0.016 ALMA HNCO observations to obtain $\sigma(0.07 \mathrm{pc}) = 4.4\pm2.2 \kms$, substantially higher than extrapolations from the large-scale measurements, again suggesting that measurement technique may dominate the uncertainties in size-linewidth measurement.

The role of turbulence in regulating star formation in the CMZ remains an active topic of research, as highlighted in the case study of G0.253+0.016, which has ongoing but limited star formation \citep{Walker2021}.
\citet{Federrath2016} measure the turbulence in G0.253+0.016, arguing that the turbulence is primarily solenoidally (as opposed to compressively) driven with Mach number $\mathcal{M}=11\pm3$.  
They argue that the solenoidal driving and high $\mathcal{M}$ reduce the star formation efficiency by a factor of 6.9 compared to Galactic disc clouds.
They attribute the overall velocity gradient in the cloud to shear rather than turbulence.
However, \citet{Henshaw2019} show that the gradient observed in the moment-1 velocity map does a poor job of capturing the kinematic structure of the cloud, and they derive a higher Mach number ($\mathcal{M}=16.45\pm0.01$).  
The debate on how turbulent CMZ gas is at different scales remains unresolved.

While most of the broad linewidths are attributable to turbulence,
some of the observed kinematic structure and line broadening may be attributed to other mechanisms.
\citet{Henshaw2016a, Henshaw2020} analyze the velocity structure of some of the streams in the CMZ, noting that they have oscillatory fluctuations in the line-of-sight velocity that may result from gravitational instability, i.e., the initiation of gravitational collapse.
Several authors have noted individual high-velocity-width features ($\sigma_v \gtrsim 50~\kms$) in the CMZ that are attributed to completely different mechanisms, such as large-scale colliding flows (EVFs, \S\ref{sec:EVFs}), cloud-cloud collisions \citep{Tanaka2015,Tsuboi2021}, intermediate-mass black holes \citep{Oka2016}, and supernova interactions \citep{Tanaka2014,Yalinewich2018}.
These broad features may represent locations where kinetic energy is added back into the molecular medium, offsetting the energy lost through the turbulent cascade.

While velocity dispersion measurements are the main driver of the argument that CMZ gas is more turbulent, there remain several controversies driven both by observational and data analysis techniques.
High spectral and spatial resolution data with the dynamic range to resolve both the driving and decay scale of turbulence are needed to resolve these issues.

\paragraph{CMZ clouds are chemically distinct from those seen in the solar neighbourhood, exhibiting much greater abundances of a wide range of complex molecules:}\label{sec:cloudchemistry} The differences in molecular chemistry are driven by the extreme
physical conditions of the CMZ: high density, CRIR, and dust temperature, strong shocks, bright X-ray emission, and higher gas metallicity.
The molecular chemistry of the CMZ means that `classic' tracers of physical processes, molecules used as signposts for physics, are unreliable in both our CMZ and, by extension, comparable extragalactic environments.

The molecular chemistry in the CMZ is so extreme that even non-star-forming regions are promising sites for discovering new molecules in the ISM.
While the Sgr B2 ``Molecular Heimat" has long been the best location to find new complex molecules \citep[e.g.][]{Belloche2013,Belloche2016,Belloche2019,Moller2021}, another site about a parsec away from Sgr B2 N has recently been the prime location for detection of unique molecules.
The G0.693-0.027 cloud has been the site of several new prebiotic molecule detections with ALMA \citep{Rivilla2020,Rivilla2021a,Rivilla2021b,Colzi2022}, yet it shows no signs of star formation \citep{Ginsburg2018b,Zeng2020}.
The extremely molecule-rich spots in Sgr B2 are not representative of the rest of the CMZ, however, and it is not clear yet to what degree the extreme chemistry in that cloud is driven by its high density, its star formation, or an overabundance of shocks.  

Commonly-used tracers of dense, star-forming gas do not exclusively trace star-forming gas in the CMZ and in galactic centres in general.
\citet{Mills2017b} analyzed the \citet{Jones2012} data in conjunction with {\em Herschel} column density maps, finding that, while HCN (1-0), a classic dense gas tracer, is well-correlated with the total dense gas mass, about 2/3 of the emission is associated with low-column-density, non-star-forming gas.
Their analysis is supported by subsequent multi-line and multi-tracer HCN and HC$_3$N observations that show the majority of gas in the CMZ is at moderate-density ($n(H_2)\sim10^3\mhyphen10^{4.5}$ \percc) and is not associated with star formation \citep{Mills2018c,Tanaka2018}.
This is of particular concern for the extragalactic community, where HCN emission is the workhorse for tracing dense molecular gas (e.g. \citealp{Usero2015, Bigiel2016, Querejeta2019, Beslic2021}). 

Chemical tracers used to highlight specific physical mechanisms, such as shocks (e.g., HNCO and SiO) or cold gas (N$_2$H$^+$), do not work in the CMZ.
While SiO and HNCO are classic shock tracers, generally seen only in regions with material interacting at relatively high-velocity ($\gtrsim10\,\kms$) in the Galactic disc \citep[e.g.,][]{Schilke1997,Jimenez-Serra2008,Gusdorf2008}, in the CMZ - and CMZs of other galaxies - they are ubiquitous and trace all components of the dense molecular medium \citep{Jones2012,Henshaw2016a,Henshaw2019,Yu2018}.
These species are still seen in outflows (see \ref{sec:outflows}), but not uniquely so.
N$_2$H$^{+}$ is prized as a tracer of cold, dense gas in the Galactic disc (e.g. \citealp{Barnes2020a}) because it is destroyed by CO in the gas phase, and therefore becomes more abundant only when CO has frozen out onto grains \citep{Flower2005,Bergin2007}.  However, in the CMZ, N$_2$H$^+$ is empirically widely distributed and does not select for cold gas (or dust); instead, it appears that the high cosmic ray ionisation rate produces a situation in which CO reactions are not important for regulating the N$_2$H$^+$ population, making it a widespread but `normal' dense gas tracer \citep{Santa-Maria2021}.
Interpretation of chemical tracers in the CMZ is complicated by both the uncertainty in abundances and by excitation; \citet{Petkova2021} show, with radiative transfer modeling applied to a hydrodynamic simulation of G0.253+0.016 \citep{Dale2019, Kruijssen2019}, that much of the observed cloud structure can be explained by varying optical depth and excitation without considering variations in abundance.  

The rich molecular chemistry of the CMZ presently renders our understanding of physical processes uncertain, but it provides substantial opportunity to improve our understanding of chemical processes in energetic environments.
The molecular lines in the CMZ are bright, and so is their future.

\paragraph{The cosmic ray density is higher in the Galactic centre than most of the disc, but their origin remains uncertain:}\label{sec:cosmicrays}
Cosmic rays are important drivers of gas chemistry and play a large role in the thermal balance of the molecular gas \citep[see][for a review of cosmic rays in star formation; they discuss the CMZ in S7.3]{Padovani2020}.

Observations of molecular ions reveal the high CRIR in both low- and moderate-density molecular gas. \citet{Indriolo2015} used {\it Herschel} observations of molecular ions (OH$^+$, H$_2$O$^+$, and H$_3$O$^+$) toward Galactic centre clouds Sgr B2 N \& M, M-0.02-0.07 (the 50 \kms\ cloud), and M-0.13-0.08 (the 20 \kms\ cloud) to infer CRIRs in the range $0.2\mhyphen1.5\times10^{-14}$\,s$^{-1}$, about 10-100 times greater than in the Galactic disc.  
Similarly, \citet{LePetit2016} and \citet{Oka2019} used H$_3^+$ absorption line observations to infer a CRIR $\sim2\times10^{-14}$\,s$^{-1}$ in the moderate-density gas in the CMZ, a value reasonably consistent with the high end of the \citet{Indriolo2015} measurements.
\citet{Ginsburg2016} argued that the cosmic ray ionisation rate must be $<10^{-14}$ s$^{-1}$ in $n\gtrsim10^5$ \percc gas based on H$_2$CO temperature measurements, since a higher CRIR would result in an equilibrium temperature in dense gas warmer than observed.
Current observations are limited to a few sightlines probing a limited range of densities, so while it is reasonable to continue using a CRIR $\sim10^{-14}~\mathrm{s}^{-1}$ for modeling work, additional observations of CR ionisation are needed.

While cosmic rays are expected to affect the chemistry of molecular gas in the CMZ, the origin of the excess CRs remains debated \citep[see][for a thorough review]{Bykov2020}.
The high supernova rate contributes substantially to the CRIR \citep{Ponti2015}, but there are other sources of CRs, including Sgr A* \citep{HESSCollaboration2016} and even \hii\ regions \citep{Meng2019,Padovani2019}.
The nonthermal filaments are sites of high-energy magnetic events, possibly magnetic reconnection events, that may also produce CRs \citep{Heywood2019, Yusef-Zadeh2019,Guenduez2020,Sofue2020, Thomas2020, Zhang2020, Coughlin2021}.
Depending on where the cosmic rays are produced, there may be significant local variations in both the intensity and spectrum of cosmic rays throughout the CMZ.

\paragraph{What about magnetic fields?}
\label{sec:magneticfields}

The magnetic field in the Galactic centre plays a critical role in several observed phenomena, especially the bright nonthermal filaments.
However, the B-field in the molecular gas, and particularly the role it plays in star formation, is only just beginning to be studied.
  
The shape and strength of the B-field in the Galactic centre has been measured at several wavelengths, but the implications of these measurements are still debated.
While there have been recent spectacular measurements of nonthermal filaments with MEERKAT \citep{Heywood2022,Yusef-Zadeh2022}, the B-field in the broader Galactic Centre appears to be decoupled from that in the dense CMZ molecular clouds \citep{Morris2015}.

On smaller scales in the dense gas, observations of dust polarization with interferometers are just beginning, and there are no published maps yet, but ground-based large single-dish telescopes with polarimeters have mapped the $\sim$pc-scale field.
\citet{Pillai2015} used SCUPOL observations of G0.253+0.016 with resolution $\sim20\arcsec$ to infer a magnetic field strength of $\sim5$ mG.  
\citet{Chuss2003} used the CSO to perform polarimetric observations within the inner $\sim50$ pc, obtaining measurements consistent with a field strength of up to 3 mG.
\citet{Crutcher1996} measured Zeeman splitting of an HI line toward Sgr B2, inferring $B=0.5$ mG.
The fields in the dense clouds tend to be loosely aligned with the B-field angles inferred from polarisation measurements made on larger scales \citep[][with PILOT at 240 \um and ACT at 1-3mm, respectively]{Mangilli2019,Guan2021}, generally supporting the hypothesis that the magnetic field is dragged along with the dense gas as it is sheared out into a toroidal shape parallel to the Galactic plane \citep{Morris2015,Hu2022b}.
Some clouds, specifically the extremely dense Sgr B2 cloud and  M-0.02-0.07 (the 50 \kms\ cloud) near the Galactic centre, show field lines perpendicular to the bulk toroidal field \citep{Guan2021}, hinting that in these cases, either global gravitational collapse or extreme optical depth have changed the apparent field orientation \citep{Morris2015}.

The observations of B-fields on small scales, in the dense gas, are so far limited, but there are several forthcoming SOFIA HAWC+ \citep[e.g.][]{Hu2022,Hu2022b} and ALMA observations that will expand our understanding of the small-scale B-field.  While these data sets are not yet broadly available, we expect a great deal of observational progress over the next few years.

\subsubsection{The CMZ as a high-$z$ analogue}
\label{sec:environmentstuff}
The CMZ molecular clouds are broadly similar to those in typical galaxies at the peak of cosmic star formation.

The kinematic, density, temperature, and magnetic structure and the cosmic ray density are similar enough in CMZ clouds to make them the best Galactic analogue of high-redshift star-forming gas. 
The two key differences are the star formation rate and the metallicity.
The SFR in the CMZ is presently low given the amount of dense gas (see \S \ref{subsec:SFR:current} \& Table \ref{tab:SFR}).
The metallicity in the CMZ \citep[$Z\sim2$;][]{Rudolph2006} is higher than in the early universe.

The comparable cloud-scale properties of CMZ clouds and those in high-$z$ galaxies helps us understand how star formation depends on gas properties over cosmic time.
\citet{Kruijssen2013} made this point specifically in the context of the size-linewidth relation, and of the relation between the gas surface density and the stellar surface density, in which the CMZ and high-$z$ galaxies occupy similar regions, but Galactic disc clouds are distinct.
\citet{Swinbank2015} confirmed this similarity in their observations of SDP 81 at $z=3.042$, in which the molecular cloud complexes fall close to the CMZ on the size-linewidth relation.

The local properties in CMZ clouds are similar to those in high-redshift `normal' star-forming galaxies.
While there are no direct measurements of cosmic ray ionisation rates in high-redshift galaxies, cosmic rays are correlated with star formation and therefore expected to be far more abundant at cosmic noon; indeed, $\gamma$-ray emission from M82, NGC 253, and Arp 220 \citep{VERITASCollaboration2009,Lacki2011,H.E.S.S.Collaboration2018,Yoast-Hull2017} confirm the close association. 
The CMZ is therefore an excellent laboratory for studying the effects of 10-100$\times$ enhanced cosmic ray energy densities \citep{Yoast-Hull2014b, Yoast-Hull2014a}.
The high gas temperatures observed in the CMZ, and the gas-dust temperature difference, are seen in nearby star-forming galaxies \citep{Mangum2013,Mangum2019}, hinting that this separation is common in galaxies with high overall SFR.
Similarly, the CMZ may serve as a template for the chemistry in `normal' galaxies that have higher temperatures and ionisation rates than the local Galactic disc; direct measurements of star-forming regions in these galaxies remain limited now \citep[e.g.][]{Meier2005,Harada2019}.

The degree to which the CMZ can be used to test models of cloud and star formation at high redshift has not been very well explored, but we suggest that CMZ clouds are a more natural starting point for comparison to the early universe than are solar neighborhood and Galactic disc clouds.

\subsection{Star formation in action} 
\label{sec:sfinaction}

Linking how the extreme properties of the CMZ influence the individual sites of star formation is crucial in developing a more general understanding of star formation as a function of galactic environment. In \S\ref{sec:incipientsf} we describe recent observational efforts to identify sites of current and potential future star formation, and in \S\ref{sec:ongoingstarformation} we discuss the classification of star-forming activity in the CMZ.

\subsubsection{Incipient star formation} 
\label{sec:incipientsf}

At the time of PPVI, the first systematic, unbiased surveys of the CMZ at far-IR and submillimetre wavelengths were being completed with single-dish telescopes (BGPS, ATLASGAL, and Hi-GAL; \citealt{Bally2010, Molinari2011, Csengeri2016}). These surveys revealed complexes of dense clumps on $\sim$1 pc scales, interconnected by fainter streams of gas \citep[\S\ref{sec:3d};][]{Bally2010}. The results from these surveys supported the initial suggestion that the SFR in the CMZ was lower than would be expected considering the amount of dense gas (\S\ref{subsec:SFR:current}).

Over the following years, piecemeal follow-up observations were made toward key CMZ clouds using submillimetre interferometers, the Submillimeter Array \citep[SMA;][]{Kauffmann2013, Kauffmann2017a, Kauffmann2017b, Kendrew2013, Johnston2014, Lu2015, Lu2019b, Walker2018} and ALMA \citep{Rathborne2014b, Ginsburg2018b, Barnes2019, Uehara2019, Miyawaki2021, Walker2021, Lu2020, Lu2021}. The Galactic Centre Molecular Cloud Survey (GCMCS) surveyed six prominent CMZ clouds (Sgr D, Sgr B1 off, G0.253+0.016, the 20 \& 50\kms~clouds, and Sgr C) at 1.1 mm with arcsecond resolution on the SMA \citep{Lu2015, Lu2019b, Kauffmann2017b, Kauffmann2017c}. In GCMCS, a total of 56 compact sources were identified, which revealed a SFR within individual clouds about a factor of ten smaller than predicted \citep{Kauffmann2017b, Lu2019b}, and that only a small fraction of the total cloud mass \citep[1-9\%;][]{Lu2019b} is contained within gravitationally bound clumps. 

The SMA CMZoom survey \citep{Battersby2020, Hatchfield2020} was the first complete high-resolution (0.1 pc) survey of dense gas ($N_\mathrm{H2}$ \textgreater\ 10$^{23}$ cm$^{-2}$) in the CMZ at submillimetre wavelengths (1.3 mm). The survey is $\sim$99\% complete to compact substructures capable of forming high-mass stars in the CMZ, and identifies 285 compact sources in a robust (high fidelity) or 816 compact sources in the high-completeness catalogue \citep{Hatchfield2020}. Of the objects detected, there is a bimodal distribution in their physical properties, with the highest mass and column density cores being located in the Sgr B2 complex. While many CMZ clouds show rich and complex sub-structure, a key result of this unbiased survey is an overall deficit in compact substructures on $0.1\mhyphen2$\,pc scales, with compact dense gas fractions (percent of total cloud mass contained within compact substructures) of less than 10\% in nearly all CMZ clouds, which is factors of several lower than in comparable Galactic disc clouds \citep{Battersby2020}. 

\citet{Lu2019b} find that the star formation efficiency on 0.2~pc scales is comparable to the Galactic disc, and hypothesise that the global deficit of star formation in the CMZ can be attributed to the low fraction of gas confined to gravitationally bound cores. Combined with the results from \citet{Battersby2020}, this suggests that the formation of compact substructure may be inhibited in the CMZ, despite the comparatively high densities of the clouds (\S\ref{sec:sfthreshold}).

The presence of dense cores, many of which are associated with signatures of active star formation, can be used to gauge the overall level of star formation activity in a cloud, or combined with an assumed star formation timescale to estimate a SFR for the cloud. The reported SFRs are highly uncertain, as both the timescale of star formation and the contribution from lower-mass stars are unconstrained.

The Sgr B2 region accounts for the majority of the present-day star formation in the CMZ, with an estimated SFR $\sim 0.08~\msun\peryr$ \citep{Schmiedeke2016,Ginsburg2018a,Barnes2017}. 
Sgr C, located on the opposite end of the CMZ is the only other very active star-forming region, containing a compact \hii\ region and 275 cores \citep{Kendrew2013,Lu2019b,Lu2020}. While the number of cores is similar to the 271 in Sgr B2 from \citet{Ginsburg2018b}, the Sgr C data from \citet{Lu2020} are more sensitive and the cores are less massive. 
Clouds in the dust ridge show moderate star formation activity from $\mathrm{SFR}\sim10^{-4}\mhyphen10^{-3}~\msun \peryr$ in G0.253+0.016 \citep{Rathborne2014a,Walker2021} to $\mathrm{SFR}\sim3\times10^{-4}~\msun\peryr$ in cloud E \citep{Lu2019b}.
The 20 \kms\ cloud is forming some stars (SFR $\sim2\times10^{-3}$) over a small area, while the 50 \kms\ cloud has very little star formation (SFR$<3\times10^{-4}$; \citealp{Lu2019b,Uehara2019,Miyawaki2021}).
Using the complete CMZoom compact structure catalogue, \citet{Hatchfield2020} estimate the maximum star formation potential of the CMZ to be $\mathrm{SFR}\,=\,0.08\mhyphen2.20~\msun\peryr$. This assumes that the compact structures will collapse to form stars with some star formation efficiency ($0.1 < \mathrm{SFE} < 0.75$) within a free-fall time of about 10$^4\mhyphen10^5$ years. Note that Sgr B2 dominates this estimate, and if excluded, the star formation potential is reduced to $0.04\mhyphen0.47~\msun\peryr$ \citep{Hatchfield2020}.

In summary, CMZ clouds have a very low fraction of their gas bound in overdensities (\textless \ 10\%), suggesting that the formation of compact substructure and subsequent star formation is inhibited. The reason for this is not fully understood, but it is likely a consequence of the extreme environmental conditions in the CMZ (\S\ref{sec:sfthreshold}). Where star formation is observed, it is contained within only several clouds, predominantly Sgr B2. In future, observations at higher angular resolution and sensitivity are needed to measure the contribution from low-mass YSOs (e.g. from the JWST and ALMA), which is presently unconstrained.

\subsubsection{Direct evidence for on-going star formation} 
\label{sec:ongoingstarformation}

\begin{figure*}
    \centering
	\includegraphics[trim=0 0cm 0 0cm, clip, width=0.95\textwidth]{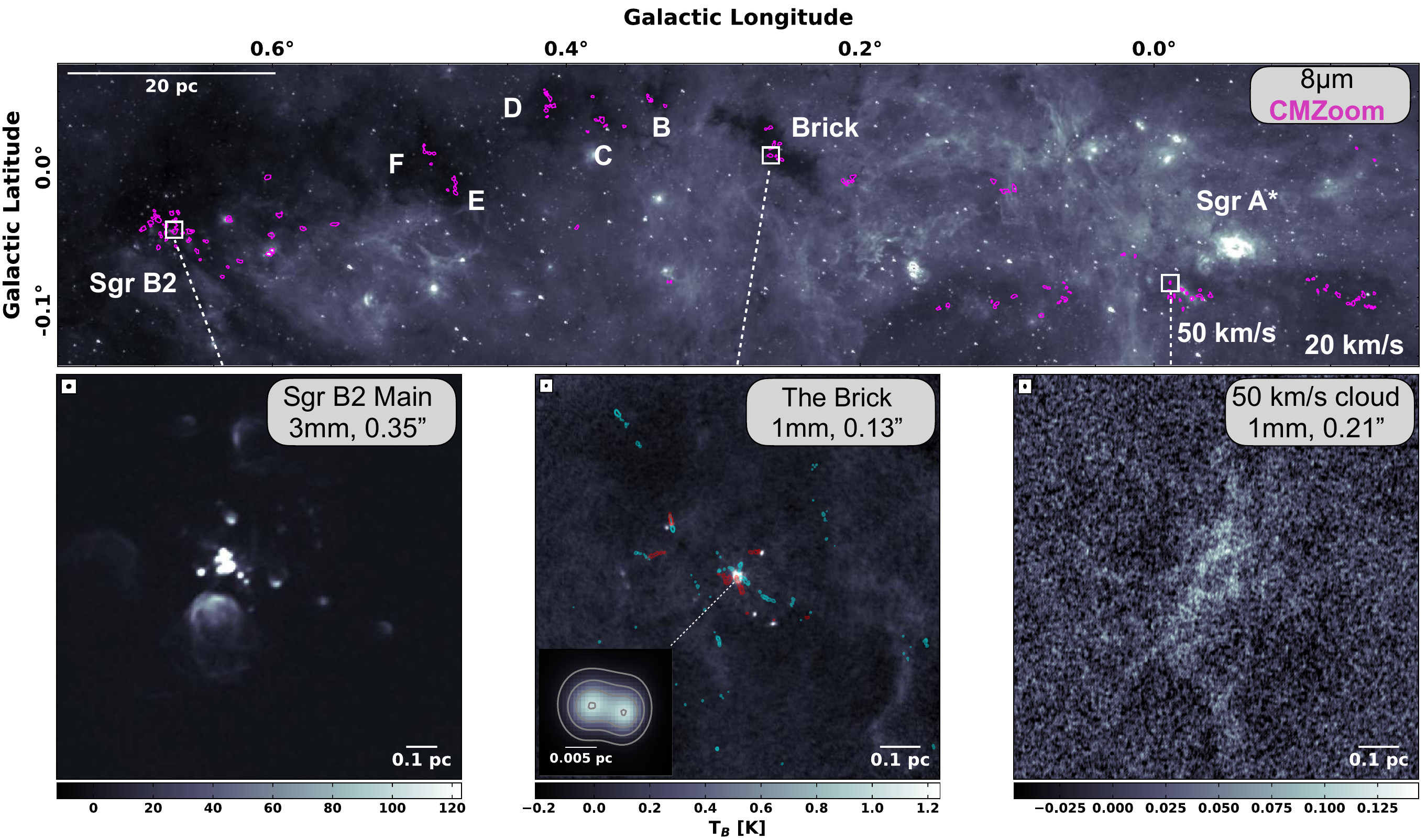}
    \caption{\textit{Top}: 8~$\mu$m map of a portion of the CMZ \citep{Churchwell2009}. Contours show the 1.3~mm source catalogue from the CMZoom survey \citep{Battersby2020}. \textit{Bottom}: ALMA observations of three regions with differing levels of star-forming activity: from the highly active Sgr B2 Main \citep[\textit{left},][]{Ginsburg2018b}, to the weakly star-forming G0.253+0.016 (the Brick) \citep[\textit{centre},][]{Walker2021}, and the 50~km~s$^{-1}$ cloud, which shows no evidence for embedded protostellar cores \citep[\textit{right},][]{Lu2020}. Blue/red contours in the centre panel show outflows via integrated blue/red-shifted SiO (5-4) emission, and the inset shows the the centre of the field, revealing a protostellar binary with separation $\sim$ 1000~AU.}
    \label{fig:high_res}
\end{figure*}

A crucial step in understanding the present-day star formation activity in the CMZ is the search for signposts of active star formation. Combined with the core populations discussed in $\S$\ref{sec:incipientsf}, these signatures (or lack thereof) provide key insights into their evolutionary phases, and ultimately a more comprehensive view of the overall SFR in the CMZ.

\paragraph{Masers:}\label{sec:masers} Maser emission is one of the most widely used tracers of early embedded star formation. Masers are particularly useful in the context of the CMZ, as their bright, compact emission signatures are more easily detectable through the complex, high-extinction line-of-sight compared to other star formation tracers. 

The two most abundant maser species found in star-forming regions are those from water (H$_{2}$O) and methanol (CH$_{3}$OH, Class II), with the latter being found exclusively towards regions of high-mass star formation \citep[e.g.][]{Ellingsen2006}. A number of Galactic plane surveys have obtained a census of these masers throughout the CMZ. The H$_{2}$O Southern Galactic Plane Survey \citep[HOPS,][]{Walsh2011} and the methanol multibeam survey \citep[MMB,][]{Green2009, Caswell2010} have surveyed the inner Galaxy for 22.2~GHz water masers and 6.7~GHz methanol masers, respectively. \citet{Walsh2011} reported an under-density of water masers in the CMZ given the amount of dense gas there as traced by NH$_{3}$ \citep{Longmore2013b}. 

\citet{Caswell2010} reported 22 methanol masers in the CMZ, 11 of which are in Sgr B2. More recently, \citet{Rickert2019} conducted a survey of 6.7~GHz methanol maser emission in the CMZ using the VLA, reporting a total of 43 masers. They note that there is an asymmetry about the Galactic centre, with more methanol masers at positive longitudes. This excess correlates with the asymmetry in the dense molecular gas in the CMZ, indicating more young high-mass star formation at positive longitudes (\S\ref{sec:gasstardistrib}). Similarly, \citet{Lu2019a} observed the 6.7~GHz methanol line over the inner 200~pc of the CMZ, reporting 23 methanol masers. Correlating these with embedded UC\hii\ regions, they find that high-mass star formation in the CMZ is limited to 7 clouds, with methanol masers detected towards only 5.
    
While the CMZ appears to be under-abundant in water and methanol masers, there is tentative evidence that it may be over-abundant in more exotic masers that trace high-mass star formation \citep{Ginsburg2015}. H$_{2}$CO masers, which appear to uniquely trace high-mass star formation, have been reported in Sgr B2, Sgr C, and dust ridge cloud C \citep{Ginsburg2015, Lu2019a}. This brings the total known Galactic star-forming regions containing H$_{2}$CO masers to 9, a third of which are in the CMZ. SiO masers have also been detected in Sgr B2 \citep{Higuchi2014} and dust ridge cloud C \citep{Ginsburg2015}. While SiO masers are common towards evolved stars, they are rare around YSOs, with 8 known high-mass star-forming regions, 2 of which are in the CMZ. $^{14}$NH$_{3}$ (2,2) maser emission has been detected in Sgr B2, which is the first reported detection in a star-forming region, and the eighth star-forming region known to contain an NH$_{3}$ maser \citep{Mills2018a}.  Additional masers from non-metastable lines of NH$_3$ have been found in Sgr B2 North \citep{Mei2020}. Though the sample is small, these results suggest that these rarer masers are more prevalent in the CMZ. The reason for this is unknown, but \citet{Ginsburg2015} speculate that either these masers trace very early stages of high-mass star formation, which would indicate that these regions are experiencing a burst of star formation, or that the comparatively extreme conditions in the CMZ could favour the production of masers.

\paragraph{Protostellar outflows:}\label{sec:outflows} Protostellar outflows are unambiguous signatures of star formation, indicating that the embedded YSO is actively accreting material. They have been ubiquitously observed both in regions of low- and high-mass star formation \citep[e.g.][and references therein]{Bally2016}. However, outflows in the CMZ have largely eluded detection until very recently. This is due to a previous lack of observations at high angular resolution and sensitivity, compounded by the kinematic complexity towards the CMZ, which makes it difficult to disentangle outflow emission. This is particularly true for the two most common outflow tracers, CO and SiO. Bright CO emission is widespread throughout the CMZ, and along the line-of-sight, resulting in absorption and complex spectra. SiO is also abundant in the gas-phase in the CMZ, likely due to turbulent shocks releasing it from grain surfaces \citep{Martin-Pintado1997}.

Until recently, the only candidate outflows in the CMZ were in Sgr B2 N and M \citep{Lis1993, Qin2008, Higuchi2014}. These outflows are extremely massive (10$^{2}$-10$^{3}$~M$_{\odot}$) and not well-collimated, and may instead represent a collective outflow from multiple high-mass protostars in the clusters (Schw\"{o}rer et al, private communication.).

With ALMA, it is now possible to resolve down to protostellar scales (\textless \ 1000~AU) at high enough sensitivity to isolate individual protostellar outflows in the CMZ. Recent ALMA observations probing these scales have unambiguously detected protostellar outflows in several CMZ clouds. \citet{Walker2021} reported at least 9 outflows in G0.253+0.016 (aka The Brick, see Fig.~\ref{fig:high_res}) as traced by SiO (5-4). \citet{Lu2021} reported a total of 43 outflows in 3 massive molecular clouds (Sgr C, Sgr B1-off, and the 20~km~s$^{-1}$ cloud) as traced by 6 different molecular lines at 1~mm (SiO, SO, CH$_{3}$OH, H$_{2}$CO, HC$_{3}$N, and HNCO). The outflows in these samples are detected towards both low- and high-mass cores. These results suggest that protostellar outflows are also ubiquitous in star-forming regions in the CMZ, and confirm that low- and high-mass star formation is occurring simultaneously in these clouds.

\paragraph{Protostars:}\label{sec:protostars} Hundreds of protostellar sources have recently been discovered in the CMZ, with a growing census as facilities push to higher resolution and sensitivity. 

The `hot cores' of Sgr B2 N and M have been known for decades, and several new smaller hot cores were recently discovered \citep[e.g.][]{Sanchez-Monge2017, Bonfand2017, Bonfand2019}. Throughout the Sgr B2 cloud, there are $\gtrsim250$ high-mass protostellar cores \citep{Ginsburg2018b}, though the published data are very shallow and sensitive only to $M\gtrsim10$~\msun.

The 20 \kms\ cloud, the Sgr B1 ``off'' region (aka dust ridge clouds E/F), and Sgr C are all forming rich groups of 100s of stars, while the 50 \kms\ cloud, with a similar mass and overall density, is not \citep{Lu2020}. \citet{Uehara2019} identify a population `cores' from H$^{13}$CO$^{+}$ and C$^{34}$S data in the 50 \kms\ cloud, though \citet{Lu2020} found that there were no overdensities on 2000~AU scales in the 1~mm dust continuum that were consistent with protostellar sources. 

G0.253+0.016 contains only one known site of ongoing star formation, with 18 clustered protostars found so far. These protostars are all low mass, but the measured outflow rates of \textgreater\ $10^{-5}\mhyphen10^{-4}$~\msun~yr$^{-1}$ towards 50~\% of the sources suggests that they may be destined to become intermediate or high-mass stars \citep{Walker2021}. There is also some indirect evidence for on-going star formation in a different region of the cloud \citep{Henshaw2022}. The rest of G0.253+0.016 appears to be totally devoid of protostars, though it has not yet been fully mapped out at high angular resolution and sensitivity. These differences in protostellar activity between prominent CMZ clouds are highlighted in Figure \ref{fig:high_res}.

The populations of prestellar and protostellar cores in the CMZ have only been catalogued towards a handful of molecular clouds so far \citep[][Fig.~\ref{fig:high_res}]{Ginsburg2018b,Lu2020,Lu2021,Walker2021}. There is tentative evidence to suggest that the distribution of core masses -- the core mass function (CMF) -- is shallow: i.e. there is an apparent excess of high-mass sources (\citealp{Lu2020}, though the source mass function based on line data only by \citealp{Uehara2019}, with coarser resolution,  has a steeper distribution). However, there remain substantial uncertainties in these measurements; a `normal' Salpeter shape can still be accommodated if the dust temperature is systematically higher for the brighter cores \citep{Lu2020}. These initial results are similar to those found in CMZ star clusters, where there is some evidence that the stellar IMF is also top-heavy (\S\ref{sec:starclusters}).

Current observations show that searching for protostars based on the presence and number density of H$_{2}$O and CH$_{3}$OH masers (\S\ref{sec:masers})
has proved efficient for identifying sites of ongoing star formation.
Most of these clustered H$_{2}$O masers have proved to be closely associated with protostars or outflows \citep{Walker2021,Lu2021}. However, there are many molecular clouds that contain sufficient dense gas that we expect ongoing star formation, yet do not contain any known masers. Thus far, these clouds have only been systematically surveyed to a depth of $M_{\mathrm{core}}>10$~\msun at an angular resolution of 3\arcsec \citep{Battersby2020,Hatchfield2020}. A deep, high resolution, and complete survey is therefore needed to determine whether these apparently quiescent cloud host populations of low-to-intermediate mass protostars.

\paragraph{HII regions:}\label{sec:hiiregions}
The intense radiation field from high-mass (O- and B-type) stars produces regions of photoionised gas known as \hii regions, which, because of the short lifetimes of such stars, are treated as a sign of recent star formation.  

The Sgr B2 complex contains the majority of compact \hii\ regions in the CMZ, ranging in size from hundreds of AU to several pc (e.g. \citealp{Gaume1995, dePree1995, dePree1996}).
\citet{Schmiedeke2016} compiled a comprehensive list of \hii\ region surveys  in this cloud, and estimate a total stellar mass content of $\sim$2-3$\times$10$^4$\,\msun.
Short timescale flux variations of ultra-compact \hii\ (\uchii) regions in Sgr B2 highlight that accretion is still ongoing \citep{DePree2015}. 

Adjacent in projection to Sgr B2 is another notable \hii\ region complex: Sgr B1 (Figure \ref{fig:rgb_main}). 
The Sgr B1 complex contains more diffuse ionised gas and less extincted infrared structure \citep{Mehringer1992}, consistent with it being at a later evolutionary stage than Sgr B2 (e.g. \citealp{Barnes2020b}).
\citet{Harris2021} argue for a physical association between Sgr B1 and B2 based on [CII] morphology.
Two scenarios have been proposed for the formation of this region: newly formed stars that are ionising their natal environment, or more evolved high-mass stars passing through, and ionising, the cloud.  
The maser emission \citep{Mehringer1993}, and YSO candidates identified towards this region (e.g \citealp{An2011,An2017}) point towards the former scenario. 
The latter scenario has been suggested by \citet{Simpson2018a, Simpson2021}, who suggest the ionising stars are several million years old based on ion line ratios. 
It is plausible that high-mass stars have drifted this far away from the nearby massive clusters \citep[][ \S\,\ref{sec:starclusters}]{Habibi2014}.
However, future tests of this hypothesis are needed, including investigating the proper motions and relative velocities of the ionised gas and stars, examining the structure of the molecular and ionised gas to search for cometary or bow-shock features \citep{Henshaw2022}, and age dating the embedded stellar populations (as in e.g. \citealp{Nogueras-Lara2020b}).

Near the projected centre of the CMZ are the Sgr A 20 \& 50 \kms\ clouds, which contain several \hii regions. \citet{Tsuboi2019} measure their recombination line emission, inferring electron temperatures $T_e\sim5000\mhyphen6000$ K, among the coolest in the Galaxy but consistent with other Galactic centre \hii regions \citep{Mills2011}.
The low electron temperatures are consistent with the high inferred metallicity in CMZ gas \citep{Balser2011}. The presence of these \hii regions is evidence for recent star formation in the 20 and 50 \kms clouds.

On the opposite side of the CMZ to Sgr B2 is the Sgr C complex (Figure~\ref{fig:rgb_main}), which contains more than $\sim$\,250\msun of ionised gas that powered by at least one O star (e.g. \citealp{Liszt1995}). To the east of the \hii\ region lies a dense molecular cloud with a mass $\sim$10$^5$\msun (e.g. \citealp{Lis1994b}), which is likely interacting with the ionised gas \citep{Lang2010}. \citet{Hankins2020} identified numerous bright mid-IR (37\micron) point sources on the boundary of the molecular cloud and \hii\ region that are candidate \uchii regions. The brightest of these is the Sgr C ``H3'' region, which has been confirmed as an \uchii region (e.g. \citealp{Forster2000, Kendrew2013, Lu2019a, Lu2019b}).

Just outside of the CMZ, there are \hii regions associated with infalling material, indicating that there is some ongoing star formation in the dust lanes.
In the far side dust-lane at negative longitudes (\S\,\ref{sec:dustlanes}), there is the Sgr E complex (e.g. \citealp{Liszt1992}). This region contains $\sim$\,60 \hii\ regions \citep{Anderson2020}, which are surrounded by a diffuse ionised gas component \citep{Langer2015}. These \hii\ regions are scattered across nearly a degree on the sky, are not centrally concentrated, and have relatively uniform sizes. Comparison with simulations suggests that the stars responsible for these \hii\ regions formed upstream in the far dust lane a few Myr ago and will overshoot the CMZ, crashing through the near dust lane in the future \citep{Anderson2020}.

While the compact \hii\ region population in the CMZ has been known and reasonably well-characterised for several decades, there have been several more recent discoveries of diffuse \hii\ regions (e.g. \citealp{Henshaw2022, Heywood2022}). This suggests that there is more to be learned through observational studies of these larger \hii\ regions and the recent ($3\mhyphen30$ Myr) star formation history of the CMZ.


\section{The impact of environment}
\label{sec:environmentofstarformation}
As discussed in \S\ref{sec:environmentstuff}, the CMZ is in many ways a good analogue of more distant and inaccessible high-$z$ regions. However, although the latter follow the SFR-dense gas relation, the present-day SFR of the CMZ is about an order of magnitude below it (\S\ref{sec:global} and Fig.~\ref{fig:sfr_main}) -- that is, even with the numerous actively star-forming regions discussed in the previous section. The origin of this puzzling phenomenon is not understood. In this section, we highlight three key areas of the star formation process that are directly impacted by the extreme environmental conditions in the CMZ. In \S\ref{sec:sfthreshold} we describe how the ISM conditions may lead to an increased density threshold for star formation in the CMZ, and in \S\ref{sec:starclusters} and \S\ref{sec:ppds} we describe how the environment may also affect the properties of star clusters and protoplanetary discs, respectively.

\subsection{An increased density threshold for star formation in the CMZ?} 
\label{sec:sfthreshold}
The concept of a critical density threshold (either column or volume) for star formation has been introduced both observationally \citep{Lada2010, Lada2012, Heiderman2010} and theoretically \citep{Krumholz2005, Hennebelle2011, Hennebelle2013, Federrath2012, Padoan2011, Padoan2014}. In the Galactic disc, a column density threshold was empirically determined from the observation that gas with a higher extinction tended to have a higher level of star formation activity; the more gas above a certain density ($\Sigma_{\rm{gas}} \sim 116$ \msun pc$^{-2}$), the more YSOs \citep[\S\ref{subsec:SFR:context}][]{Lada2010, Lada2012}. 
Volumetric star formation relations were developed analytically, based on the premise that star-forming cores result from gravitationally unstable perturbations caused by supersonic turbulence in molecular clouds.
The Probability Distribution Function (PDF) of isothermal supersonic turbulent density fluctuations has an approximately log-normal profile \citep[e.g.][]{Nordlund1999}, but additional physical mechanisms, such as self-gravity, may result in a departure from this log-normal shape \citep[e.g.][]{Kainulainen2014, Burkhart2019}. The SFR can be calculated from the fraction of material above the critical density, $\rho_\mathrm{crit}$, above which self-gravity dominates.

The value of this critical density varies amongst the different theories and depends upon the balance of turbulent, gravitational, and magnetic energy in the molecular cloud. 
In the solar neighbourhood, the predicted critical densities \citep[e.g.][]{Padoan2014} are roughly consistent with observations \citep[e.g.][]{Lada2010}.
The environmental conditions in the CMZ are comparatively extreme (\S\,\ref{sec:extremeclouds}), and in particular the high turbulent energy may play a dual role -- both enhancing the density contrast in clouds and elevating the critical density threshold for star formation \citep[e.g.\ ][]{Kruijssen2014a, Burkhart2019} relative to that in the Galactic disc. 

Current observational evidence suggests that there is a higher density threshold for star formation in the CMZ.
\citet{Rathborne2014b, Rathborne2015} measured the 3 mm dust continuum emission toward G0.253+0.016 (the Brick) with ALMA. With these data, they derived a column density PDF width, placing a lower limit on the critical density of collapse, and finding that their observations were consistent with an environmentally-dependent density threshold for star formation which is several orders of magnitude higher than the threshold derived for solar neighbourhood clouds. A similar result was found by \citet{Johnston2014} using 1~mm SMA observations of G0.253+0.016. \citet{Ginsburg2018b} confirmed that the apparent threshold also holds in Sgr B2, in which all observed star formation occurs at column densities above $\Sigma\gtrsim1$ g \persc, similar to that in G0.253+0.016. 
As described in \ref{sec:incipientsf}, many studies note a surprising lack of star-forming cores within dense regions of the CMZ, further supporting the notion that stars only form above an increased critical density threshold \citep{Kauffmann2013, Kauffmann2017a, Walker2018, Barnes2019, Lu2019b, Lu2020, Battersby2020}.

Turbulent star formation theories based on a density threshold do not correctly and uniquely predict the cloud star formation rates with Galactic disc parameters; they require tuning to match observations.
\citet{Federrath2016} builds on the data from \citet{Rathborne2014b} and demonstrates that, based on the observed physical properties of G0.253+0.016, its SFE per free-fall time from these analytic theories is expected to be about 4 $\pm$ 3 \%, which is consistent with later estimates from \citet{Kauffmann2017a} and \citet{Barnes2017}, only if the turbulence is driven more solenoidally in the CMZ than in the disc.
\citet{Barnes2017} compare estimates of the SFE toward dust ridge clouds (\S\ref{subsec:SFR:current}) with analytic predictions from \citet{Krumholz2005}, \citet{Padoan2011}, and \citet{ Hennebelle2013}.
Assuming that the dust ridge clouds have physical conditions that are similar to G0.253+0.016, these authors suggest that only the latter two models are in agreement with the observed SFEs, though tighter constraints on the physical properties and free parameters included within these volumetric star formation models are ultimately required before any can be verified (or falsified).

It is also worth noting that the CMZ is also a comparatively high-pressure environment, owing to the high turbulent energy, strong magnetic field, and its location near the the minimum of the Galactic gravitational potential. The interplay of these extreme conditions results in external pressures in the CMZ of P/$k$ \textgreater \ 10$^{7}$~K~cm$^{-3}$ \citep[e.g. ][]{Rathborne2014a, Myers2022}, compared to typical values of $\sim$ 10$^{5}$~K~cm$^{-3}$ in the Galactic disc \citep[e.g.][]{Blitz1993, Schruba2019}. While such high pressures are important in the context of the density and confinement of clouds \citep[e.g.\ ][]{Longmore2014, Rathborne2014b, Walker2018}, studies since PPVI have not explicitly considered the role of pressure in regulating the critical density threshold for star formation.

Current observational evidence is roughly consistent with the hypothesis that there is an environmentally-dependent critical density threshold for star formation. Similar results have been found in nearby galaxies also \citep[e.g.][]{Usero2015, Bigiel2016, Querejeta2019, Jimenez-Donaire2019, Beslic2021, Eibensteiner2022}. However, the exact nature and magnitude of this dependence remains an unresolved question. If indeed an environmentally-dependent threshold exists, then simple dense gas scaling relationships are not sufficient to predict the SFR even in our own Galactic centre, never mind distant parts in our cosmos with even more varied properties. 
A more complex star formation prescription, which depends on not just the gas density, but also other intrinsic gas properties such as turbulent and magnetic energy, is required. 

\subsection{The formation and evolution of star clusters} 
\label{sec:starclusters}
The fraction of stars forming in clusters is an essential descriptor of star forming environment, since clusters - especially more massive gravitationally bound clusters - are affected by physical processes including dynamical interactions and high-mass stellar feedback.
Despite hosting a small number of observed clusters, the CMZ appears to form a higher fraction of stars in bound clusters compared to the Galactic disc.
The known CMZ clusters have reported top-heavy stellar initial mass functions, highlighting that these environmental differences are likely substantial.

The Arches and Quintuplet clusters are the only known star clusters in the CMZ
\citep[we leave out the Young Nuclear Cluster, YNC, as it is at $R<10$ pc and its formation mechanism is likely entirely different from these clusters, see e.g.][]{Genzel2010, Neumayer2020},
and they are among the most massive in the Galaxy with $M\sim10^4$~\msun.
The Arches cluster is the younger of the two at 2$\mhyphen$3 Myr old \citep{Lohr2018, Clark2019}.
The Quintuplet cluster is older, with an age $\sim$\,5\,Myr \citep{Liermann2012,Schneider2014,Rui2019}.
The effects of feedback surrounding this region are also clearly seen as \hii\ region(s) in the infrared and radio (e.g. \citealp{Lang2005,Hankins2020}), which are potentially interacting with molecular gas within the region \citep{Butterfield2018}.
Prominent features in the vicinity of the Quintuplet cluster include the ``sickle" region, which contains finger-like features reminiscent of an eroding photodissociation region like the ``Pillars of Creation" in the Eagle Nebula \citep{Hankins2020}. 
The ``helix” region that appears to extend from a potential run-away a potential Quintuplet cluster member Wolf-Rayet star, WR102c \citep{Lau2016,Steinke2016}. 
Indeed, some of the massive stars spread throughout the CMZ may be runaways from these two clusters, ejected during the cluster's dynamical evolution \citep{Habibi2014,Dong2015}.

The IMF in these clusters appears to be more top-heavy than the typical Salpeter slope.
\citet{Hosek2019} used multi-epoch HST data in conjunction with K-band spectroscopy to determine that the Arches cluster has an initial mass function inconsistent with a single slope.
They find a power-law slope $\alpha_\mathrm{IMF}\approx1.80^{+0.05}_{-0.05}$ $\mhyphen$ 2.0$^{+0.14}_{-0.19}$ depending on the functional form fitted (a steeper slope with more power-law breaks is consistent with the data), in any case, shallower than the typical $\alpha_\mathrm{IMF}=2.35$.
The shallow top-end IMF is confirmed with radio measurements that are sensitive to the mass loss rate of high-mass stars \citep{Gallego-Calvente2021a,Gallego-Calvente2021b}.
It remains unclear if this shallower slope is a unique feature of the CMZ, or if it is instead a common feature of high-mass star clusters, since other Galactic clusters \citep[e.g., Wd1, NGC 3603,][]{Pang2013,Lim2013,Andersen2017} have slopes consistent with that seen in the Arches.
There are also hints that this shallow IMF may be seen in an earlier core mass function stage (\S \ref{sec:protostars}).
However, there are many possible systematic errors that affect IMF measurements, and while many are accounted for in these works, the measurements need to be treated with caution \citep[e.g.][]{Bastian2010}. 

While the Arches and Quintuplet formed 3-5 Myr ago, and no additional similarly massive clusters have formed more recently, it is clear that the CMZ is still actively forming new high-mass clusters. Indeed, young proto-clusters are seen embedded in the molecular clouds (\S\,\ref{sec:sfinaction}).
The small number of clusters, and lack of older clusters, is consistent with expectations that cluster lifetimes are shorter, at a given mass, in the high-density CMZ than in the Galactic disc \citep{Kruijssen2012}.

Formation models for high-mass clusters range from sudden monolithic collapse to a more gradual `conveyor belt' buildup, in which the evolution of the proto-cluster is defined by concurrent star formation and gravitational collapse of the cloud \citep[e.g.,][]{Longmore2014, Vazquez-Semadeni2019, Krumholz2019, Krumholz2020}.
Within the CMZ, the protocluster clouds, as seen by \emph{Herschel} in the dust and single-dish line data, are less centrally condensed than the final clusters \citep{Walker2015}.
The lack of centrally condensed protocluster clumps capable of rapid monolithic collapse, combined with measurements of the virial parameter showing that the clouds are globally gravitationally unstable, led \citet{Walker2016} and \citet{Barnes2019} to conclude that the clusters must form from the more distributed `conveyor belt' mechanism (see also \citealp{Schworer2019}).
The Sgr B2 cloud contains two of these protoclusters, Sgr B2 M and Sgr B2 N, the former being older and star-dominated while the latter is still gas-dominated \citep{Schmiedeke2016,Ginsburg2018b}.  
\citet{Barnes2019} propose that dust ridge clouds D and E are undergoing collapse to form one or more star clusters based on the low observed virial parameters, but at present they contain little compact substructure and are relatively starless.

All of the presently-observed clusters \& protoclusters are high-mass, $M\gtrsim10^4$ \msun.
While observational biases may play a role here, sensitive surveys in the infrared and millimeter have not yet turned up additional clusters, suggesting that the lack of low-mass clusters is physical.
Models suggest that the high gas densities and strong shear in galactic nuclei should result in clusters with elevated minimum masses and an initial cluster mass function narrower than other galactic environments \citep{Trujillo-Gomez2019}.
Theories also predict that a systematically higher fraction of stars will form in bound clusters at high gas surface densities.
The higher density and overall star formation efficiency result in larger regions of gas becoming globally self-gravitating and forming bound clusters of objects \citep{Kruijssen2012}.
This theory is backed by hydrodynamic simulations \citep[e.g.][]{Grudic2021}.
The CMZ is an ideal place to test these theories, since it is the only region within our Galaxy with an order-of-magnitude higher gas surface density than the solar neighbourhood on $\sim100$\,pc scales .
\citet{Ginsburg2018a} counted the fraction of young stars forming in the high-mass bound clusters Sgr B2 N and M compared to the number forming in the surrounding cloud, finding that the fraction in bound clusters was $\Gamma_\mathrm{CFE}\approx37\%$, much higher than the $\Gamma_\mathrm{CFE}\sim7\mhyphen10\%$ seen in the solar neighbourhood \citep[e.g.][]{Lada2003}.
This finding supports the theory and suggests that bound clusters play a major, possibly dominant role in Galactic Centre star formation.

If the high $\Gamma_\mathrm{CFE}$ and the noted shallower IMF slope $\alpha_\mathrm{IMF}$ in CMZ star clusters both hold, the overall IMF in the CMZ is top-heavier than that in the Galactic disc.
Such environmental dependencies of star and cluster formation are important ingredients in galaxy formation models and highlight the critical role of the CMZ in understanding star formation on a cosmic scale.

\subsection{Protoplanetary discs} 
\label{sec:ppds}

Accretion discs are ubiquitous around forming YSOs. With the advent of recent facilities, observations of CMZ clouds are regularly approaching physical resolutions $\sim$ 1000~AU. While this resolution is too coarse to resolve low-mass discs, it ought to be sufficient to detect disc candidates around high-mass YSOs \citep[e.g.][]{Ahmadi2019}. However, accretion discs around CMZ YSOs have not been directly observed on 1000~AU scales, despite high sensitivity observations towards Sgr B2, Sgr C, the dust ridge, and the 20/50~km~s$^{-1}$ clouds \citep[e.g.][]{Schworer2019, Lu2021, Walker2021}. There are hints of discs in new ALMA observations probing scales $\sim$ 200~AU towards disc candidates in the CMZ. Initial analysis shows direct evidence of accretion discs around YSOs in some CMZ clouds (Xing Lu, private communication), but they are not clearly detected in all comparable data sets (Adam Ginsburg, private communication). 

While there is currently little direct evidence of accretion discs in the CMZ, the growing sample of collimated outflows suggests that they are common (\S \ref{sec:outflows}). These outflows are detected towards both low- and high-mass protostellar cores, indicating the presence of accretion discs throughout the protostellar mass range. Given this growing evidence, it is important to consider whether such discs might form and evolve differently in the comparatively extreme environment of the CMZ. 

Theoretical work suggests that protostars that form in higher stellar density environments experience stronger far-ultraviolet (FUV) radiation fields, leading to photoevaporation and dispersal of protoplanetary discs \citep[PPDs,][]{Winter2018, Winter2020}. At the typical gas densities in the CMZ, ram pressure stripping also plays a more significant role in dispersing discs.
Tidal truncation of PPDs due to dynamical encounters with neighbouring stars can also lead to PPD mass loss. These effects are compounded in regions of very high stellar density, such as young stellar clusters, where PPDs may be destroyed on $\sim$ Myr timescales, with FUV photoevaporation being the dominant mechanism \citep[e.g.][and references therein]{Winter2018}. 

Given the high stellar (\S \ref{sec:potential}) and gas densities (\S \ref{subsubsec:densitystructure}) in the CMZ, as well as the high cluster formation efficiency in Sgr B2 \citep{Ginsburg2018a}, discs in the CMZ may have short lifetimes.
Indeed, \citet{Winter2020} estimate PPD lifetimes that are at least 5 times shorter compared to the Solar neighbourhood, with a predicted $\sim$90\% of CMZ PPDs being destroyed within 1~Myr.

Though PPDs have not yet been directly observed in the CMZ, circumstellar discs have been catalogued in the Arches and Quintuplet clusters via mid-infrared excess emission \citep{Stolte2010, Stolte2015}. These catalogues report low disc fractions (9\% and 4\%, respectively), with a decreasing fraction towards the centre of the Arches cluster, suggesting that discs are dispersed more rapidly in the dense central regions of the cluster. No radial dependence was found in the older, less dense Quintuplet cluster. While these results are consistent with the expected rapid dispersal of discs in these extremely dense stellar clusters, the fact that there is still a small population of discs is surprising. As the Arches and Quintuplet are several Myr old (\S\ref{sec:starclusters}), this would suggest that these discs are long-lived. Investigating this, \citet{Stolte2015} suggest that the detected discs are likely secondary mass-transfer discs in high-mass binary stellar systems. However, further investigation is required to conclude their origin.

In summary, while the field of PPD formation and evolution in the CMZ is still in its infancy, new and forthcoming results are beginning to detect signatures of accretion discs around YSOs. Theoretical work suggests that discs should be short-lived in this extreme environment, which could have significant implications for both the IMF and planetary systems of the resulting stellar population. Future high angular resolution observations of CMZ disc candidates will be crucial in determining any variation in disc properties as a function of Galactic environment.


\section{Summary and outlook}\label{sec:summary}

In the context of Galactic star formation, the CMZ is an environment like no other. It hosts the supermassive black hole, Sgr\,A$^{*}$, some of the closest and most massive young Galactic star clusters, the largest number of supernovae per unit volume, and the most concentrated reservoir of dense gas in the Milky Way. It is, furthermore, the only galactic nucleus in which it is currently possible to resolve the multi-scale physics of star formation down to the scales of protoplanetary discs. The recent discovery that the CMZ is underproducing stars relative to expectations based on its vast reservoir of dense gas has inspired a resurgence in observational and theoretical efforts to understand the star formation process in this complex environment, and has raised the important question of if (and how) the physics of star formation depends on local environmental conditions.

The factors responsible for the observed low present-day SFR of the CMZ are still the subject of intense scrutiny. Two forefront explanations have emerged, which may operate in unison. Either star formation is directly related to the macroscopic evolution of the system as a whole, possibly via a feedback-driven ``boom and bust'' duty cycle or via discrete large accretion events (\S\ref{sec:global}), and/or the extreme local ISM conditions (\S\ref{sec:cloudtodisc}) elevate the critical density for star formation (\S\ref{sec:environmentofstarformation}). Both explanations are supported by theoretical work and consistent with current observations, and both have implications that reach beyond the field of star and planet formation into that of galaxy formation and evolution. 
For the former of these scenarios, further theoretical work is needed to identify a specific variability mechanism and timescale. Higher-angular resolution observations of nearby galaxy centres will help to address whether the star formation in galactic nuclei occurs in discrete bursts or quasi-continuously. For the latter, tighter constraints on CMZ cloud properties, including the density structure, turbulence, and magnetic field strengths are needed to test predictions from star formation theory, in particular of a turbulence-regulated density threshold, and whether or not they hold under the extreme conditions in the CMZ.  

The observational future is bright. 
The observing capabilities offered by Large Programs with current (e.g. ALMA, VLA, SOFIA) and next-generation facilities such as the {\em James Webb Space Telescope} ({\em JWST}), Square Kilometre Array (SKA) and the next-generation VLA (ngVLA) will likely reshape our view of the CMZ. 
For example, the upcoming ALMA CMZ Exploration Survey, or  ``ACES'', will provide an unparalleled insight into the physical and kinematic structure of the CMZ, while polarisation measurements and deep, large spectral coverage surveys will allow us to further probe the magnetic field structure (\S\ref{sec:magneticfields}) and complex chemistry (\S\ref{sec:cloudchemistry}), respectively.
Combining with further observational and modelling work across the electromagnetic spectrum, from long-term VLBI maser monitoring to tracing X-ray flares, will add key constraints on the 3D geometry (\S\ref{sec:3d}).
Going beyond what was possible with {\it Spitzer} in the IR regime, the recently launched {\em JWST} will uncover hidden star formation in the CMZ, extending the YSO-counting methods used to determine the SFR in nearby clouds to the CMZ. 
This will directly address whether the low SFR in the CMZ is the result of a presently undetected low-mass YSO population, as well as whether the mass distribution of the YSOs is consistent with or different from a standard IMF (\S\ref{sec:starclusters}). 
High-angular-resolution observations of protostars will furthermore provide measurements of primordial binary statistics in the extreme environment of the CMZ (e.g., the Offner et al. chapter).
High-angular resolution observations (e.g.\ Fig~\ref{fig:high_res}) will also be critical in uncovering protoplanetary discs, with potentially important implications for our understanding of planet formation in extreme, cosmologically-representative environments (\S\ref{sec:ppds}).
A final key avenue for future observations is to go beyond the single sample size of our CMZ, and investigate the nuclear rings in other nearby galaxy centers with resolved observations from e.g. ALMA and {\em JWST}. 
Such observations will provide context and perspective for many of the open questions in our CMZ, including its geometry and the possibility of a ``boom and bust'' duty cycle for star formation in galaxy centers (\S\ref{subsec:SFR:starformationhistory}).

On the theoretical side, continued improvement to simulations of the inner regions of the Milky Way will be critical to understanding the global cycle of matter and energy in the CMZ. Global simulations  will help to understand the origin of turbulence in the CMZ gas (\S\ref{sec:turbulentdriving}), the transport of gas towards the central black hole (\S\ref{sec:cndinflow}), whether there are preferred locations for star formation in the CMZ (\S\ref{sec:sfhotspots}), and what drives its episodic nature (\S\ref{subsec:SFR:timeevolution}). Simulations that zoom-in at much higher resolution on individual molecular clouds will allow us to understand their formation self-consistently from the large-scale flow, and to follow the evolution of the clouds and of their embedded star formation as they orbit in the Galactic Centre. Zoom-in simulations will also probe the properties of dense gas, which is unresolved in current simulations, down to scales of $<0.1$ \pc, and post-processing of these simulations with radiative transfer tools will enable the creation of synthetic observations for direct comparisons with the plethora of observational data on the horizon. Finally, simulations of Milky Way-like galaxies in a cosmological context will provide further insight on how the evolution of galactic nuclei correlates with the evolution of their host galaxies across a wide range of galaxy properties.

Despite decades of observational and theoretical work, many foundational questions about the nature, context, and future of our CMZ remain unsettled. With decisive observational programs in the works and a resurgence of theoretical interest, we expect the next few years to be an exciting and productive time for CMZ research that will lead to major breakthroughs in our understanding of the star and planet formation process in extreme galactic environments and the role of galaxy centers in global evolution of galaxies.
\\

\noindent\textit{Acknowledgements: } The authors would like to thank Alyssa Goodman, and the other anonymous referee for their insightful comments that have helped to strengthen this review. We further would like to thank James Binney, Maïca Clavel, Filippo Fraternali, Dimitri Gadotti, Simon Glover, Jouni Kainulainen, Melanie Kaasinen, Allison Kirkpatrick, Diederik Kruijssen, Mark Krumholz, Adam Leroy, Steven Longmore, Xing (Walker) Lu, Mattis Magg, Betsy Mills, Desika Narayanan, Tomoharu Oka, Miguel Querejeta, Rainer Schödel, Yoshiaki Sofue, and Jiayi Sun for helpful discussions and comments. 
We thank the speakers of the CMZoom talk series (\url{https://sites.google.com/view/cmzsftalkseries/home}), Thushara Pillai, Kunihiko Tanaka, Pei-Ying Hsieh, Farhad Yusef-Zadeh, Betsy Mills, Jesus Salas, Mark Krumholz, Alvaro Sanchez-Monge, Francisco Nogueras Lara, Maïca Clavel, Mark Morris, Matt Hosek, and Melisse Bonfand-Caldeira.

\bibliographystyle{pp7}
\bibliography{references, references_noads}

\begin{thebibliography}{550}
\parskip=0pt \itemsep=0pt \small \baselineskip=11pt
\providecommand{\natexlab}[1]{#1}

\bibitem[\protect\astroncite{\emph{{Ackermann} et~al.}}{2014}]{Ackermann2014}
{Ackermann} M. et~al., 2014 \emph{\apj}, \emph{793}, 1, 64.

\bibitem[\protect\astroncite{\emph{{Aguirre} et~al.}}{2011}]{Aguirre2011}
{Aguirre} J.~E. et~al., 2011 \emph{\apjs}, \emph{192}, 1, 4.

\bibitem[\protect\astroncite{\emph{{Ahmadi} et~al.}}{2019}]{Ahmadi2019}
{Ahmadi} A. et~al., 2019 \emph{\aap}, \emph{632}, A50.

\bibitem[\protect\astroncite{\emph{{Akhter} et~al.}}{2021}]{Akhter2021}
{Akhter} S. et~al., 2021 \emph{\mnras}, \emph{502}, 4, 5896.

\bibitem[\protect\astroncite{\emph{{Alard}}}{2001}]{Alard2001}
{Alard} C., 2001 \emph{\aap}, \emph{379}, L44.

\bibitem[\protect\astroncite{\emph{{Allard} et~al.}}{2006}]{Allard2006}
{Allard} E.~L. et~al., 2006 \emph{\mnras}, \emph{371}, 3, 1087.

\bibitem[\protect\astroncite{\emph{{An} et~al.}}{2009}]{An2009}
{An} D. et~al., 2009 \emph{\apjl}, \emph{702}, 2, L128.

\bibitem[\protect\astroncite{\emph{{An} et~al.}}{2011}]{An2011}
{An} D. et~al., 2011 \emph{\apj}, \emph{736}, 2, 133.

\bibitem[\protect\astroncite{\emph{{An} et~al.}}{2017}]{An2017}
{An} D. et~al., 2017 \emph{\apjl}, \emph{843}, 2, L36.

\bibitem[\protect\astroncite{\emph{{Andersen} et~al.}}{2017}]{Andersen2017}
{Andersen} M. et~al., 2017 \emph{\aap}, \emph{602}, A22.

\bibitem[\protect\astroncite{\emph{{Anderson} et~al.}}{2020}]{Anderson2020}
{Anderson} L.~D. et~al., 2020 \emph{\apj}, \emph{901}, 1, 51.

\bibitem[\protect\astroncite{\emph{{Andr{\'e}} et~al.}}{2014}]{Andre2014}
{Andr{\'e}} P. et~al., 2014 \emph{Protostars and Planets VI} (H.~{Beuther},
  R.~S. {Klessen}, C.~P. {Dullemond}, and T.~{Henning}), p.~27.

\bibitem[\protect\astroncite{\emph{{Ao} et~al.}}{2013}]{Ao2013}
{Ao} Y. et~al., 2013 \emph{\aap}, \emph{550}, A135.

\bibitem[\protect\astroncite{\emph{{Arendt} et~al.}}{2019}]{Arendt2019}
{Arendt} R.~G. et~al., 2019 \emph{\apj}, \emph{885}, 1, 71.

\bibitem[\protect\astroncite{\emph{{Armijos-Abenda{\~n}o}
  et~al.}}{2020}]{Armijos-Abendano2020}
{Armijos-Abenda{\~n}o} J. et~al., 2020 \emph{\mnras}, \emph{499}, 4, 4918.

\bibitem[\protect\astroncite{\emph{{Armillotta} et~al.}}{2019}]{Armillotta2019}
{Armillotta} L. et~al., 2019 \emph{\mnras}, \emph{490}, 3, 4401.

\bibitem[\protect\astroncite{\emph{{Armillotta} et~al.}}{2020}]{Armillotta2020}
{Armillotta} L. et~al., 2020 \emph{\mnras}, \emph{493}, 4, 5273.

\bibitem[\protect\astroncite{\emph{{Athanassoula}}}{1992{\natexlab{a}}}]{Athanassoula1992a}
{Athanassoula} E., 1992{\natexlab{a}} \emph{\mnras}, \emph{259}, 328.

\bibitem[\protect\astroncite{\emph{{Athanassoula}}}{1992{\natexlab{b}}}]{Athanassoula1992b}
{Athanassoula} E., 1992{\natexlab{b}} \emph{\mnras}, \emph{259}, 345.

\bibitem[\protect\astroncite{\emph{{Audibert} et~al.}}{2019}]{Audibert2019}
{Audibert} A. et~al., 2019 \emph{\aap}, \emph{632}, A33.

\bibitem[\protect\astroncite{\emph{{Baba} and {Kawata}}}{2020}]{Baba2020}
{Baba} J. and {Kawata} D., 2020 \emph{\mnras}, \emph{492}, 3, 4500.

\bibitem[\protect\astroncite{\emph{{Baba} et~al.}}{2010}]{Baba2010}
{Baba} J. et~al., 2010 \emph{\pasj}, \emph{62}, 1413.

\bibitem[\protect\astroncite{\emph{{Balbus} and {Hawley}}}{1998}]{Balbus1998}
{Balbus} S.~A. and {Hawley} J.~F., 1998 \emph{Reviews of Modern Physics},
  \emph{70}, 1, 1.

\bibitem[\protect\astroncite{\emph{{Ballone} et~al.}}{2019}]{Ballone2019}
{Ballone} A. et~al., 2019 \emph{\mnras}, \emph{488}, 4, 5802.

\bibitem[\protect\astroncite{\emph{{Bally}}}{2016}]{Bally2016}
{Bally} J., 2016 \emph{\araa}, \emph{54}, 491.

\bibitem[\protect\astroncite{\emph{{Bally} et~al.}}{1987}]{Bally1987}
{Bally} J. et~al., 1987 \emph{\apjs}, \emph{65}, 13.

\bibitem[\protect\astroncite{\emph{{Bally} et~al.}}{1988}]{Bally1988}
{Bally} J. et~al., 1988 \emph{\apj}, \emph{324}, 223.

\bibitem[\protect\astroncite{\emph{{Bally} et~al.}}{2010}]{Bally2010}
{Bally} J. et~al., 2010 \emph{\apj}, \emph{721}, 1, 137.

\bibitem[\protect\astroncite{\emph{{Balser} et~al.}}{2011}]{Balser2011}
{Balser} D.~S. et~al., 2011 \emph{\apj}, \emph{738}, 1, 27.

\bibitem[\protect\astroncite{\emph{{Bania}}}{1977}]{Bania1977}
{Bania} T.~M., 1977 \emph{\apj}, \emph{216}, 381.

\bibitem[\protect\astroncite{\emph{{Barnes} et~al.}}{2017}]{Barnes2017}
{Barnes} A.~T. et~al., 2017 \emph{\mnras}, \emph{469}, 2, 2263.

\bibitem[\protect\astroncite{\emph{{Barnes} et~al.}}{2019}]{Barnes2019}
{Barnes} A.~T. et~al., 2019 \emph{\mnras}, \emph{486}, 1, 283.

\bibitem[\protect\astroncite{\emph{{Barnes}
  et~al.}}{2020{\natexlab{a}}}]{Barnes2020b}
{Barnes} A.~T. et~al., 2020{\natexlab{a}} \emph{\mnras}, \emph{497}, 2, 1972.

\bibitem[\protect\astroncite{\emph{{Barnes}
  et~al.}}{2020{\natexlab{b}}}]{Barnes2020a}
{Barnes} A.~T. et~al., 2020{\natexlab{b}} \emph{\mnras}, \emph{498}, 4, 4906.

\bibitem[\protect\astroncite{\emph{{Bastian} et~al.}}{2010}]{Bastian2010}
{Bastian} N. et~al., 2010 \emph{\araa}, \emph{48}, 339.

\bibitem[\protect\astroncite{\emph{{Battersby} et~al.}}{2011}]{Battersby2011}
{Battersby} C. et~al., 2011 \emph{\aap}, \emph{535}, A128.

\bibitem[\protect\astroncite{\emph{{Battersby} et~al.}}{2020}]{Battersby2020}
{Battersby} C. et~al., 2020 \emph{\apjs}, \emph{249}, 2, 35.

\bibitem[\protect\astroncite{\emph{{Belloche} et~al.}}{2013}]{Belloche2013}
{Belloche} A. et~al., 2013 \emph{\aap}, \emph{559}, A47.

\bibitem[\protect\astroncite{\emph{{Belloche} et~al.}}{2016}]{Belloche2016}
{Belloche} A. et~al., 2016 \emph{\aap}, \emph{587}, A91.

\bibitem[\protect\astroncite{\emph{{Belloche} et~al.}}{2019}]{Belloche2019}
{Belloche} A. et~al., 2019 \emph{\aap}, \emph{628}, A10.

\bibitem[\protect\astroncite{\emph{{Bergin} and {Tafalla}}}{2007}]{Bergin2007}
{Bergin} E.~A. and {Tafalla} M., 2007 \emph{\araa}, \emph{45}, 1, 339.

\bibitem[\protect\astroncite{\emph{{Be{\v{s}}li{\'c}}
  et~al.}}{2021}]{Beslic2021}
{Be{\v{s}}li{\'c}} I. et~al., 2021 \emph{\mnras}, \emph{506}, 1, 963.

\bibitem[\protect\astroncite{\emph{{Bigiel} et~al.}}{2008}]{Bigiel2008}
{Bigiel} F. et~al., 2008 \emph{\aj}, \emph{136}, 6, 2846.

\bibitem[\protect\astroncite{\emph{{Bigiel} et~al.}}{2016}]{Bigiel2016}
{Bigiel} F. et~al., 2016 \emph{\apjl}, \emph{822}, 2, L26.

\bibitem[\protect\astroncite{\emph{{Binney} and {Tremaine}}}{2008}]{Binney2008}
{Binney} J. and {Tremaine} S., 2008 \emph{{Galactic Dynamics: Second Edition}}.

\bibitem[\protect\astroncite{\emph{{Binney} et~al.}}{1991}]{Binney1991}
{Binney} J. et~al., 1991 \emph{\mnras}, \emph{252}, 210.

\bibitem[\protect\astroncite{\emph{{Bland-Hawthorn} and
  {Cohen}}}{2003}]{Bland-Hawthorn2003}
{Bland-Hawthorn} J. and {Cohen} M., 2003 \emph{\apj}, \emph{582}, 1, 246.

\bibitem[\protect\astroncite{\emph{{Bland-Hawthorn} and
  {Gerhard}}}{2016}]{Bland-Hawthorn2016}
{Bland-Hawthorn} J. and {Gerhard} O., 2016 \emph{\araa}, \emph{54}, 529.

\bibitem[\protect\astroncite{\emph{{Blitz}}}{1993}]{Blitz1993}
{Blitz} L., 1993 \emph{Protostars and Planets III} (E.~H. {Levy} and J.~I.
  {Lunine}), p. 125.

\bibitem[\protect\astroncite{\emph{{Blitz} and {Spergel}}}{1991}]{Blitz1991}
{Blitz} L. and {Spergel} D.~N., 1991 \emph{\apj}, \emph{379}, 631.

\bibitem[\protect\astroncite{\emph{{B{\"o}ker} et~al.}}{2008}]{Boker2008}
{B{\"o}ker} T. et~al., 2008 \emph{\aj}, \emph{135}, 2, 479.

\bibitem[\protect\astroncite{\emph{{Bolatto} et~al.}}{2015}]{Bolatto2015}
{Bolatto} A.~D. et~al., 2015 \emph{\apj}, \emph{809}, 2, 175.

\bibitem[\protect\astroncite{\emph{{Bonfand} et~al.}}{2017}]{Bonfand2017}
{Bonfand} M. et~al., 2017 \emph{\aap}, \emph{604}, A60.

\bibitem[\protect\astroncite{\emph{{Bonfand} et~al.}}{2019}]{Bonfand2019}
{Bonfand} M. et~al., 2019 \emph{\aap}, \emph{628}, A27.

\bibitem[\protect\astroncite{\emph{{Bordoloi} et~al.}}{2017}]{Bordoloi2017}
{Bordoloi} R. et~al., 2017 \emph{\apj}, \emph{834}, 2, 191.

\bibitem[\protect\astroncite{\emph{{Boyce} et~al.}}{1989}]{Boyce1989}
{Boyce} P.~J. et~al., 1989 \emph{\mnras}, \emph{239}, 1013.

\bibitem[\protect\astroncite{\emph{{Bradford} et~al.}}{2005}]{Bradford2005}
{Bradford} C.~M. et~al., 2005 \emph{\apj}, \emph{623}, 2, 866.

\bibitem[\protect\astroncite{\emph{{Bryant} and {Krabbe}}}{2021}]{Bryant2021}
{Bryant} A. and {Krabbe} A., 2021 \emph{\nar}, \emph{93}, 101630.

\bibitem[\protect\astroncite{\emph{{Burkhart} and {Mocz}}}{2019}]{Burkhart2019}
{Burkhart} B. and {Mocz} P., 2019 \emph{\apj}, \emph{879}, 2, 129.

\bibitem[\protect\astroncite{\emph{{Buta} and {Combes}}}{1996}]{Buta1996}
{Buta} R. and {Combes} F., 1996 \emph{\fcp}, \emph{17}, 95.

\bibitem[\protect\astroncite{\emph{{Butterfield}
  et~al.}}{2018}]{Butterfield2018}
{Butterfield} N. et~al., 2018 \emph{\apj}, \emph{852}, 1, 11.

\bibitem[\protect\astroncite{\emph{{Bykov} et~al.}}{2020}]{Bykov2020}
{Bykov} A.~M. et~al., 2020 \emph{\ssr}, \emph{216}, 3, 42.

\bibitem[\protect\astroncite{\emph{{Callanan} et~al.}}{2021}]{Callanan2021}
{Callanan} D. et~al., 2021 \emph{\mnras}, \emph{505}, 3, 4310.

\bibitem[\protect\astroncite{\emph{{Carey} et~al.}}{2009}]{Carey2009}
{Carey} S.~J. et~al., 2009 \emph{\pasp}, \emph{121}, 875, 76.

\bibitem[\protect\astroncite{\emph{{Carretti} et~al.}}{2013}]{Carretti2013}
{Carretti} E. et~al., 2013 \emph{\nat}, \emph{493}, 7430, 66.

\bibitem[\protect\astroncite{\emph{{Caswell} et~al.}}{2010}]{Caswell2010}
{Caswell} J.~L. et~al., 2010 \emph{\mnras}, \emph{404}, 2, 1029.

\bibitem[\protect\astroncite{\emph{{Chambers} et~al.}}{2009}]{Chambers2009}
{Chambers} E.~T. et~al., 2009 \emph{\apjs}, \emph{181}, 2, 360.

\bibitem[\protect\astroncite{\emph{{Chapman} et~al.}}{2011}]{Chapman2011}
{Chapman} N.~L. et~al., 2011 \emph{\apj}, \emph{741}, 1, 21.

\bibitem[\protect\astroncite{\emph{{Chen} et~al.}}{2016}]{Chen2016}
{Chen} M. C.-Y. et~al., 2016 \emph{\apj}, \emph{826}, 1, 95.

\bibitem[\protect\astroncite{\emph{{Christopher}
  et~al.}}{2005}]{Christopher2005}
{Christopher} M.~H. et~al., 2005 \emph{\apj}, \emph{622}, 1, 346.

\bibitem[\protect\astroncite{\emph{{Chuard} et~al.}}{2018}]{Chuard2018}
{Chuard} D. et~al., 2018 \emph{\aap}, \emph{610}, A34.

\bibitem[\protect\astroncite{\emph{{Churazov}
  et~al.}}{2017{\natexlab{a}}}]{Churazov2017b}
{Churazov} E. et~al., 2017{\natexlab{a}} \emph{\mnras}, \emph{465}, 1, 45.

\bibitem[\protect\astroncite{\emph{{Churazov}
  et~al.}}{2017{\natexlab{b}}}]{Churazov2017a}
{Churazov} E. et~al., 2017{\natexlab{b}} \emph{\mnras}, \emph{468}, 1, 165.

\bibitem[\protect\astroncite{\emph{{Churchwell} et~al.}}{2009}]{Churchwell2009}
{Churchwell} E. et~al., 2009 \emph{\pasp}, \emph{121}, 877, 213.

\bibitem[\protect\astroncite{\emph{{Chuss} et~al.}}{2003}]{Chuss2003}
{Chuss} D.~T. et~al., 2003 \emph{\apj}, \emph{599}, 2, 1116.

\bibitem[\protect\astroncite{\emph{{Clark} et~al.}}{2019}]{Clark2019}
{Clark} J.~S. et~al., 2019 \emph{\aap}, \emph{623}, A84.

\bibitem[\protect\astroncite{\emph{{Clark} et~al.}}{2021}]{Clark2021}
{Clark} J.~S. et~al., 2021 \emph{\aap}, \emph{649}, A43.

\bibitem[\protect\astroncite{\emph{{Clark} et~al.}}{2013}]{Clark2013}
{Clark} P.~C. et~al., 2013 \emph{\apjl}, \emph{768}, 2, L34.

\bibitem[\protect\astroncite{\emph{{Clavel} et~al.}}{2013}]{Clavel2013}
{Clavel} M. et~al., 2013 \emph{\aap}, \emph{558}, A32.

\bibitem[\protect\astroncite{\emph{{Cohen} and {Davies}}}{1976}]{Cohen1976}
{Cohen} R.~J. and {Davies} R.~D., 1976 \emph{\mnras}, \emph{175}, 1.

\bibitem[\protect\astroncite{\emph{{Coil} and {Ho}}}{1999}]{Coil1999}
{Coil} A.~L. and {Ho} P. T.~P., 1999 \emph{\apj}, \emph{513}, 2, 752.

\bibitem[\protect\astroncite{\emph{{Coil} and {Ho}}}{2000}]{Coil2000}
{Coil} A.~L. and {Ho} P. T.~P., 2000 \emph{\apj}, \emph{533}, 1, 245.

\bibitem[\protect\astroncite{\emph{{Colzi} et~al.}}{2022}]{Colzi2022}
{Colzi} L. et~al., 2022 \emph{\apjl}, \emph{926}, 2, L22.

\bibitem[\protect\astroncite{\emph{{Combes}}}{1988}]{Combes1988}
{Combes} F., 1988 \emph{Galactic and Extragalactic Star Formation}, vol. 232 of
  \emph{NATO Advanced Study Institute (ASI) Series C} (R.~E. {Pudritz} and
  M.~{Fich}), p. 475.

\bibitem[\protect\astroncite{\emph{{Combes}}}{1991}]{Combes1991}
{Combes} F., 1991 \emph{\araa}, \emph{29}, 195.

\bibitem[\protect\astroncite{\emph{{Combes}}}{2001}]{Combes2001}
{Combes} F., 2001 \emph{Advanced Lectures on the Starburst-AGN} (I.~{Aretxaga},
  D.~{Kunth}, and R.~{M{\'u}jica}), p. 223.

\bibitem[\protect\astroncite{\emph{{Combes}}}{2017{\natexlab{a}}}]{Combes2017a}
{Combes} F., 2017{\natexlab{a}} \emph{Frontiers in Astronomy and Space
  Sciences}, \emph{4}, 10.

\bibitem[\protect\astroncite{\emph{{Combes}}}{2017{\natexlab{b}}}]{Combes2017b}
{Combes} F., 2017{\natexlab{b}} \emph{The Multi-Messenger Astrophysics of the
  Galactic Centre}, vol. 322 (R.~M. {Crocker}, S.~N. {Longmore}, and G.~V.
  {Bicknell}), pp. 245--252.

\bibitem[\protect\astroncite{\emph{{Comer{\'o}n} et~al.}}{2009}]{Comeron2009}
{Comer{\'o}n} S. et~al., 2009 \emph{\apjl}, \emph{706}, 2, L256.

\bibitem[\protect\astroncite{\emph{{Comer{\'o}n} et~al.}}{2010}]{Comeron2010}
{Comer{\'o}n} S. et~al., 2010 \emph{\mnras}, \emph{402}, 4, 2462.

\bibitem[\protect\astroncite{\emph{{Contopoulos} and
  {Grosbol}}}{1989}]{Contopoulos1989}
{Contopoulos} G. and {Grosbol} P., 1989 \emph{\aapr}, \emph{1}, 3-4, 261.

\bibitem[\protect\astroncite{\emph{{Coughlin} et~al.}}{2021}]{Coughlin2021}
{Coughlin} E.~R. et~al., 2021 \emph{\mnras}, \emph{501}, 2, 1868.

\bibitem[\protect\astroncite{\emph{{Crocker}}}{2012}]{Crocker2012}
{Crocker} R.~M., 2012 \emph{\mnras}, \emph{423}, 4, 3512.

\bibitem[\protect\astroncite{\emph{{Crocker} et~al.}}{2011}]{Crocker2011a}
{Crocker} R.~M. et~al., 2011 \emph{\mnras}, \emph{413}, 2, 763.

\bibitem[\protect\astroncite{\emph{{Crocker} et~al.}}{2015}]{Crocker2015}
{Crocker} R.~M. et~al., 2015 \emph{\apj}, \emph{808}, 2, 107.

\bibitem[\protect\astroncite{\emph{{Crutcher} et~al.}}{1996}]{Crutcher1996}
{Crutcher} R.~M. et~al., 1996 \emph{\apjl}, \emph{462}, L79.

\bibitem[\protect\astroncite{\emph{{Csengeri} et~al.}}{2016}]{Csengeri2016}
{Csengeri} T. et~al., 2016 \emph{\aap}, \emph{585}, A104.

\bibitem[\protect\astroncite{\emph{{Cyganowski} et~al.}}{2008}]{Cyganowski2008}
{Cyganowski} C.~J. et~al., 2008 \emph{\aj}, \emph{136}, 6, 2391.

\bibitem[\protect\astroncite{\emph{{Dahmen} et~al.}}{1998}]{Dahmen1998}
{Dahmen} G. et~al., 1998 \emph{\aap}, \emph{331}, 959.

\bibitem[\protect\astroncite{\emph{{Dale} et~al.}}{2019}]{Dale2019}
{Dale} J.~E. et~al., 2019 \emph{\mnras}, \emph{486}, 3, 3307.

\bibitem[\protect\astroncite{\emph{{Dame} et~al.}}{2001}]{Dame2001}
{Dame} T.~M. et~al., 2001 \emph{\apj}, \emph{547}, 2, 792.

\bibitem[\protect\astroncite{\emph{{Davies} et~al.}}{2007}]{Davies2007}
{Davies} R.~I. et~al., 2007 \emph{\apj}, \emph{671}, 2, 1388.

\bibitem[\protect\astroncite{\emph{{de Pree} et~al.}}{1995}]{dePree1995}
{de Pree} C.~G. et~al., 1995 \emph{\apj}, \emph{451}, 284.

\bibitem[\protect\astroncite{\emph{{de Pree} et~al.}}{1996}]{dePree1996}
{de Pree} C.~G. et~al., 1996 \emph{\apj}, \emph{464}, 788.

\bibitem[\protect\astroncite{\emph{{De Pree} et~al.}}{2015}]{DePree2015}
{De Pree} C.~G. et~al., 2015 \emph{\apj}, \emph{815}, 2, 123.

\bibitem[\protect\astroncite{\emph{{de Vaucouleurs}}}{1964}]{deVaucouleurs1964}
{de Vaucouleurs} G., 1964 \emph{The Galaxy and the Magellanic Clouds}, vol.~20
  (F.~J. {Kerr}), p. 195.

\bibitem[\protect\astroncite{\emph{{Di Teodoro} et~al.}}{2018}]{DiTeodoro2018}
{Di Teodoro} E.~M. et~al., 2018 \emph{\apj}, \emph{855}, 1, 33.

\bibitem[\protect\astroncite{\emph{{Di Teodoro} et~al.}}{2020}]{Diteodoro2020}
{Di Teodoro} E.~M. et~al., 2020 \emph{\nat}, \emph{584}, 7821, 364.

\bibitem[\protect\astroncite{\emph{{Dinh} et~al.}}{2021}]{Dinh2021}
{Dinh} C.~K. et~al., 2021 \emph{\apj}, \emph{920}, 2, 79.

\bibitem[\protect\astroncite{\emph{{Dokara} et~al.}}{2021}]{Dokara2021}
{Dokara} R. et~al., 2021 \emph{\aap}, \emph{651}, A86.

\bibitem[\protect\astroncite{\emph{{Dong} et~al.}}{2015}]{Dong2015}
{Dong} H. et~al., 2015 \emph{\mnras}, \emph{446}, 1, 842.

\bibitem[\protect\astroncite{\emph{{Ebisawa} et~al.}}{2001}]{Ebisawa2001}
{Ebisawa} K. et~al., 2001 \emph{Science}, \emph{293}, 5535, 1633.

\bibitem[\protect\astroncite{\emph{{Eden} et~al.}}{2020}]{Eden2020}
{Eden} D.~J. et~al., 2020 \emph{\mnras}, \emph{498}, 4, 5936.

\bibitem[\protect\astroncite{\emph{{Eibensteiner}
  et~al.}}{2022}]{Eibensteiner2022}
{Eibensteiner} C. et~al., 2022 \emph{arXiv e-prints}, arXiv:2201.02209.

\bibitem[\protect\astroncite{\emph{{Ekers} et~al.}}{1983}]{Ekers1983}
{Ekers} R.~D. et~al., 1983 \emph{\aap}, \emph{122}, 143.

\bibitem[\protect\astroncite{\emph{{Ellingsen}}}{2006}]{Ellingsen2006}
{Ellingsen} S.~P., 2006 \emph{\apj}, \emph{638}, 1, 241.

\bibitem[\protect\astroncite{\emph{{Elmegreen}}}{1994}]{Elmegreen1994}
{Elmegreen} B.~G., 1994 \emph{\apjl}, \emph{425}, L73.

\bibitem[\protect\astroncite{\emph{{Emsellem} et~al.}}{2015}]{Emsellem2015}
{Emsellem} E. et~al., 2015 \emph{\mnras}, \emph{446}, 3, 2468.

\bibitem[\protect\astroncite{\emph{{Englmaier} and
  {Gerhard}}}{1997}]{Englmaier1997}
{Englmaier} P. and {Gerhard} O., 1997 \emph{\mnras}, \emph{287}, 1, 57.

\bibitem[\protect\astroncite{\emph{{Enokiya} et~al.}}{2021}]{Enokiya2021}
{Enokiya} R. et~al., 2021 \emph{\pasj}, \emph{73}, S75.

\bibitem[\protect\astroncite{\emph{{Etxaluze} et~al.}}{2011}]{Etxaluze2011}
{Etxaluze} M. et~al., 2011 \emph{\aj}, \emph{142}, 4, 134.

\bibitem[\protect\astroncite{\emph{{Evans} et~al.}}{2009}]{Evans2009}
{Evans} Neal~J. I. et~al., 2009 \emph{\apjs}, \emph{181}, 2, 321.

\bibitem[\protect\astroncite{\emph{{Federrath} and
  {Klessen}}}{2012}]{Federrath2012}
{Federrath} C. and {Klessen} R.~S., 2012 \emph{\apj}, \emph{761}, 2, 156.

\bibitem[\protect\astroncite{\emph{{Federrath} et~al.}}{2016}]{Federrath2016}
{Federrath} C. et~al., 2016 \emph{\apj}, \emph{832}, 2, 143.

\bibitem[\protect\astroncite{\emph{{Feldmeier-Krause}
  et~al.}}{2017}]{Feldmeier-Krause2017}
{Feldmeier-Krause} A. et~al., 2017 \emph{\mnras}, \emph{466}, 4, 4040.

\bibitem[\protect\astroncite{\emph{{Ferri{\`e}re}}}{2012}]{Ferriere2012}
{Ferri{\`e}re} K., 2012 \emph{\aap}, \emph{540}, A50.

\bibitem[\protect\astroncite{\emph{{Ferri{\`e}re} et~al.}}{2007}]{Ferriere2007}
{Ferri{\`e}re} K. et~al., 2007 \emph{\aap}, \emph{467}, 2, 611.

\bibitem[\protect\astroncite{\emph{{Field} et~al.}}{2011}]{Field2011}
{Field} G.~B. et~al., 2011 \emph{\mnras}, \emph{416}, 1, 710.

\bibitem[\protect\astroncite{\emph{{Figer} et~al.}}{2004}]{Figer2004}
{Figer} D.~F. et~al., 2004 \emph{\apj}, \emph{601}, 1, 319.

\bibitem[\protect\astroncite{\emph{{Flower} et~al.}}{2005}]{Flower2005}
{Flower} D.~R. et~al., 2005 \emph{\aap}, \emph{436}, 3, 933.

\bibitem[\protect\astroncite{\emph{{Forster} and
  {Caswell}}}{2000}]{Forster2000}
{Forster} J.~R. and {Caswell} J.~L., 2000 \emph{\apj}, \emph{530}, 1, 371.

\bibitem[\protect\astroncite{\emph{{Freundlich} et~al.}}{2013}]{Freundlich2013}
{Freundlich} J. et~al., 2013 \emph{\aap}, \emph{553}, A130.

\bibitem[\protect\astroncite{\emph{{Friesen} et~al.}}{2017}]{Friesen2017}
{Friesen} R.~K. et~al., 2017 \emph{\apj}, \emph{843}, 1, 63.

\bibitem[\protect\astroncite{\emph{{Fujishita} et~al.}}{2009}]{Fujishita2009}
{Fujishita} M. et~al., 2009 \emph{\pasj}, \emph{61}, 1039.

\bibitem[\protect\astroncite{\emph{{Fukui} et~al.}}{2006}]{Fukui2006}
{Fukui} Y. et~al., 2006 \emph{Science}, \emph{314}, 5796, 106.

\bibitem[\protect\astroncite{\emph{{Fukui} et~al.}}{2021}]{Fukui2021}
{Fukui} Y. et~al., 2021 \emph{\pasj}, \emph{73}, S1.

\bibitem[\protect\astroncite{\emph{{Fux}}}{1999}]{Fux1999}
{Fux} R., 1999 \emph{\aap}, \emph{345}, 787.

\bibitem[\protect\astroncite{\emph{{Gadotti} et~al.}}{2019}]{Gadotti2019}
{Gadotti} D.~A. et~al., 2019 \emph{\mnras}, \emph{482}, 1, 506.

\bibitem[\protect\astroncite{\emph{{Gallego-Calvente}
  et~al.}}{2021{\natexlab{a}}}]{Gallego-Calvente2021b}
{Gallego-Calvente} A.~T. et~al., 2021{\natexlab{a}} \emph{\aap}, \emph{647},
  A110.

\bibitem[\protect\astroncite{\emph{{Gallego-Calvente}
  et~al.}}{2021{\natexlab{b}}}]{Gallego-Calvente2021a}
{Gallego-Calvente} A.~T. et~al., 2021{\natexlab{b}} \emph{arXiv e-prints},
  arXiv:2107.14481.

\bibitem[\protect\astroncite{\emph{{Gao} and {Solomon}}}{2004}]{Gao2004}
{Gao} Y. and {Solomon} P.~M., 2004 \emph{\apj}, \emph{606}, 1, 271.

\bibitem[\protect\astroncite{\emph{{Garc{\'\i}a-Burillo}
  et~al.}}{2005}]{Garcia-Burillo2005}
{Garc{\'\i}a-Burillo} S. et~al., 2005 \emph{\aap}, \emph{441}, 3, 1011.

\bibitem[\protect\astroncite{\emph{{Gaume} et~al.}}{1995}]{Gaume1995}
{Gaume} R.~A. et~al., 1995 \emph{\apj}, \emph{449}, 663.

\bibitem[\protect\astroncite{\emph{{Genzel} et~al.}}{1985}]{Genzel1985}
{Genzel} R. et~al., 1985 \emph{\apj}, \emph{297}, 766.

\bibitem[\protect\astroncite{\emph{{Genzel} et~al.}}{1994}]{Genzel1994}
{Genzel} R. et~al., 1994 \emph{Reports on Progress in Physics}, \emph{57}, 5,
  417.

\bibitem[\protect\astroncite{\emph{{Genzel} et~al.}}{2010}]{Genzel2010}
{Genzel} R. et~al., 2010 \emph{Reviews of Modern Physics}, \emph{82}, 4, 3121.

\bibitem[\protect\astroncite{\emph{{Gerhard} and
  {Martinez-Valpuesta}}}{2012}]{Gerhard2012}
{Gerhard} O. and {Martinez-Valpuesta} I., 2012 \emph{\apjl}, \emph{744}, 1, L8.

\bibitem[\protect\astroncite{\emph{{Gerhard}}}{1992}]{Gerhard1992}
{Gerhard} O.~E., 1992 \emph{Reviews in Modern Astronomy}, \emph{5}, 174.

\bibitem[\protect\astroncite{\emph{{Ghez} et~al.}}{2008}]{Ghez2008}
{Ghez} A.~M. et~al., 2008 \emph{\apj}, \emph{689}, 2, 1044.

\bibitem[\protect\astroncite{\emph{{Gillessen} et~al.}}{2009}]{Gillessen2009}
{Gillessen} S. et~al., 2009 \emph{\apj}, \emph{692}, 2, 1075.

\bibitem[\protect\astroncite{\emph{{Ginsburg} and
  {Kruijssen}}}{2018}]{Ginsburg2018a}
{Ginsburg} A. and {Kruijssen} J.~M.~D., 2018 \emph{\apjl}, \emph{864}, 1, L17.

\bibitem[\protect\astroncite{\emph{{Ginsburg} et~al.}}{2013}]{Ginsburg2013}
{Ginsburg} A. et~al., 2013 \emph{\apjs}, \emph{208}, 2, 14.

\bibitem[\protect\astroncite{\emph{{Ginsburg} et~al.}}{2015}]{Ginsburg2015}
{Ginsburg} A. et~al., 2015 \emph{\aap}, \emph{584}, L7.

\bibitem[\protect\astroncite{\emph{{Ginsburg} et~al.}}{2016}]{Ginsburg2016}
{Ginsburg} A. et~al., 2016 \emph{\aap}, \emph{586}, A50.

\bibitem[\protect\astroncite{\emph{{Ginsburg} et~al.}}{2018}]{Ginsburg2018b}
{Ginsburg} A. et~al., 2018 \emph{\apj}, \emph{853}, 2, 171.

\bibitem[\protect\astroncite{\emph{{Ginsburg} et~al.}}{2020}]{Ginsburg2020}
{Ginsburg} A. et~al., 2020 \emph{\apjs}, \emph{248}, 2, 24.

\bibitem[\protect\astroncite{\emph{{Goicoechea} et~al.}}{2018}]{Goicoechea2018}
{Goicoechea} J.~R. et~al., 2018 \emph{\aap}, \emph{618}, A35.

\bibitem[\protect\astroncite{\emph{{Gravity Collaboration}
  et~al.}}{2019}]{GravityCollaboration2019}
{Gravity Collaboration} et~al., 2019 \emph{\aap}, \emph{625}, L10.

\bibitem[\protect\astroncite{\emph{{Green}}}{2019}]{Green2019}
{Green} D.~A., 2019 \emph{Journal of Astrophysics and Astronomy}, \emph{40}, 4,
  36.

\bibitem[\protect\astroncite{\emph{{Green} et~al.}}{2009}]{Green2009}
{Green} J.~A. et~al., 2009 \emph{\mnras}, \emph{392}, 2, 783.

\bibitem[\protect\astroncite{\emph{{Grudi{\'c}} et~al.}}{2021}]{Grudic2021}
{Grudi{\'c}} M.~Y. et~al., 2021 \emph{\mnras}, \emph{506}, 3, 3239.

\bibitem[\protect\astroncite{\emph{{Guan} et~al.}}{2021}]{Guan2021}
{Guan} Y. et~al., 2021 \emph{\apj}, \emph{920}, 1, 6.

\bibitem[\protect\astroncite{\emph{{Guenduez} et~al.}}{2020}]{Guenduez2020}
{Guenduez} M. et~al., 2020 \emph{\aap}, \emph{644}, A71.

\bibitem[\protect\astroncite{\emph{{Guesten} et~al.}}{1987}]{Guesten1987}
{Guesten} R. et~al., 1987 \emph{\apj}, \emph{318}, 124.

\bibitem[\protect\astroncite{\emph{{Guo} and {Mathews}}}{2012}]{Guo2012}
{Guo} F. and {Mathews} W.~G., 2012 \emph{\apj}, \emph{756}, 2, 181.

\bibitem[\protect\astroncite{\emph{{Gusdorf} et~al.}}{2008}]{Gusdorf2008}
{Gusdorf} A. et~al., 2008 \emph{\aap}, \emph{482}, 3, 809.

\bibitem[\protect\astroncite{\emph{{H.~E.~S.~S. Collaboration}
  et~al.}}{2018}]{H.E.S.S.Collaboration2018}
{H.~E.~S.~S. Collaboration} et~al., 2018 \emph{\aap}, \emph{617}, A73.

\bibitem[\protect\astroncite{\emph{{Habibi} et~al.}}{2014}]{Habibi2014}
{Habibi} M. et~al., 2014 \emph{\aap}, \emph{566}, A6.

\bibitem[\protect\astroncite{\emph{{Hankins} et~al.}}{2020}]{Hankins2020}
{Hankins} M.~J. et~al., 2020 \emph{\apj}, \emph{894}, 1, 55.

\bibitem[\protect\astroncite{\emph{{Harada} et~al.}}{2019}]{Harada2019}
{Harada} N. et~al., 2019 \emph{\apj}, \emph{884}, 2, 100.

\bibitem[\protect\astroncite{\emph{{Harris} et~al.}}{2021}]{Harris2021}
{Harris} A.~I. et~al., 2021 \emph{\apj}, \emph{921}, 1, 33.

\bibitem[\protect\astroncite{\emph{{Harris} et~al.}}{2001}]{Harris2001}
{Harris} J. et~al., 2001 \emph{\aj}, \emph{122}, 6, 3046.

\bibitem[\protect\astroncite{\emph{{Hasegawa} et~al.}}{1994}]{Hasegawa1994}
{Hasegawa} T. et~al., 1994 \emph{\apjl}, \emph{429}, L77.

\bibitem[\protect\astroncite{\emph{{Hatchfield} et~al.}}{2020}]{Hatchfield2020}
{Hatchfield} H.~P. et~al., 2020 \emph{\apjs}, \emph{251}, 1, 14.

\bibitem[\protect\astroncite{\emph{{Hatchfield} et~al.}}{2021}]{Hatchfield2021}
{Hatchfield} H.~P. et~al., 2021 \emph{\apj}, \emph{922}, 1, 79.

\bibitem[\protect\astroncite{\emph{{Hawley} et~al.}}{1995}]{Hawley1995}
{Hawley} J.~F. et~al., 1995 \emph{\apj}, \emph{440}, 742.

\bibitem[\protect\astroncite{\emph{{Haworth} et~al.}}{2015}]{Haworth2015}
{Haworth} T.~J. et~al., 2015 \emph{\mnras}, \emph{450}, 1, 10.

\bibitem[\protect\astroncite{\emph{{Heiderman} et~al.}}{2010}]{Heiderman2010}
{Heiderman} A. et~al., 2010 \emph{\apj}, \emph{723}, 2, 1019.

\bibitem[\protect\astroncite{\emph{{Hennebelle} and
  {Chabrier}}}{2011}]{Hennebelle2011}
{Hennebelle} P. and {Chabrier} G., 2011 \emph{\apjl}, \emph{743}, 2, L29.

\bibitem[\protect\astroncite{\emph{{Hennebelle} and
  {Chabrier}}}{2013}]{Hennebelle2013}
{Hennebelle} P. and {Chabrier} G., 2013 \emph{\apj}, \emph{770}, 2, 150.

\bibitem[\protect\astroncite{\emph{{Hennig} et~al.}}{2018}]{Hennig2018}
{Hennig} M.~G. et~al., 2018 \emph{\mnras}, \emph{477}, 1, 1086.

\bibitem[\protect\astroncite{\emph{{Henshaw}
  et~al.}}{2016{\natexlab{a}}}]{Henshaw2016b}
{Henshaw} J.~D. et~al., 2016{\natexlab{a}} \emph{\mnras}, \emph{457}, 3, 2675.

\bibitem[\protect\astroncite{\emph{{Henshaw}
  et~al.}}{2016{\natexlab{b}}}]{Henshaw2016a}
{Henshaw} J.~D. et~al., 2016{\natexlab{b}} \emph{\mnras}, \emph{463}, 1, L122.

\bibitem[\protect\astroncite{\emph{{Henshaw} et~al.}}{2017}]{Henshaw2017}
{Henshaw} J.~D. et~al., 2017 \emph{\mnras}, \emph{466}, 1, L13.

\bibitem[\protect\astroncite{\emph{{Henshaw} et~al.}}{2019}]{Henshaw2019}
{Henshaw} J.~D. et~al., 2019 \emph{\mnras}, \emph{485}, 2, 2457.

\bibitem[\protect\astroncite{\emph{{Henshaw} et~al.}}{2020}]{Henshaw2020}
{Henshaw} J.~D. et~al., 2020 \emph{Nature Astronomy}, \emph{4}, 1064.

\bibitem[\protect\astroncite{\emph{{Henshaw} et~al.}}{2022}]{Henshaw2022}
{Henshaw} J.~D. et~al., 2022 \emph{\mnras}, \emph{509}, 4, 4758.

\bibitem[\protect\astroncite{\emph{{Herrnstein} and
  {Ho}}}{2005}]{Herrnstein2005}
{Herrnstein} R.~M. and {Ho} P. T.~P., 2005 \emph{\apj}, \emph{620}, 1, 287.

\bibitem[\protect\astroncite{\emph{{HESS Collaboration}
  et~al.}}{2016}]{HESSCollaboration2016}
{HESS Collaboration} et~al., 2016 \emph{\nat}, \emph{531}, 7595, 476.

\bibitem[\protect\astroncite{\emph{{Heyer} and {Dame}}}{2015}]{Heyer2015}
{Heyer} M. and {Dame} T.~M., 2015 \emph{\araa}, \emph{53}, 583.

\bibitem[\protect\astroncite{\emph{{Heywood} et~al.}}{2019}]{Heywood2019}
{Heywood} I. et~al., 2019 \emph{\nat}, \emph{573}, 7773, 235.

\bibitem[\protect\astroncite{\emph{{Heywood} et~al.}}{2022}]{Heywood2022}
{Heywood} I. et~al., 2022 \emph{\apj}, \emph{925}, 2, 165.

\bibitem[\protect\astroncite{\emph{{Higuchi} et~al.}}{2014}]{Higuchi2014}
{Higuchi} A.~E. et~al., 2014 \emph{\aj}, \emph{147}, 6, 141.

\bibitem[\protect\astroncite{\emph{{Ho} et~al.}}{1991}]{Ho1991}
{Ho} P. T.~P. et~al., 1991 \emph{\nat}, \emph{350}, 6316, 309.

\bibitem[\protect\astroncite{\emph{{Hosek} et~al.}}{2019}]{Hosek2019}
{Hosek} Matthew~W. J. et~al., 2019 \emph{\apj}, \emph{870}, 1, 44.

\bibitem[\protect\astroncite{\emph{{Hsieh} et~al.}}{2017}]{Hsieh2017}
{Hsieh} P.-Y. et~al., 2017 \emph{\apj}, \emph{847}, 1, 3.

\bibitem[\protect\astroncite{\emph{{Hsieh} et~al.}}{2019}]{Hsieh2019}
{Hsieh} P.-Y. et~al., 2019 \emph{\apjl}, \emph{885}, 1, L20.

\bibitem[\protect\astroncite{\emph{{Hsieh} et~al.}}{2021}]{Hsieh2021}
{Hsieh} P.-Y. et~al., 2021 \emph{\apj}, \emph{913}, 2, 94.

\bibitem[\protect\astroncite{\emph{{Hu} et~al.}}{2022{\natexlab{a}}}]{Hu2022b}
{Hu} Y. et~al., 2022{\natexlab{a}} \emph{arXiv e-prints}, arXiv:2201.07970.

\bibitem[\protect\astroncite{\emph{{Hu} et~al.}}{2022{\natexlab{b}}}]{Hu2022}
{Hu} Y. et~al., 2022{\natexlab{b}} \emph{\mnras}, \emph{511}, 1, 829.

\bibitem[\protect\astroncite{\emph{{Huettemeister}
  et~al.}}{1998}]{Huettemeister1998}
{Huettemeister} S. et~al., 1998 \emph{\aap}, \emph{334}, 646.

\bibitem[\protect\astroncite{\emph{{Hunt} et~al.}}{2008}]{Hunt2008}
{Hunt} L.~K. et~al., 2008 \emph{\aap}, \emph{482}, 1, 133.

\bibitem[\protect\astroncite{\emph{{Immer}
  et~al.}}{2012{\natexlab{a}}}]{Immer2012a}
{Immer} K. et~al., 2012{\natexlab{a}} \emph{\aap}, \emph{548}, A120.

\bibitem[\protect\astroncite{\emph{{Immer}
  et~al.}}{2012{\natexlab{b}}}]{Immer2012b}
{Immer} K. et~al., 2012{\natexlab{b}} \emph{\aap}, \emph{537}, A121.

\bibitem[\protect\astroncite{\emph{{Immer} et~al.}}{2016}]{Immer2016}
{Immer} K. et~al., 2016 \emph{\aap}, \emph{595}, A94.

\bibitem[\protect\astroncite{\emph{{Immer} et~al.}}{2020}]{Immer2020}
{Immer} K. et~al., 2020 \emph{New Horizons in Galactic Center Astronomy and
  Beyond}, vol. 528 (M.~{Tsuboi} and T.~{Oka}).

\bibitem[\protect\astroncite{\emph{{Indriolo} et~al.}}{2015}]{Indriolo2015}
{Indriolo} N. et~al., 2015 \emph{\apj}, \emph{800}, 1, 40.

\bibitem[\protect\astroncite{\emph{{Jackson} et~al.}}{1993}]{Jackson1993}
{Jackson} J.~M. et~al., 1993 \emph{\apj}, \emph{402}, 173.

\bibitem[\protect\astroncite{\emph{{James} et~al.}}{2021}]{James2021}
{James} T.~A. et~al., 2021 \emph{\apj}, \emph{916}, 2, 69.

\bibitem[\protect\astroncite{\emph{{Jeffreson} et~al.}}{2018}]{Jeffreson2018a}
{Jeffreson} S.~M.~R. et~al., 2018 \emph{\mnras}, \emph{478}, 3, 3380.

\bibitem[\protect\astroncite{\emph{{Jiang} et~al.}}{2020}]{Jiang2020}
{Jiang} X.-J. et~al., 2020 \emph{\mnras}, \emph{494}, 1, 1276.

\bibitem[\protect\astroncite{\emph{{Jim{\'e}nez-Donaire}
  et~al.}}{2019}]{Jimenez-Donaire2019}
{Jim{\'e}nez-Donaire} M.~J. et~al., 2019 \emph{\apj}, \emph{880}, 2, 127.

\bibitem[\protect\astroncite{\emph{{Jim{\'e}nez-Serra}
  et~al.}}{2008}]{Jimenez-Serra2008}
{Jim{\'e}nez-Serra} I. et~al., 2008 \emph{\aap}, \emph{482}, 2, 549.

\bibitem[\protect\astroncite{\emph{{Johnston} et~al.}}{2014}]{Johnston2014}
{Johnston} K.~G. et~al., 2014 \emph{\aap}, \emph{568}, A56.

\bibitem[\protect\astroncite{\emph{{Jones} et~al.}}{2012}]{Jones2012}
{Jones} P.~A. et~al., 2012 \emph{\mnras}, \emph{419}, 4, 2961.

\bibitem[\protect\astroncite{\emph{{Jones} et~al.}}{2013}]{Jones2013}
{Jones} P.~A. et~al., 2013 \emph{\mnras}, \emph{433}, 1, 221.

\bibitem[\protect\astroncite{\emph{{Kaifu} et~al.}}{1972}]{Kaifu1972}
{Kaifu} N. et~al., 1972 \emph{Nature Physical Science}, \emph{238}, 85, 105.

\bibitem[\protect\astroncite{\emph{{Kainulainen}
  et~al.}}{2014}]{Kainulainen2014}
{Kainulainen} J. et~al., 2014 \emph{Science}, \emph{344}, 6180, 183.

\bibitem[\protect\astroncite{\emph{{Kauffmann} et~al.}}{2013}]{Kauffmann2013}
{Kauffmann} J. et~al., 2013 \emph{\apjl}, \emph{765}, 2, L35.

\bibitem[\protect\astroncite{\emph{{Kauffmann}
  et~al.}}{2017{\natexlab{a}}}]{Kauffmann2017a}
{Kauffmann} J. et~al., 2017{\natexlab{a}} \emph{\aap}, \emph{605}, L5.

\bibitem[\protect\astroncite{\emph{{Kauffmann}
  et~al.}}{2017{\natexlab{b}}}]{Kauffmann2017b}
{Kauffmann} J. et~al., 2017{\natexlab{b}} \emph{\aap}, \emph{603}, A89.

\bibitem[\protect\astroncite{\emph{{Kauffmann}
  et~al.}}{2017{\natexlab{c}}}]{Kauffmann2017c}
{Kauffmann} J. et~al., 2017{\natexlab{c}} \emph{\aap}, \emph{603}, A90.

\bibitem[\protect\astroncite{\emph{{Kendrew} et~al.}}{2013}]{Kendrew2013}
{Kendrew} S. et~al., 2013 \emph{\apjl}, \emph{775}, 2, L50.

\bibitem[\protect\astroncite{\emph{{Kennicutt}}}{1998}]{Kennicutt1998}
{Kennicutt} Robert~C. J., 1998 \emph{\apj}, \emph{498}, 2, 541.

\bibitem[\protect\astroncite{\emph{{Kennicutt} and
  {Evans}}}{2012}]{Kennicutt2012}
{Kennicutt} R.~C. and {Evans} N.~J., 2012 \emph{\araa}, \emph{50}, 531.

\bibitem[\protect\astroncite{\emph{{Kim} and {Elmegreen}}}{2017}]{Kim2017}
{Kim} W.-T. and {Elmegreen} B.~G., 2017 \emph{\apjl}, \emph{841}, 1, L4.

\bibitem[\protect\astroncite{\emph{{Kim} and {Stone}}}{2012}]{Kim2012b}
{Kim} W.-T. and {Stone} J.~M., 2012 \emph{\apj}, \emph{751}, 2, 124.

\bibitem[\protect\astroncite{\emph{{Kim}
  et~al.}}{2012{\natexlab{a}}}]{Kim2012c}
{Kim} W.-T. et~al., 2012{\natexlab{a}} \emph{\apj}, \emph{747}, 1, 60.

\bibitem[\protect\astroncite{\emph{{Kim}
  et~al.}}{2012{\natexlab{b}}}]{Kim2012a}
{Kim} W.-T. et~al., 2012{\natexlab{b}} \emph{\apj}, \emph{758}, 1, 14.

\bibitem[\protect\astroncite{\emph{{Klessen} and
  {Hennebelle}}}{2010}]{Klessen2010}
{Klessen} R.~S. and {Hennebelle} P., 2010 \emph{\aap}, \emph{520}, A17.

\bibitem[\protect\astroncite{\emph{{Knapen} et~al.}}{2002}]{Knapen2002}
{Knapen} J.~H. et~al., 2002 \emph{\mnras}, \emph{337}, 3, 808.

\bibitem[\protect\astroncite{\emph{{Koepferl} et~al.}}{2015}]{Koepferl2015}
{Koepferl} C.~M. et~al., 2015 \emph{\apj}, \emph{799}, 1, 53.

\bibitem[\protect\astroncite{\emph{{Kormendy} and
  {Kennicutt}}}{2004}]{Kormendy2004}
{Kormendy} J. and {Kennicutt} Robert~C. J., 2004 \emph{\araa}, \emph{42}, 1,
  603.

\bibitem[\protect\astroncite{\emph{{Koyama} et~al.}}{2009}]{Koyama2009}
{Koyama} K. et~al., 2009 \emph{\pasj}, \emph{61}, S255.

\bibitem[\protect\astroncite{\emph{{Krieger} et~al.}}{2017}]{Krieger2017}
{Krieger} N. et~al., 2017 \emph{\apj}, \emph{850}, 1, 77.

\bibitem[\protect\astroncite{\emph{{Krieger} et~al.}}{2020}]{Krieger2020}
{Krieger} N. et~al., 2020 \emph{\apj}, \emph{899}, 2, 158.

\bibitem[\protect\astroncite{\emph{{Krishnarao}
  et~al.}}{2020}]{Krishnarao2020b}
{Krishnarao} D. et~al., 2020 \emph{Science Advances}, \emph{6}, 27, 9711.

\bibitem[\protect\astroncite{\emph{{Krugel} and {Tutukov}}}{1993}]{Krugel1993}
{Krugel} E. and {Tutukov} A.~V., 1993 \emph{\aap}, \emph{275}, 416.

\bibitem[\protect\astroncite{\emph{{Kruijssen}}}{2012}]{Kruijssen2012}
{Kruijssen} J.~M.~D., 2012 \emph{\mnras}, \emph{426}, 4, 3008.

\bibitem[\protect\astroncite{\emph{{Kruijssen}}}{2017}]{Kruijssen2017}
{Kruijssen} J.~M.~D., 2017 \emph{The Multi-Messenger Astrophysics of the
  Galactic Centre}, vol. 322 (R.~M. {Crocker}, S.~N. {Longmore}, and G.~V.
  {Bicknell}), pp. 64--74.

\bibitem[\protect\astroncite{\emph{{Kruijssen} and
  {Longmore}}}{2013}]{Kruijssen2013}
{Kruijssen} J.~M.~D. and {Longmore} S.~N., 2013 \emph{\mnras}, \emph{435}, 3,
  2598.

\bibitem[\protect\astroncite{\emph{{Kruijssen} et~al.}}{2014}]{Kruijssen2014a}
{Kruijssen} J.~M.~D. et~al., 2014 \emph{\mnras}, \emph{440}, 4, 3370.

\bibitem[\protect\astroncite{\emph{{Kruijssen} et~al.}}{2015}]{Kruijssen2015}
{Kruijssen} J.~M.~D. et~al., 2015 \emph{\mnras}, \emph{447}, 2, 1059.

\bibitem[\protect\astroncite{\emph{{Kruijssen} et~al.}}{2019}]{Kruijssen2019}
{Kruijssen} J.~M.~D. et~al., 2019 \emph{\mnras}, \emph{484}, 4, 5734.

\bibitem[\protect\astroncite{\emph{{Krumholz}}}{2014}]{Krumholz2014a}
{Krumholz} M.~R., 2014 \emph{\physrep}, \emph{539}, 49.

\bibitem[\protect\astroncite{\emph{{Krumholz} and
  {Kruijssen}}}{2015}]{Krumholz2015}
{Krumholz} M.~R. and {Kruijssen} J.~M.~D., 2015 \emph{\mnras}, \emph{453}, 1,
  739.

\bibitem[\protect\astroncite{\emph{{Krumholz} and
  {McKee}}}{2005}]{Krumholz2005}
{Krumholz} M.~R. and {McKee} C.~F., 2005 \emph{\apj}, \emph{630}, 1, 250.

\bibitem[\protect\astroncite{\emph{{Krumholz} and
  {McKee}}}{2020}]{Krumholz2020}
{Krumholz} M.~R. and {McKee} C.~F., 2020 \emph{\mnras}, \emph{494}, 1, 624.

\bibitem[\protect\astroncite{\emph{{Krumholz} et~al.}}{2017}]{Krumholz2017}
{Krumholz} M.~R. et~al., 2017 \emph{\mnras}, \emph{466}, 1, 1213.

\bibitem[\protect\astroncite{\emph{{Krumholz} et~al.}}{2019}]{Krumholz2019}
{Krumholz} M.~R. et~al., 2019 \emph{\araa}, \emph{57}, 227.

\bibitem[\protect\astroncite{\emph{{Lacki}}}{2014}]{Lacki2014}
{Lacki} B.~C., 2014 \emph{\mnras}, \emph{444}, L39.

\bibitem[\protect\astroncite{\emph{{Lacki} et~al.}}{2011}]{Lacki2011}
{Lacki} B.~C. et~al., 2011 \emph{\apj}, \emph{734}, 2, 107.

\bibitem[\protect\astroncite{\emph{{Lada} and {Lada}}}{2003}]{Lada2003}
{Lada} C.~J. and {Lada} E.~A., 2003 \emph{\araa}, \emph{41}, 57.

\bibitem[\protect\astroncite{\emph{{Lada} et~al.}}{2010}]{Lada2010}
{Lada} C.~J. et~al., 2010 \emph{\apj}, \emph{724}, 1, 687.

\bibitem[\protect\astroncite{\emph{{Lada} et~al.}}{2012}]{Lada2012}
{Lada} C.~J. et~al., 2012 \emph{\apj}, \emph{745}, 2, 190.

\bibitem[\protect\astroncite{\emph{{Lang} et~al.}}{2005}]{Lang2005}
{Lang} C.~C. et~al., 2005 \emph{\aj}, \emph{130}, 5, 2185.

\bibitem[\protect\astroncite{\emph{{Lang} et~al.}}{2010}]{Lang2010}
{Lang} C.~C. et~al., 2010 \emph{\apjs}, \emph{191}, 2, 275.

\bibitem[\protect\astroncite{\emph{{Langer} et~al.}}{2015}]{Langer2015}
{Langer} W.~D. et~al., 2015 \emph{\aap}, \emph{576}, A1.

\bibitem[\protect\astroncite{\emph{{LaRosa} et~al.}}{2000}]{LaRosa2000}
{LaRosa} T.~N. et~al., 2000 \emph{\aj}, \emph{119}, 1, 207.

\bibitem[\protect\astroncite{\emph{{Lau} et~al.}}{2013}]{Lau2013}
{Lau} R.~M. et~al., 2013 \emph{\apj}, \emph{775}, 1, 37.

\bibitem[\protect\astroncite{\emph{{Lau} et~al.}}{2016}]{Lau2016}
{Lau} R.~M. et~al., 2016 \emph{\apj}, \emph{818}, 2, 117.

\bibitem[\protect\astroncite{\emph{{Launhardt} et~al.}}{2002}]{Launhardt2002}
{Launhardt} R. et~al., 2002 \emph{\aap}, \emph{384}, 112.

\bibitem[\protect\astroncite{\emph{{Law}}}{2010}]{Law2010}
{Law} C.~J., 2010 \emph{\apj}, \emph{708}, 1, 474.

\bibitem[\protect\astroncite{\emph{{Law} et~al.}}{2008}]{Law2008}
{Law} C.~J. et~al., 2008 \emph{\apjs}, \emph{177}, 1, 255.

\bibitem[\protect\astroncite{\emph{{Law} et~al.}}{2009}]{Law2009}
{Law} C.~J. et~al., 2009 \emph{\apj}, \emph{695}, 2, 1070.

\bibitem[\protect\astroncite{\emph{{Le Petit} et~al.}}{2016}]{LePetit2016}
{Le Petit} F. et~al., 2016 \emph{\aap}, \emph{585}, A105.

\bibitem[\protect\astroncite{\emph{{Lee} et~al.}}{2012}]{Lee2012}
{Lee} E.~J. et~al., 2012 \emph{\apj}, \emph{752}, 2, 146.

\bibitem[\protect\astroncite{\emph{{Lee} et~al.}}{2022}]{Lee2022}
{Lee} J.~C. et~al., 2022 \emph{\apjs}, \emph{258}, 1, 10.

\bibitem[\protect\astroncite{\emph{{Lee} et~al.}}{2008}]{Lee2008}
{Lee} S. et~al., 2008 \emph{\apj}, \emph{674}, 1, 247.

\bibitem[\protect\astroncite{\emph{{Leitherer} et~al.}}{2014}]{Leitherer2014}
{Leitherer} C. et~al., 2014 \emph{\apjs}, \emph{212}, 1, 14.

\bibitem[\protect\astroncite{\emph{{Leroy} et~al.}}{2021}]{Leroy2021}
{Leroy} A.~K. et~al., 2021 \emph{\apjs}, \emph{257}, 2, 43.

\bibitem[\protect\astroncite{\emph{{Lesch} et~al.}}{1990}]{Lesch1990}
{Lesch} H. et~al., 1990 \emph{\mnras}, \emph{242}, 194.

\bibitem[\protect\astroncite{\emph{{Leslie} et~al.}}{2020}]{Leslie2020}
{Leslie} S.~K. et~al., 2020 \emph{\apj}, \emph{899}, 1, 58.

\bibitem[\protect\astroncite{\emph{{Li} et~al.}}{2015}]{Li2015}
{Li} Z. et~al., 2015 \emph{\apj}, \emph{806}, 2, 150.

\bibitem[\protect\astroncite{\emph{{Li} et~al.}}{2020}]{Li2020b}
{Li} Z. et~al., 2020 \emph{\apj}, \emph{889}, 2, 88.

\bibitem[\protect\astroncite{\emph{{Liermann} et~al.}}{2012}]{Liermann2012}
{Liermann} A. et~al., 2012 \emph{\aap}, \emph{540}, A14.

\bibitem[\protect\astroncite{\emph{{Lim} et~al.}}{2013}]{Lim2013}
{Lim} B. et~al., 2013 \emph{\aj}, \emph{145}, 2, 46.

\bibitem[\protect\astroncite{\emph{{Lis} and {Menten}}}{1998}]{Lis1998}
{Lis} D.~C. and {Menten} K.~M., 1998 \emph{\apj}, \emph{507}, 2, 794.

\bibitem[\protect\astroncite{\emph{{Lis} et~al.}}{1993}]{Lis1993}
{Lis} D.~C. et~al., 1993 \emph{\apj}, \emph{402}, 238.

\bibitem[\protect\astroncite{\emph{{Lis} et~al.}}{1994}]{Lis1994b}
{Lis} D.~C. et~al., 1994 \emph{\apjl}, \emph{423}, L39.

\bibitem[\protect\astroncite{\emph{{Lis} et~al.}}{2001}]{Lis2001}
{Lis} D.~C. et~al., 2001 \emph{\apj}, \emph{550}, 2, 761.

\bibitem[\protect\astroncite{\emph{{Liszt}}}{1992}]{Liszt1992}
{Liszt} H.~S., 1992 \emph{\apjs}, \emph{82}, 495.

\bibitem[\protect\astroncite{\emph{{Liszt}}}{2006}]{Liszt2006}
{Liszt} H.~S., 2006 \emph{\aap}, \emph{447}, 2, 533.

\bibitem[\protect\astroncite{\emph{{Liszt}}}{2008}]{Liszt2008}
{Liszt} H.~S., 2008 \emph{\aap}, \emph{486}, 2, 467.

\bibitem[\protect\astroncite{\emph{{Liszt} and {Burton}}}{1978}]{Liszt1978}
{Liszt} H.~S. and {Burton} W.~B., 1978 \emph{\apj}, \emph{226}, 790.

\bibitem[\protect\astroncite{\emph{{Liszt} and {Spiker}}}{1995}]{Liszt1995}
{Liszt} H.~S. and {Spiker} R.~W., 1995 \emph{\apjs}, \emph{98}, 259.

\bibitem[\protect\astroncite{\emph{{Lo} and {Claussen}}}{1983}]{Lo1983}
{Lo} K.~Y. and {Claussen} M.~J., 1983 \emph{\nat}, \emph{306}, 5944, 647.

\bibitem[\protect\astroncite{\emph{{Lohr} et~al.}}{2018}]{Lohr2018}
{Lohr} M.~E. et~al., 2018 \emph{\aap}, \emph{617}, A66.

\bibitem[\protect\astroncite{\emph{{Longmore}
  et~al.}}{2013{\natexlab{a}}}]{Longmore2013a}
{Longmore} S.~N. et~al., 2013{\natexlab{a}} \emph{\mnras}, \emph{433}, L15.

\bibitem[\protect\astroncite{\emph{{Longmore}
  et~al.}}{2013{\natexlab{b}}}]{Longmore2013b}
{Longmore} S.~N. et~al., 2013{\natexlab{b}} \emph{\mnras}, \emph{429}, 2, 987.

\bibitem[\protect\astroncite{\emph{{Longmore} et~al.}}{2014}]{Longmore2014}
{Longmore} S.~N. et~al., 2014 \emph{Protostars and Planets VI} (H.~{Beuther},
  R.~S. {Klessen}, C.~P. {Dullemond}, and T.~{Henning}), p. 291.

\bibitem[\protect\astroncite{\emph{{Longmore} et~al.}}{2017}]{Longmore2017}
{Longmore} S.~N. et~al., 2017 \emph{\mnras}, \emph{470}, 2, 1462.

\bibitem[\protect\astroncite{\emph{{Loose} et~al.}}{1982}]{Loose1982}
{Loose} H.~H. et~al., 1982 \emph{\aap}, \emph{105}, 2, 342.

\bibitem[\protect\astroncite{\emph{{Lu} et~al.}}{2013}]{Lu2013}
{Lu} J.~R. et~al., 2013 \emph{\apj}, \emph{764}, 2, 155.

\bibitem[\protect\astroncite{\emph{{Lu} et~al.}}{2015}]{Lu2015}
{Lu} X. et~al., 2015 \emph{\apjl}, \emph{814}, 2, L18.

\bibitem[\protect\astroncite{\emph{{Lu} et~al.}}{2019{\natexlab{a}}}]{Lu2019a}
{Lu} X. et~al., 2019{\natexlab{a}} \emph{\apjs}, \emph{244}, 2, 35.

\bibitem[\protect\astroncite{\emph{{Lu} et~al.}}{2019{\natexlab{b}}}]{Lu2019b}
{Lu} X. et~al., 2019{\natexlab{b}} \emph{\apj}, \emph{872}, 2, 171.

\bibitem[\protect\astroncite{\emph{{Lu} et~al.}}{2020}]{Lu2020}
{Lu} X. et~al., 2020 \emph{\apjl}, \emph{894}, 2, L14.

\bibitem[\protect\astroncite{\emph{{Lu} et~al.}}{2021}]{Lu2021}
{Lu} X. et~al., 2021 \emph{\apj}, \emph{909}, 2, 177.

\bibitem[\protect\astroncite{\emph{{Lynden-Bell} and
  {Pringle}}}{1974}]{Lynden-Bell1974}
{Lynden-Bell} D. and {Pringle} J.~E., 1974 \emph{\mnras}, \emph{168}, 603.

\bibitem[\protect\astroncite{\emph{{Mac Low} and {Klessen}}}{2004}]{MacLow2004}
{Mac Low} M.-M. and {Klessen} R.~S., 2004 \emph{Reviews of Modern Physics},
  \emph{76}, 1, 125.

\bibitem[\protect\astroncite{\emph{{Machida} et~al.}}{2009}]{Machida2009}
{Machida} M. et~al., 2009 \emph{\pasj}, \emph{61}, 411.

\bibitem[\protect\astroncite{\emph{{Maiolino} et~al.}}{2008}]{Maiolino2008}
{Maiolino} R. et~al., 2008 \emph{\aap}, \emph{488}, 2, 463.

\bibitem[\protect\astroncite{\emph{{Mangilli} et~al.}}{2019}]{Mangilli2019}
{Mangilli} A. et~al., 2019 \emph{\aap}, \emph{630}, A74.

\bibitem[\protect\astroncite{\emph{{Mangum} et~al.}}{2013}]{Mangum2013}
{Mangum} J.~G. et~al., 2013 \emph{\apj}, \emph{779}, 1, 33.

\bibitem[\protect\astroncite{\emph{{Mangum} et~al.}}{2019}]{Mangum2019}
{Mangum} J.~G. et~al., 2019 \emph{\apj}, \emph{871}, 2, 170.

\bibitem[\protect\astroncite{\emph{{Mapelli} and {Trani}}}{2016}]{Mapelli2016}
{Mapelli} M. and {Trani} A.~A., 2016 \emph{\aap}, \emph{585}, A161.

\bibitem[\protect\astroncite{\emph{{Marsh} et~al.}}{2016}]{Marsh2016}
{Marsh} K.~A. et~al., 2016 \emph{\mnras}, \emph{461}, 1, L16.

\bibitem[\protect\astroncite{\emph{{Marshall} et~al.}}{2008}]{Marshall2008}
{Marshall} D.~J. et~al., 2008 \emph{\aap}, \emph{477}, 2, L21.

\bibitem[\protect\astroncite{\emph{{Mart{\'\i}n} et~al.}}{2012}]{Martin2012}
{Mart{\'\i}n} S. et~al., 2012 \emph{\aap}, \emph{539}, A29.

\bibitem[\protect\astroncite{\emph{{Mart{\'\i}n-Pintado}
  et~al.}}{1997}]{Martin-Pintado1997}
{Mart{\'\i}n-Pintado} J. et~al., 1997 \emph{\apjl}, \emph{482}, 1, L45.

\bibitem[\protect\astroncite{\emph{{Martini}
  et~al.}}{2003{\natexlab{a}}}]{Martini2003a}
{Martini} P. et~al., 2003{\natexlab{a}} \emph{\apjs}, \emph{146}, 2, 353.

\bibitem[\protect\astroncite{\emph{{Martini}
  et~al.}}{2003{\natexlab{b}}}]{Martini2003b}
{Martini} P. et~al., 2003{\natexlab{b}} \emph{\apj}, \emph{589}, 2, 774.

\bibitem[\protect\astroncite{\emph{{Matsunaga} et~al.}}{2011}]{Matsunaga2011}
{Matsunaga} N. et~al., 2011 \emph{\nat}, \emph{477}, 7363, 188.

\bibitem[\protect\astroncite{\emph{{Mazzuca} et~al.}}{2008}]{Mazzuca2008}
{Mazzuca} L.~M. et~al., 2008 \emph{\apjs}, \emph{174}, 2, 337.

\bibitem[\protect\astroncite{\emph{{McClure-Griffiths}
  et~al.}}{2012}]{McClure-Griffiths2012}
{McClure-Griffiths} N.~M. et~al., 2012 \emph{\apjs}, \emph{199}, 1, 12.

\bibitem[\protect\astroncite{\emph{{McGary} et~al.}}{2001}]{McGary2001}
{McGary} R.~S. et~al., 2001 \emph{\apj}, \emph{559}, 1, 326.

\bibitem[\protect\astroncite{\emph{{McKee} et~al.}}{2015}]{McKee2015}
{McKee} C.~F. et~al., 2015 \emph{\apj}, \emph{814}, 1, 13.

\bibitem[\protect\astroncite{\emph{{Mehringer} et~al.}}{1992}]{Mehringer1992}
{Mehringer} D.~M. et~al., 1992 \emph{\apj}, \emph{401}, 168.

\bibitem[\protect\astroncite{\emph{{Mehringer} et~al.}}{1993}]{Mehringer1993}
{Mehringer} D.~M. et~al., 1993 \emph{\apjl}, \emph{402}, L69.

\bibitem[\protect\astroncite{\emph{{Mei} et~al.}}{2020}]{Mei2020}
{Mei} Y. et~al., 2020 \emph{\apj}, \emph{898}, 2, 157.

\bibitem[\protect\astroncite{\emph{{Meier} and {Turner}}}{2005}]{Meier2005}
{Meier} D.~S. and {Turner} J.~L., 2005 \emph{\apj}, \emph{618}, 1, 259.

\bibitem[\protect\astroncite{\emph{{Meijerink} et~al.}}{2011}]{Meijerink2011}
{Meijerink} R. et~al., 2011 \emph{\aap}, \emph{525}, A119.

\bibitem[\protect\astroncite{\emph{{Meng} et~al.}}{2019}]{Meng2019}
{Meng} F. et~al., 2019 \emph{\aap}, \emph{630}, A73.

\bibitem[\protect\astroncite{\emph{{Mihos} and {Hernquist}}}{1994}]{Mihos1994}
{Mihos} J.~C. and {Hernquist} L., 1994 \emph{\apjl}, \emph{425}, L13.

\bibitem[\protect\astroncite{\emph{{Miller} and {Bregman}}}{2016}]{Miller2016}
{Miller} M.~J. and {Bregman} J.~N., 2016 \emph{\apj}, \emph{829}, 1, 9.

\bibitem[\protect\astroncite{\emph{{Mills} et~al.}}{2011}]{Mills2011}
{Mills} E. et~al., 2011 \emph{\apj}, \emph{735}, 2, 84.

\bibitem[\protect\astroncite{\emph{{Mills} and {Battersby}}}{2017}]{Mills2017b}
{Mills} E. A.~C. and {Battersby} C., 2017 \emph{\apj}, \emph{835}, 1, 76.

\bibitem[\protect\astroncite{\emph{{Mills} et~al.}}{2013}]{Mills2013}
{Mills} E.~A.~C. et~al., 2013 \emph{\apj}, \emph{779}, 1, 47.

\bibitem[\protect\astroncite{\emph{{Mills} et~al.}}{2017}]{Mills2017a}
{Mills} E. A.~C. et~al., 2017 \emph{\apj}, \emph{850}, 2, 192.

\bibitem[\protect\astroncite{\emph{{Mills}
  et~al.}}{2018{\natexlab{a}}}]{Mills2018a}
{Mills} E.~A.~C. et~al., 2018{\natexlab{a}} \emph{\apj}, \emph{869}, 2, 121.

\bibitem[\protect\astroncite{\emph{{Mills}
  et~al.}}{2018{\natexlab{b}}}]{Mills2018c}
{Mills} E.~A.~C. et~al., 2018{\natexlab{b}} \emph{\apj}, \emph{868}, 1, 7.

\bibitem[\protect\astroncite{\emph{{Miyawaki} et~al.}}{2021}]{Miyawaki2021}
{Miyawaki} R. et~al., 2021 \emph{\pasj}, \emph{73}, 4, 943.

\bibitem[\protect\astroncite{\emph{{Molinari} et~al.}}{2010}]{Molinari2010}
{Molinari} S. et~al., 2010 \emph{\pasp}, \emph{122}, 889, 314.

\bibitem[\protect\astroncite{\emph{{Molinari} et~al.}}{2011}]{Molinari2011}
{Molinari} S. et~al., 2011 \emph{\apjl}, \emph{735}, 2, L33.

\bibitem[\protect\astroncite{\emph{{M{\"o}ller} et~al.}}{2021}]{Moller2021}
{M{\"o}ller} T. et~al., 2021 \emph{\aap}, \emph{651}, A9.

\bibitem[\protect\astroncite{\emph{{Montenegro} et~al.}}{1999}]{Montenegro1999}
{Montenegro} L.~E. et~al., 1999 \emph{\apj}, \emph{520}, 2, 592.

\bibitem[\protect\astroncite{\emph{{Montero-Casta{\~n}o}
  et~al.}}{2009}]{Montero-Castano2009}
{Montero-Casta{\~n}o} M. et~al., 2009 \emph{\apj}, \emph{695}, 2, 1477.

\bibitem[\protect\astroncite{\emph{{Moon} et~al.}}{2021}]{Moon2021}
{Moon} S. et~al., 2021 \emph{\apj}, \emph{914}, 1, 9.

\bibitem[\protect\astroncite{\emph{{Moon} et~al.}}{2022}]{Moon2022}
{Moon} S. et~al., 2022 \emph{\apj}, \emph{925}, 1, 99.

\bibitem[\protect\astroncite{\emph{{Morris} and {Serabyn}}}{1996}]{Morris1996}
{Morris} M. and {Serabyn} E., 1996 \emph{\araa}, \emph{34}, 645.

\bibitem[\protect\astroncite{\emph{{Morris}}}{2015}]{Morris2015}
{Morris} M.~R., 2015 \emph{Lessons from the Local Group: A Conference in honor
  of David Block and Bruce Elmegreen}, p. 391.

\bibitem[\protect\astroncite{\emph{{Murray} and {Rahman}}}{2010}]{Murray2010}
{Murray} N. and {Rahman} M., 2010 \emph{\apj}, \emph{709}, 1, 424.

\bibitem[\protect\astroncite{\emph{{Myers} et~al.}}{2022}]{Myers2022}
{Myers} P.~C. et~al., 2022 \emph{arXiv e-prints}, arXiv:2202.13987.

\bibitem[\protect\astroncite{\emph{{Nandakumar} et~al.}}{2018}]{Nandakumar2018}
{Nandakumar} G. et~al., 2018 \emph{\aap}, \emph{609}, A109.

\bibitem[\protect\astroncite{\emph{{Neufeld} and
  {Wolfire}}}{2017}]{Neufeld2017}
{Neufeld} D.~A. and {Wolfire} M.~G., 2017 \emph{\apj}, \emph{845}, 2, 163.

\bibitem[\protect\astroncite{\emph{{Neumayer} et~al.}}{2020}]{Neumayer2020}
{Neumayer} N. et~al., 2020 \emph{\aapr}, \emph{28}, 1, 4.

\bibitem[\protect\astroncite{\emph{{Nguyen} et~al.}}{2021}]{Nguyen2021}
{Nguyen} H. et~al., 2021 \emph{\aap}, \emph{651}, A88.

\bibitem[\protect\astroncite{\emph{{Nishiyama} et~al.}}{2006}]{Nishiyama2006}
{Nishiyama} S. et~al., 2006 \emph{\apj}, \emph{638}, 2, 839.

\bibitem[\protect\astroncite{\emph{{Nishiyama} et~al.}}{2013}]{Nishiyama2013}
{Nishiyama} S. et~al., 2013 \emph{\apjl}, \emph{769}, 2, L28.

\bibitem[\protect\astroncite{\emph{{Nitschai} et~al.}}{2020}]{Nitschai2020a}
{Nitschai} M.~S. et~al., 2020 \emph{\apj}, \emph{896}, 1, 68.

\bibitem[\protect\astroncite{\emph{{Nogueras-Lara}
  et~al.}}{2020}]{Nogueras-Lara2020b}
{Nogueras-Lara} F. et~al., 2020 \emph{Nature Astronomy}, \emph{4}, 377.

\bibitem[\protect\astroncite{\emph{{Nogueras-Lara}
  et~al.}}{2021}]{Nogueras-Lara2021}
{Nogueras-Lara} F. et~al., 2021 \emph{\aap}, \emph{647}, L6.

\bibitem[\protect\astroncite{\emph{{Nordlund} and
  {Padoan}}}{1999}]{Nordlund1999}
{Nordlund} {\r{A}}.~K. and {Padoan} P., 1999 \emph{Interstellar Turbulence}
  (J.~{Franco} and A.~{Carraminana}), p. 218.

\bibitem[\protect\astroncite{\emph{{Offner} et~al.}}{2014}]{Offner2014}
{Offner} S.~S.~R. et~al., 2014 \emph{Protostars and Planets VI} (H.~{Beuther},
  R.~S. {Klessen}, C.~P. {Dullemond}, and T.~{Henning}), p.~53.

\bibitem[\protect\astroncite{\emph{{Oka} and {Geballe}}}{2020}]{Oka2020}
{Oka} T. and {Geballe} T.~R., 2020 \emph{\apj}, \emph{902}, 1, 9.

\bibitem[\protect\astroncite{\emph{{Oka} et~al.}}{1998}]{Oka1998b}
{Oka} T. et~al., 1998 \emph{\apj}, \emph{493}, 2, 730.

\bibitem[\protect\astroncite{\emph{{Oka} et~al.}}{1999}]{Oka1999}
{Oka} T. et~al., 1999 \emph{\apj}, \emph{515}, 1, 249.

\bibitem[\protect\astroncite{\emph{{Oka} et~al.}}{2007}]{Oka2007}
{Oka} T. et~al., 2007 \emph{\pasj}, \emph{59}, 15.

\bibitem[\protect\astroncite{\emph{{Oka} et~al.}}{2011}]{Oka2011}
{Oka} T. et~al., 2011 \emph{\apj}, \emph{732}, 2, 120.

\bibitem[\protect\astroncite{\emph{{Oka} et~al.}}{2016}]{Oka2016}
{Oka} T. et~al., 2016 \emph{\apjl}, \emph{816}, 1, L7.

\bibitem[\protect\astroncite{\emph{{Oka} et~al.}}{2017}]{Oka2017}
{Oka} T. et~al., 2017 \emph{Nature Astronomy}, \emph{1}, 709.

\bibitem[\protect\astroncite{\emph{{Oka} et~al.}}{2019}]{Oka2019}
{Oka} T. et~al., 2019 \emph{\apj}, \emph{883}, 1, 54.

\bibitem[\protect\astroncite{\emph{{Oort}}}{1977}]{Oort1977}
{Oort} J.~H., 1977 \emph{\araa}, \emph{15}, 295.

\bibitem[\protect\astroncite{\emph{{Oort} et~al.}}{1958}]{Oort1958}
{Oort} J.~H. et~al., 1958 \emph{\mnras}, \emph{118}, 379.

\bibitem[\protect\astroncite{\emph{{Orr} et~al.}}{2021}]{Orr2021}
{Orr} M.~E. et~al., 2021 \emph{\apjl}, \emph{908}, 2, L31.

\bibitem[\protect\astroncite{\emph{{Padoan} and {Nordlund}}}{2011}]{Padoan2011}
{Padoan} P. and {Nordlund} {\r{A}}., 2011 \emph{\apj}, \emph{730}, 1, 40.

\bibitem[\protect\astroncite{\emph{{Padoan} et~al.}}{2014}]{Padoan2014}
{Padoan} P. et~al., 2014 \emph{Protostars and Planets VI} (H.~{Beuther}, R.~S.
  {Klessen}, C.~P. {Dullemond}, and T.~{Henning}), p.~77.

\bibitem[\protect\astroncite{\emph{{Padovani} et~al.}}{2019}]{Padovani2019}
{Padovani} M. et~al., 2019 \emph{\aap}, \emph{630}, A72.

\bibitem[\protect\astroncite{\emph{{Padovani} et~al.}}{2020}]{Padovani2020}
{Padovani} M. et~al., 2020 \emph{\ssr}, \emph{216}, 2, 29.

\bibitem[\protect\astroncite{\emph{{Pan} et~al.}}{2013}]{Pan2013}
{Pan} H.-A. et~al., 2013 \emph{\apj}, \emph{768}, 1, 57.

\bibitem[\protect\astroncite{\emph{{Pang} et~al.}}{2013}]{Pang2013}
{Pang} X. et~al., 2013 \emph{\apj}, \emph{764}, 1, 73.

\bibitem[\protect\astroncite{\emph{{Parker}}}{1966}]{Parker1966}
{Parker} E.~N., 1966 \emph{\apj}, \emph{145}, 811.

\bibitem[\protect\astroncite{\emph{{Parmentier} and
  {Pasquali}}}{2020}]{Parmentier2020}
{Parmentier} G. and {Pasquali} A., 2020 \emph{\apj}, \emph{903}, 1, 56.

\bibitem[\protect\astroncite{\emph{{Parsons} et~al.}}{2018}]{Parsons2018}
{Parsons} H. et~al., 2018 \emph{\apjs}, \emph{234}, 2, 22.

\bibitem[\protect\astroncite{\emph{{Patsis} and
  {Athanassoula}}}{2000}]{Patsis2000}
{Patsis} P.~A. and {Athanassoula} E., 2000 \emph{\aap}, \emph{358}, 45.

\bibitem[\protect\astroncite{\emph{{Peeples} and
  {Martini}}}{2006}]{Peeples2006}
{Peeples} M.~S. and {Martini} P., 2006 \emph{\apj}, \emph{652}, 2, 1097.

\bibitem[\protect\astroncite{\emph{{Peretto} et~al.}}{2010}]{Peretto2010}
{Peretto} N. et~al., 2010 \emph{\aap}, \emph{518}, L98.

\bibitem[\protect\astroncite{\emph{{Pessa} et~al.}}{2021}]{Pessa2021}
{Pessa} I. et~al., 2021 \emph{\aap}, \emph{650}, A134.

\bibitem[\protect\astroncite{\emph{{Petkova} et~al.}}{2021}]{Petkova2021}
{Petkova} M.~A. et~al., 2021 \emph{arXiv e-prints}, arXiv:2104.09558.

\bibitem[\protect\astroncite{\emph{{Pillai} et~al.}}{2015}]{Pillai2015}
{Pillai} T. et~al., 2015 \emph{\apj}, \emph{799}, 1, 74.

\bibitem[\protect\astroncite{\emph{{Ponti} et~al.}}{2010}]{Ponti2010}
{Ponti} G. et~al., 2010 \emph{\apj}, \emph{714}, 1, 732.

\bibitem[\protect\astroncite{\emph{{Ponti} et~al.}}{2013}]{Ponti2013}
{Ponti} G. et~al., 2013 \emph{Cosmic Rays in Star-Forming Environments},
  vol.~34 of \emph{Astrophysics and Space Science Proceedings} (D.~F. {Torres}
  and O.~{Reimer}), p. 331.

\bibitem[\protect\astroncite{\emph{{Ponti} et~al.}}{2015}]{Ponti2015}
{Ponti} G. et~al., 2015 \emph{\mnras}, \emph{453}, 1, 172.

\bibitem[\protect\astroncite{\emph{{Ponti} et~al.}}{2019}]{Ponti2019}
{Ponti} G. et~al., 2019 \emph{\nat}, \emph{567}, 7748, 347.

\bibitem[\protect\astroncite{\emph{{Ponti} et~al.}}{2021}]{Ponti2021}
{Ponti} G. et~al., 2021 \emph{\aap}, \emph{646}, A66.

\bibitem[\protect\astroncite{\emph{{Portail} et~al.}}{2017}]{Portail2017}
{Portail} M. et~al., 2017 \emph{\mnras}, \emph{465}, 2, 1621.

\bibitem[\protect\astroncite{\emph{{Pound} and
  {Yusef-Zadeh}}}{2018}]{Pound2018}
{Pound} M.~W. and {Yusef-Zadeh} F., 2018 \emph{\mnras}, \emph{473}, 3, 2899.

\bibitem[\protect\astroncite{\emph{{Predehl} et~al.}}{2020}]{Predehl2020}
{Predehl} P. et~al., 2020 \emph{\nat}, \emph{588}, 7837, 227.

\bibitem[\protect\astroncite{\emph{{Prendergast}}}{1983}]{Prendergast1983}
{Prendergast} K.~H., 1983 \emph{Internal Kinematics and Dynamics of Galaxies},
  vol. 100 (E.~{Athanassoula}), pp. 215--220.

\bibitem[\protect\astroncite{\emph{{Priestley} and
  {Whitworth}}}{2021}]{Priestley2021}
{Priestley} F.~D. and {Whitworth} A.~P., 2021 \emph{\mnras}, \emph{506}, 1,
  775.

\bibitem[\protect\astroncite{\emph{{Prieto} et~al.}}{2019}]{Prieto2019}
{Prieto} M.~A. et~al., 2019 \emph{\mnras}, \emph{485}, 3, 3264.

\bibitem[\protect\astroncite{\emph{{Purcell} et~al.}}{2012}]{Purcell2012}
{Purcell} C.~R. et~al., 2012 \emph{\mnras}, \emph{426}, 3, 1972.

\bibitem[\protect\astroncite{\emph{{Qin} et~al.}}{2008}]{Qin2008}
{Qin} S.-L. et~al., 2008 \emph{\apj}, \emph{677}, 1, 353.

\bibitem[\protect\astroncite{\emph{{Querejeta} et~al.}}{2015}]{Querejeta2015}
{Querejeta} M. et~al., 2015 \emph{\apjs}, \emph{219}, 1, 5.

\bibitem[\protect\astroncite{\emph{{Querejeta} et~al.}}{2019}]{Querejeta2019}
{Querejeta} M. et~al., 2019 \emph{\aap}, \emph{625}, A19.

\bibitem[\protect\astroncite{\emph{{Ram{\'\i}rez} et~al.}}{2008}]{Ramirez2008}
{Ram{\'\i}rez} S.~V. et~al., 2008 \emph{\apjs}, \emph{175}, 1, 147.

\bibitem[\protect\astroncite{\emph{{Rathborne}
  et~al.}}{2014{\natexlab{a}}}]{Rathborne2014b}
{Rathborne} J.~M. et~al., 2014{\natexlab{a}} \emph{\apj}, \emph{786}, 2, 140.

\bibitem[\protect\astroncite{\emph{{Rathborne}
  et~al.}}{2014{\natexlab{b}}}]{Rathborne2014a}
{Rathborne} J.~M. et~al., 2014{\natexlab{b}} \emph{\apjl}, \emph{795}, 2, L25.

\bibitem[\protect\astroncite{\emph{{Rathborne} et~al.}}{2015}]{Rathborne2015}
{Rathborne} J.~M. et~al., 2015 \emph{\apj}, \emph{802}, 2, 125.

\bibitem[\protect\astroncite{\emph{{Regan} and {Teuben}}}{2003}]{Regan2003}
{Regan} M.~W. and {Teuben} P., 2003 \emph{\apj}, \emph{582}, 2, 723.

\bibitem[\protect\astroncite{\emph{{Reid} et~al.}}{2009}]{Reid2009}
{Reid} M.~J. et~al., 2009 \emph{\apj}, \emph{705}, 2, 1548.

\bibitem[\protect\astroncite{\emph{{Requena-Torres}
  et~al.}}{2012}]{Requena-Torres2012}
{Requena-Torres} M.~A. et~al., 2012 \emph{\aap}, \emph{542}, L21.

\bibitem[\protect\astroncite{\emph{{Revnivtsev} et~al.}}{2009}]{Revnivtsev2009}
{Revnivtsev} M. et~al., 2009 \emph{\nat}, \emph{458}, 7242, 1142.

\bibitem[\protect\astroncite{\emph{{Rickert} et~al.}}{2019}]{Rickert2019}
{Rickert} M. et~al., 2019 \emph{\mnras}, \emph{482}, 4, 5349.

\bibitem[\protect\astroncite{\emph{{Ridley} et~al.}}{2017}]{Ridley2017}
{Ridley} M. G.~L. et~al., 2017 \emph{\mnras}, \emph{469}, 2, 2251.

\bibitem[\protect\astroncite{\emph{{Rieke} et~al.}}{2009}]{Rieke2009}
{Rieke} G.~H. et~al., 2009 \emph{\apj}, \emph{692}, 1, 556.

\bibitem[\protect\astroncite{\emph{{Riquelme} et~al.}}{2018}]{Riquelme2018}
{Riquelme} D. et~al., 2018 \emph{\aap}, \emph{613}, A42.

\bibitem[\protect\astroncite{\emph{{Rivilla} et~al.}}{2020}]{Rivilla2020}
{Rivilla} V.~M. et~al., 2020 \emph{\apjl}, \emph{899}, 2, L28.

\bibitem[\protect\astroncite{\emph{{Rivilla}
  et~al.}}{2021{\natexlab{a}}}]{Rivilla2021a}
{Rivilla} V.~M. et~al., 2021{\natexlab{a}} \emph{\mnras}, \emph{506}, 1, L79.

\bibitem[\protect\astroncite{\emph{{Rivilla}
  et~al.}}{2021{\natexlab{b}}}]{Rivilla2021b}
{Rivilla} V.~M. et~al., 2021{\natexlab{b}} \emph{Proceedings of the National
  Academy of Science}, \emph{118}, 22, 2101314118.

\bibitem[\protect\astroncite{\emph{{Robitaille} et~al.}}{2006}]{Robitaille2006}
{Robitaille} T.~P. et~al., 2006 \emph{\apjs}, \emph{167}, 2, 256.

\bibitem[\protect\astroncite{\emph{{Robitaille} et~al.}}{2007}]{Robitaille2007}
{Robitaille} T.~P. et~al., 2007 \emph{\apjs}, \emph{169}, 2, 328.

\bibitem[\protect\astroncite{\emph{{Rodriguez-Fernandez} and
  {Combes}}}{2008}]{Rodriguez-Fernandez2008}
{Rodriguez-Fernandez} N.~J. and {Combes} F., 2008 \emph{\aap}, \emph{489}, 1,
  115.

\bibitem[\protect\astroncite{\emph{{Rodriguez-Fernandez}
  et~al.}}{2006}]{Rodriguez-Fernandez2006}
{Rodriguez-Fernandez} N.~J. et~al., 2006 \emph{\aap}, \emph{455}, 3, 963.

\bibitem[\protect\astroncite{\emph{{Roman-Duval}
  et~al.}}{2016}]{Roman-Duval2016}
{Roman-Duval} J. et~al., 2016 \emph{\apj}, \emph{818}, 2, 144.

\bibitem[\protect\astroncite{\emph{{Rosolowsky} et~al.}}{2010}]{Rosolowsky2010}
{Rosolowsky} E. et~al., 2010 \emph{\apjs}, \emph{188}, 1, 123.

\bibitem[\protect\astroncite{\emph{{Rougoor} and {Oort}}}{1960}]{Rougoor1960}
{Rougoor} G.~W. and {Oort} J.~H., 1960 \emph{Proceedings of the National
  Academy of Science}, \emph{46}, 1, 1.

\bibitem[\protect\astroncite{\emph{{Rudolph} et~al.}}{2006}]{Rudolph2006}
{Rudolph} A.~L. et~al., 2006 \emph{\apjs}, \emph{162}, 2, 346.

\bibitem[\protect\astroncite{\emph{{Rui} et~al.}}{2019}]{Rui2019}
{Rui} N.~Z. et~al., 2019 \emph{\apj}, \emph{877}, 1, 37.

\bibitem[\protect\astroncite{\emph{{Ryder} et~al.}}{2001}]{Ryder2001}
{Ryder} S.~D. et~al., 2001 \emph{\mnras}, \emph{323}, 3, 663.

\bibitem[\protect\astroncite{\emph{{Salas} et~al.}}{2020}]{Salas2020}
{Salas} J.~M. et~al., 2020 \emph{arXiv e-prints}, arXiv:2010.04170.

\bibitem[\protect\astroncite{\emph{{S{\'a}nchez-Monge}
  et~al.}}{2017}]{Sanchez-Monge2017}
{S{\'a}nchez-Monge} {\'A}. et~al., 2017 \emph{\aap}, \emph{604}, A6.

\bibitem[\protect\astroncite{\emph{{Sandage}}}{1961}]{Sandage1961}
{Sandage} A., 1961 \emph{{The Hubble Atlas of Galaxies}}.

\bibitem[\protect\astroncite{\emph{{Santa-Maria}
  et~al.}}{2021}]{Santa-Maria2021}
{Santa-Maria} M.~G. et~al., 2021 \emph{\aap}, \emph{649}, A32.

\bibitem[\protect\astroncite{\emph{{Sarzi} et~al.}}{2007}]{Sarzi2007}
{Sarzi} M. et~al., 2007 \emph{\mnras}, \emph{380}, 3, 949.

\bibitem[\protect\astroncite{\emph{{Sato} et~al.}}{2000}]{Sato2000}
{Sato} F. et~al., 2000 \emph{\apj}, \emph{535}, 2, 857.

\bibitem[\protect\astroncite{\emph{{Sawada} et~al.}}{2004}]{Sawada2004}
{Sawada} T. et~al., 2004 \emph{\mnras}, \emph{349}, 4, 1167.

\bibitem[\protect\astroncite{\emph{{Schilke} et~al.}}{1997}]{Schilke1997}
{Schilke} P. et~al., 1997 \emph{\aap}, \emph{321}, 293.

\bibitem[\protect\astroncite{\emph{{Schinnerer} et~al.}}{2002}]{Schinnerer2002}
{Schinnerer} E. et~al., 2002 \emph{\apj}, \emph{575}, 2, 826.

\bibitem[\protect\astroncite{\emph{{Schmidt}}}{1959}]{Schmidt1959}
{Schmidt} M., 1959 \emph{\apj}, \emph{129}, 243.

\bibitem[\protect\astroncite{\emph{{Schmiedeke} et~al.}}{2016}]{Schmiedeke2016}
{Schmiedeke} A. et~al., 2016 \emph{\aap}, \emph{588}, A143.

\bibitem[\protect\astroncite{\emph{{Schneider} et~al.}}{2014}]{Schneider2014}
{Schneider} F.~R.~N. et~al., 2014 \emph{\apj}, \emph{780}, 2, 117.

\bibitem[\protect\astroncite{\emph{{Sch{\"o}del} et~al.}}{2014}]{Schodel2014b}
{Sch{\"o}del} R. et~al., 2014 \emph{\aap}, \emph{566}, A47.

\bibitem[\protect\astroncite{\emph{{Sch{\"o}nrich}
  et~al.}}{2015}]{Schonrich2015}
{Sch{\"o}nrich} R. et~al., 2015 \emph{\apjl}, \emph{812}, 2, L21.

\bibitem[\protect\astroncite{\emph{{Schruba} et~al.}}{2019}]{Schruba2019}
{Schruba} A. et~al., 2019 \emph{\apj}, \emph{883}, 1, 2.

\bibitem[\protect\astroncite{\emph{{Schuller} et~al.}}{2009}]{Schuller2009}
{Schuller} F. et~al., 2009 \emph{\aap}, \emph{504}, 2, 415.

\bibitem[\protect\astroncite{\emph{{Schultheis} et~al.}}{2021}]{Schultheis2021}
{Schultheis} M. et~al., 2021 \emph{\aap}, \emph{650}, A191.

\bibitem[\protect\astroncite{\emph{{Schw{\"o}rer} et~al.}}{2019}]{Schworer2019}
{Schw{\"o}rer} A. et~al., 2019 \emph{\aap}, \emph{628}, A6.

\bibitem[\protect\astroncite{\emph{{Scoville}}}{1972}]{Scoville1972}
{Scoville} N.~Z., 1972 \emph{\apjl}, \emph{175}, L127.

\bibitem[\protect\astroncite{\emph{{Scoville} et~al.}}{1974}]{Scoville1974}
{Scoville} N.~Z. et~al., 1974 \emph{\apjl}, \emph{187}, L63.

\bibitem[\protect\astroncite{\emph{{Sellwood} and
  {Balbus}}}{1999}]{Sellwood1999}
{Sellwood} J.~A. and {Balbus} S.~A., 1999 \emph{\apj}, \emph{511}, 2, 660.

\bibitem[\protect\astroncite{\emph{{Sellwood} and
  {Wilkinson}}}{1993}]{Sellwood1993}
{Sellwood} J.~A. and {Wilkinson} A., 1993 \emph{Reports on Progress in
  Physics}, \emph{56}, 2, 173.

\bibitem[\protect\astroncite{\emph{{Seo} et~al.}}{2019}]{Seo2019}
{Seo} W.-Y. et~al., 2019 \emph{\apj}, \emph{872}, 1, 5.

\bibitem[\protect\astroncite{\emph{{Serabyn} et~al.}}{1992}]{Serabyn1992}
{Serabyn} E. et~al., 1992 \emph{\apj}, \emph{395}, 166.

\bibitem[\protect\astroncite{\emph{{Shetty} et~al.}}{2012}]{Shetty2012}
{Shetty} R. et~al., 2012 \emph{\mnras}, \emph{425}, 1, 720.

\bibitem[\protect\astroncite{\emph{{Shlosman} et~al.}}{1989}]{Shlosman1989}
{Shlosman} I. et~al., 1989 \emph{\nat}, \emph{338}, 6210, 45.

\bibitem[\protect\astroncite{\emph{{Shlosman} et~al.}}{1990}]{Shlosman1990}
{Shlosman} I. et~al., 1990 \emph{\nat}, \emph{345}, 6277, 679.

\bibitem[\protect\astroncite{\emph{{Simpson} et~al.}}{2018}]{Simpson2018a}
{Simpson} J.~P. et~al., 2018 \emph{\apjl}, \emph{867}, 1, L13.

\bibitem[\protect\astroncite{\emph{{Simpson} et~al.}}{2021}]{Simpson2021}
{Simpson} J.~P. et~al., 2021 \emph{\apj}, \emph{910}, 1, 59.

\bibitem[\protect\astroncite{\emph{{Sjouwerman} and
  {Pihlstr{\"o}m}}}{2008}]{Sjouwerman2008}
{Sjouwerman} L.~O. and {Pihlstr{\"o}m} Y.~M., 2008 \emph{\apj}, \emph{681}, 2,
  1287.

\bibitem[\protect\astroncite{\emph{{Sofue}}}{1995{\natexlab{a}}}]{Sofue1995a}
{Sofue} Y., 1995{\natexlab{a}} \emph{\pasj}, \emph{47}, 527.

\bibitem[\protect\astroncite{\emph{{Sofue}}}{1995{\natexlab{b}}}]{sofue1995b}
{Sofue} Y., 1995{\natexlab{b}} \emph{\pasj}, \emph{47}, 551.

\bibitem[\protect\astroncite{\emph{{Sofue}}}{2000}]{Sofue2000}
{Sofue} Y., 2000 \emph{\apj}, \emph{540}, 1, 224.

\bibitem[\protect\astroncite{\emph{{Sofue}}}{2017{\natexlab{a}}}]{Sofue2017b}
{Sofue} Y., 2017{\natexlab{a}} \emph{\mnras}, \emph{469}, 2, 1647.

\bibitem[\protect\astroncite{\emph{{Sofue}}}{2017{\natexlab{b}}}]{Sofue2017a}
{Sofue} Y., 2017{\natexlab{b}} \emph{\mnras}, \emph{470}, 2, 1982.

\bibitem[\protect\astroncite{\emph{{Sofue}}}{2020}]{Sofue2020}
{Sofue} Y., 2020 \emph{\pasj}, \emph{72}, 2, L4.

\bibitem[\protect\astroncite{\emph{{Sofue} and {Handa}}}{1984}]{Sofue1984}
{Sofue} Y. and {Handa} T., 1984 \emph{\nat}, \emph{310}, 5978, 568.

\bibitem[\protect\astroncite{\emph{{Sofue} and {Nakanishi}}}{2017}]{Sofue2017c}
{Sofue} Y. and {Nakanishi} H., 2017 \emph{\pasj}, \emph{69}, 2, 19.

\bibitem[\protect\astroncite{\emph{{Sofue} et~al.}}{1987}]{Sofue1987}
{Sofue} Y. et~al., 1987 \emph{\pasj}, \emph{39}, 95.

\bibitem[\protect\astroncite{\emph{{Sofue} et~al.}}{2016}]{Sofue2016}
{Sofue} Y. et~al., 2016 \emph{\mnras}, \emph{459}, 1, 108.

\bibitem[\protect\astroncite{\emph{{Sormani} and
  {Barnes}}}{2019}]{Sormani2019b}
{Sormani} M.~C. and {Barnes} A.~T., 2019 \emph{\mnras}, \emph{484}, 1, 1213.

\bibitem[\protect\astroncite{\emph{{Sormani} and {Li}}}{2020}]{Sormani2020c}
{Sormani} M.~C. and {Li} Z., 2020 \emph{\mnras}, \emph{494}, 4, 6030.

\bibitem[\protect\astroncite{\emph{{Sormani}
  et~al.}}{2015{\natexlab{a}}}]{Sormani2015c}
{Sormani} M.~C. et~al., 2015{\natexlab{a}} \emph{\mnras}, \emph{449}, 3, 2421.

\bibitem[\protect\astroncite{\emph{{Sormani}
  et~al.}}{2015{\natexlab{b}}}]{Sormani2015b}
{Sormani} M.~C. et~al., 2015{\natexlab{b}} \emph{\mnras}, \emph{451}, 4, 3437.

\bibitem[\protect\astroncite{\emph{{Sormani}
  et~al.}}{2015{\natexlab{c}}}]{Sormani2015a}
{Sormani} M.~C. et~al., 2015{\natexlab{c}} \emph{\mnras}, \emph{454}, 2, 1818.

\bibitem[\protect\astroncite{\emph{{Sormani}
  et~al.}}{2018{\natexlab{a}}}]{Sormani2018a}
{Sormani} M.~C. et~al., 2018{\natexlab{a}} \emph{\mnras}, \emph{481}, 1, 2.

\bibitem[\protect\astroncite{\emph{{Sormani}
  et~al.}}{2018{\natexlab{b}}}]{Sormani2018b}
{Sormani} M.~C. et~al., 2018{\natexlab{b}} \emph{\mnras}, \emph{475}, 2, 2383.

\bibitem[\protect\astroncite{\emph{{Sormani} et~al.}}{2019}]{Sormani2019a}
{Sormani} M.~C. et~al., 2019 \emph{\mnras}, \emph{488}, 4, 4663.

\bibitem[\protect\astroncite{\emph{{Sormani}
  et~al.}}{2020{\natexlab{a}}}]{Sormani2020a}
{Sormani} M.~C. et~al., 2020{\natexlab{a}} \emph{\mnras}, \emph{499}, 1, 7.

\bibitem[\protect\astroncite{\emph{{Sormani}
  et~al.}}{2020{\natexlab{b}}}]{Sormani2020b}
{Sormani} M.~C. et~al., 2020{\natexlab{b}} \emph{\mnras}, \emph{497}, 4, 5024.

\bibitem[\protect\astroncite{\emph{{Sormani} et~al.}}{2021}]{Sormani2021}
{Sormani} M.~C. et~al., 2021 \emph{arXiv e-prints}, arXiv:2111.12713.

\bibitem[\protect\astroncite{\emph{{Spilker} et~al.}}{2021}]{Spilker2021}
{Spilker} A. et~al., 2021 \emph{\aap}, \emph{653}, A63.

\bibitem[\protect\astroncite{\emph{{Staguhn} et~al.}}{2019}]{Staguhn2019}
{Staguhn} J. et~al., 2019 \emph{\apj}, \emph{885}, 1, 72.

\bibitem[\protect\astroncite{\emph{{Stark} and {Bania}}}{1986}]{Stark1986}
{Stark} A.~A. and {Bania} T.~M., 1986 \emph{\apjl}, \emph{306}, L17.

\bibitem[\protect\astroncite{\emph{{Stark} et~al.}}{2004}]{Stark2004}
{Stark} A.~A. et~al., 2004 \emph{\apjl}, \emph{614}, 1, L41.

\bibitem[\protect\astroncite{\emph{{Steinke} et~al.}}{2016}]{Steinke2016}
{Steinke} M. et~al., 2016 \emph{\aap}, \emph{588}, A9.

\bibitem[\protect\astroncite{\emph{{Stolte} et~al.}}{2010}]{Stolte2010}
{Stolte} A. et~al., 2010 \emph{\apj}, \emph{718}, 2, 810.

\bibitem[\protect\astroncite{\emph{{Stolte} et~al.}}{2015}]{Stolte2015}
{Stolte} A. et~al., 2015 \emph{\aap}, \emph{578}, A4.

\bibitem[\protect\astroncite{\emph{{Strickland} and
  {Stevens}}}{2000}]{Strickland2000}
{Strickland} D.~K. and {Stevens} I.~R., 2000 \emph{\mnras}, \emph{314}, 3, 511.

\bibitem[\protect\astroncite{\emph{{Su} et~al.}}{2010}]{Su2010}
{Su} M. et~al., 2010 \emph{\apj}, \emph{724}, 2, 1044.

\bibitem[\protect\astroncite{\emph{{Sun} et~al.}}{2020}]{Sun2020b}
{Sun} J. et~al., 2020 \emph{\apj}, \emph{892}, 2, 148.

\bibitem[\protect\astroncite{\emph{{Swinbank} et~al.}}{2015}]{Swinbank2015}
{Swinbank} A.~M. et~al., 2015 \emph{\apjl}, \emph{806}, 1, L17.

\bibitem[\protect\astroncite{\emph{{Tacconi} et~al.}}{2020}]{Tacconi2020}
{Tacconi} L.~J. et~al., 2020 \emph{\araa}, \emph{58}, 157.

\bibitem[\protect\astroncite{\emph{{Takekawa} et~al.}}{2017}]{Takekawa2017b}
{Takekawa} S. et~al., 2017 \emph{\apj}, \emph{834}, 2, 121.

\bibitem[\protect\astroncite{\emph{{Tanaka} et~al.}}{2011}]{Tanaka2011}
{Tanaka} K. et~al., 2011 \emph{\apjl}, \emph{743}, 2, L39.

\bibitem[\protect\astroncite{\emph{{Tanaka} et~al.}}{2014}]{Tanaka2014}
{Tanaka} K. et~al., 2014 \emph{\apj}, \emph{783}, 1, 62.

\bibitem[\protect\astroncite{\emph{{Tanaka} et~al.}}{2015}]{Tanaka2015}
{Tanaka} K. et~al., 2015 \emph{\apj}, \emph{806}, 1, 130.

\bibitem[\protect\astroncite{\emph{{Tanaka} et~al.}}{2018}]{Tanaka2018}
{Tanaka} K. et~al., 2018 \emph{\apjs}, \emph{236}, 2, 40.

\bibitem[\protect\astroncite{\emph{{Tanaka} et~al.}}{2020}]{Tanaka2020}
{Tanaka} K. et~al., 2020 \emph{\apj}, \emph{903}, 2, 111.

\bibitem[\protect\astroncite{\emph{{Tanaka} et~al.}}{2021}]{Tanaka2021}
{Tanaka} K. et~al., 2021 \emph{\apj}, \emph{915}, 2, 79.

\bibitem[\protect\astroncite{\emph{{Tang}
  et~al.}}{2021{\natexlab{a}}}]{Tang2021a}
{Tang} Y. et~al., 2021{\natexlab{a}} \emph{\mnras}, \emph{505}, 2, 2392.

\bibitem[\protect\astroncite{\emph{{Tang}
  et~al.}}{2021{\natexlab{b}}}]{Tang2021b}
{Tang} Y. et~al., 2021{\natexlab{b}} \emph{\mnras}, \emph{505}, 2, 2377.

\bibitem[\protect\astroncite{\emph{{Terrier} et~al.}}{2018}]{Terrier2018}
{Terrier} R. et~al., 2018 \emph{\aap}, \emph{612}, A102.

\bibitem[\protect\astroncite{\emph{{Thomas} et~al.}}{2020}]{Thomas2020}
{Thomas} T. et~al., 2020 \emph{\apjl}, \emph{890}, 2, L18.

\bibitem[\protect\astroncite{\emph{{Tokuyama} et~al.}}{2019}]{Tokuyama2019}
{Tokuyama} S. et~al., 2019 \emph{\pasj}, \emph{71}, S19.

\bibitem[\protect\astroncite{\emph{{Torii} et~al.}}{2010}]{Torii2010}
{Torii} K. et~al., 2010 \emph{\pasj}, \emph{62}, 1307.

\bibitem[\protect\astroncite{\emph{{Torrey} et~al.}}{2017}]{Torrey2017}
{Torrey} P. et~al., 2017 \emph{\mnras}, \emph{467}, 2, 2301.

\bibitem[\protect\astroncite{\emph{{Tress} et~al.}}{2020}]{Tress2020}
{Tress} R.~G. et~al., 2020 \emph{\mnras}, \emph{499}, 3, 4455.

\bibitem[\protect\astroncite{\emph{{Trujillo-Gomez}
  et~al.}}{2019}]{Trujillo-Gomez2019}
{Trujillo-Gomez} S. et~al., 2019 \emph{\mnras}, \emph{488}, 3, 3972.

\bibitem[\protect\astroncite{\emph{{Tsuboi} et~al.}}{1999}]{Tsuboi1999}
{Tsuboi} M. et~al., 1999 \emph{\apjs}, \emph{120}, 1, 1.

\bibitem[\protect\astroncite{\emph{{Tsuboi}
  et~al.}}{2015{\natexlab{a}}}]{Tsuboi2015a}
{Tsuboi} M. et~al., 2015{\natexlab{a}} \emph{\pasj}, \emph{67}, 6, 109.

\bibitem[\protect\astroncite{\emph{{Tsuboi}
  et~al.}}{2015{\natexlab{b}}}]{Tsuboi2015b}
{Tsuboi} M. et~al., 2015{\natexlab{b}} \emph{\pasj}, \emph{67}, 5, 90.

\bibitem[\protect\astroncite{\emph{{Tsuboi} et~al.}}{2018}]{Tsuboi2018}
{Tsuboi} M. et~al., 2018 \emph{\pasj}, \emph{70}, 5, 85.

\bibitem[\protect\astroncite{\emph{{Tsuboi} et~al.}}{2019}]{Tsuboi2019}
{Tsuboi} M. et~al., 2019 \emph{\pasj}, \emph{71}, 6, 128.

\bibitem[\protect\astroncite{\emph{{Tsuboi} et~al.}}{2021}]{Tsuboi2021}
{Tsuboi} M. et~al., 2021 \emph{\pasj}, \emph{73}, S91.

\bibitem[\protect\astroncite{\emph{{Tsujimoto} et~al.}}{2018}]{Tsujimoto2018}
{Tsujimoto} S. et~al., 2018 \emph{\apj}, \emph{856}, 2, 91.

\bibitem[\protect\astroncite{\emph{{Uchida} et~al.}}{1994}]{Uchida1994}
{Uchida} K.~I. et~al., 1994 \emph{\apj}, \emph{421}, 505.

\bibitem[\protect\astroncite{\emph{{Uchiyama} et~al.}}{2013}]{Uchiyama2013}
{Uchiyama} H. et~al., 2013 \emph{\pasj}, \emph{65}, 19.

\bibitem[\protect\astroncite{\emph{{Uehara} et~al.}}{2019}]{Uehara2019}
{Uehara} K. et~al., 2019 \emph{\apj}, \emph{872}, 2, 121.

\bibitem[\protect\astroncite{\emph{{Usero} et~al.}}{2015}]{Usero2015}
{Usero} A. et~al., 2015 \emph{\aj}, \emph{150}, 4, 115.

\bibitem[\protect\astroncite{\emph{{Valenti} et~al.}}{2016}]{Valenti2016}
{Valenti} E. et~al., 2016 \emph{\aap}, \emph{587}, L6.

\bibitem[\protect\astroncite{\emph{{van Albada} and
  {Sanders}}}{1982}]{VanAlbada1982}
{van Albada} T.~S. and {Sanders} R.~H., 1982 \emph{\mnras}, \emph{201}, 303.

\bibitem[\protect\astroncite{\emph{{van de Ven} and
  {Fathi}}}{2010}]{vandeVen2010}
{van de Ven} G. and {Fathi} K., 2010 \emph{\apj}, \emph{723}, 1, 767.

\bibitem[\protect\astroncite{\emph{{van Dokkum} et~al.}}{2010}]{vanDokkum2010}
{van Dokkum} P.~G. et~al., 2010 \emph{\apj}, \emph{709}, 2, 1018.

\bibitem[\protect\astroncite{\emph{{V{\'a}zquez-Semadeni}
  et~al.}}{2019}]{Vazquez-Semadeni2019}
{V{\'a}zquez-Semadeni} E. et~al., 2019 \emph{\mnras}, \emph{490}, 3, 3061.

\bibitem[\protect\astroncite{\emph{{Veilleux} et~al.}}{2020}]{Veilleux2020}
{Veilleux} S. et~al., 2020 \emph{\aapr}, \emph{28}, 1, 2.

\bibitem[\protect\astroncite{\emph{{VERITAS Collaboration}
  et~al.}}{2009}]{VERITASCollaboration2009}
{VERITAS Collaboration} et~al., 2009 \emph{\nat}, \emph{462}, 7274, 770.

\bibitem[\protect\astroncite{\emph{{Viti} et~al.}}{2014}]{Viti2014}
{Viti} S. et~al., 2014 \emph{\aap}, \emph{570}, A28.

\bibitem[\protect\astroncite{\emph{{Wakker} and {van
  Woerden}}}{1997}]{Wakker1997}
{Wakker} B.~P. and {van Woerden} H., 1997 \emph{\araa}, \emph{35}, 217.

\bibitem[\protect\astroncite{\emph{{Walker} et~al.}}{2015}]{Walker2015}
{Walker} D.~L. et~al., 2015 \emph{\mnras}, \emph{449}, 1, 715.

\bibitem[\protect\astroncite{\emph{{Walker} et~al.}}{2016}]{Walker2016}
{Walker} D.~L. et~al., 2016 \emph{\mnras}, \emph{457}, 4, 4536.

\bibitem[\protect\astroncite{\emph{{Walker} et~al.}}{2018}]{Walker2018}
{Walker} D.~L. et~al., 2018 \emph{\mnras}, \emph{474}, 2, 2373.

\bibitem[\protect\astroncite{\emph{{Walker} et~al.}}{2021}]{Walker2021}
{Walker} D.~L. et~al., 2021 \emph{\mnras}, \emph{503}, 1, 77.

\bibitem[\protect\astroncite{\emph{{Walls} et~al.}}{2016}]{Walls2016}
{Walls} M. et~al., 2016 \emph{\mnras}, \emph{463}, 3, 2893.

\bibitem[\protect\astroncite{\emph{{Walsh} et~al.}}{2008}]{Walsh2008}
{Walsh} A.~J. et~al., 2008 \emph{\pasa}, \emph{25}, 2, 105.

\bibitem[\protect\astroncite{\emph{{Walsh} et~al.}}{2011}]{Walsh2011}
{Walsh} A.~J. et~al., 2011 \emph{\mnras}, \emph{416}, 3, 1764.

\bibitem[\protect\astroncite{\emph{{Wang} et~al.}}{2002}]{Wang2002}
{Wang} Q.~D. et~al., 2002 \emph{\nat}, \emph{415}, 6868, 148.

\bibitem[\protect\astroncite{\emph{{Ward-Thompson}
  et~al.}}{2007}]{Ward-Thompson2007}
{Ward-Thompson} D. et~al., 2007 \emph{Protostars and Planets V} (B.~{Reipurth},
  D.~{Jewitt}, and K.~{Keil}), p.~33.

\bibitem[\protect\astroncite{\emph{{Wegg} and {Gerhard}}}{2013}]{Wegg2013}
{Wegg} C. and {Gerhard} O., 2013 \emph{\mnras}, \emph{435}, 3, 1874.

\bibitem[\protect\astroncite{\emph{{Winter} et~al.}}{2018}]{Winter2018}
{Winter} A.~J. et~al., 2018 \emph{\mnras}, \emph{478}, 2, 2700.

\bibitem[\protect\astroncite{\emph{{Winter} et~al.}}{2020}]{Winter2020}
{Winter} A.~J. et~al., 2020 \emph{\mnras}, \emph{491}, 1, 903.

\bibitem[\protect\astroncite{\emph{{Yalinewich} and
  {Beniamini}}}{2018}]{Yalinewich2018}
{Yalinewich} A. and {Beniamini} P., 2018 \emph{\aap}, \emph{612}, L9.

\bibitem[\protect\astroncite{\emph{{Yan} et~al.}}{2017}]{Yan2017}
{Yan} Q.-Z. et~al., 2017 \emph{\mnras}, \emph{471}, 3, 2523.

\bibitem[\protect\astroncite{\emph{{Yang} et~al.}}{2022}]{Yang2022}
{Yang} H. Y.~K. et~al., 2022 \emph{arXiv e-prints}, arXiv:2203.02526.

\bibitem[\protect\astroncite{\emph{{Yoast-Hull}
  et~al.}}{2014{\natexlab{a}}}]{Yoast-Hull2014b}
{Yoast-Hull} T. et~al., 2014{\natexlab{a}} \emph{The Galactic Center: Feeding
  and Feedback in a Normal Galactic Nucleus}, vol. 303 (L.~O. {Sjouwerman},
  C.~C. {Lang}, and J.~{Ott}), pp. 153--155.

\bibitem[\protect\astroncite{\emph{{Yoast-Hull}
  et~al.}}{2014{\natexlab{b}}}]{Yoast-Hull2014a}
{Yoast-Hull} T.~M. et~al., 2014{\natexlab{b}} \emph{\apj}, \emph{790}, 2, 86.

\bibitem[\protect\astroncite{\emph{{Yoast-Hull} et~al.}}{2017}]{Yoast-Hull2017}
{Yoast-Hull} T.~M. et~al., 2017 \emph{\mnras}, \emph{469}, 1, L89.

\bibitem[\protect\astroncite{\emph{{Yu} et~al.}}{2018}]{Yu2018}
{Yu} N.-P. et~al., 2018 \emph{Research in Astronomy and Astrophysics},
  \emph{18}, 2, 015.

\bibitem[\protect\astroncite{\emph{{Yusef-Zadeh} and
  {Wardle}}}{2019}]{Yusef-Zadeh2019}
{Yusef-Zadeh} F. and {Wardle} M., 2019 \emph{\mnras}, \emph{490}, 1, L1.

\bibitem[\protect\astroncite{\emph{{Yusef-Zadeh}
  et~al.}}{1999}]{Yusef-Zadeh1999a}
{Yusef-Zadeh} F. et~al., 1999 \emph{\apj}, \emph{527}, 1, 172.

\bibitem[\protect\astroncite{\emph{{Yusef-Zadeh}
  et~al.}}{2009}]{Yusef-Zadeh2009}
{Yusef-Zadeh} F. et~al., 2009 \emph{\apj}, \emph{702}, 1, 178.

\bibitem[\protect\astroncite{\emph{{Yusef-Zadeh}
  et~al.}}{2013}]{Yusef-Zadeh2013a}
{Yusef-Zadeh} F. et~al., 2013 \emph{\apjl}, \emph{767}, 2, L32.

\bibitem[\protect\astroncite{\emph{{Yusef-Zadeh}
  et~al.}}{2015}]{Yusef-Zadeh2015b}
{Yusef-Zadeh} F. et~al., 2015 \emph{\apj}, \emph{808}, 1, 97.

\bibitem[\protect\astroncite{\emph{{Yusef-Zadeh}
  et~al.}}{2022}]{Yusef-Zadeh2022}
{Yusef-Zadeh} F. et~al., 2022 \emph{\apjl}, \emph{925}, 2, L18.

\bibitem[\protect\astroncite{\emph{{Zeng} et~al.}}{2020}]{Zeng2020}
{Zeng} S. et~al., 2020 \emph{\mnras}, \emph{497}, 4, 4896.

\bibitem[\protect\astroncite{\emph{{Zhang} et~al.}}{2020}]{Zhang2020}
{Zhang} S. et~al., 2020 \emph{\apj}, \emph{893}, 1, 3.

\bibitem[\protect\astroncite{\emph{{Zoccali} et~al.}}{2021}]{Zoccali2021}
{Zoccali} M. et~al., 2021 \emph{\mnras}, \emph{502}, 1, 1246.

\bibitem[\protect\astroncite{\emph{{Zubovas} et~al.}}{2011}]{Zubovas2011}
{Zubovas} K. et~al., 2011 \emph{\mnras}, \emph{415}, 1, L21.

\bibitem[\protect\astroncite{\emph{{Zylka} et~al.}}{1990}]{Zylka1990}
{Zylka} R. et~al., 1990 \emph{\aap}, \emph{234}, 133.

\end{thebibliography}
\end{document}